\documentclass{amsart}
\usepackage{graphicx}
\theoremstyle{definition}

\theoremstyle{remark}

\numberwithin{equation}{section}

\newcommand{\A}{\mathcal{A}}
\begin{document}

\title{Multiset Estimates and Combinatorial Synthesis
 }
\author{Mark Sh. Levin}%

 \address{
 Mail address: Mark Sh. Levin, Sumskoy Proezd 5-1-103, Moscow 117208, Russia;
 Http: //www.mslevin.iitp.ru/}%
\email{mslevin@acm.org}%

 %

\begin{abstract}
 The paper addresses an approach to
 ordinal assessment of alternatives based on
 assignment of elements into an ordinal scale.
%
 Basic versions of the
 assessment problems are formulated while
 taking into account
 the number of levels at a basic ordinal scale [1,2,...,l]
 and the number of assigned elements (e.g., 1,2,3).
 The obtained estimates are multisets (or bags)
 (cardinality of the multiset equals a constant).
 Scale-posets for the examined assessment problems
 are presented.
 ``Interval multiset estimates'' are suggested.
%
%
%
 Further, operations over multiset estimates
%
 are examined:
 (a) integration of multiset estimates,
 (b) proximity for multiset estimates,
 (c) comparison of multiset estimates,
 (d) aggregation of multiset estimates, and
%
 (e) alignment of multiset estimates.
%
 Combinatorial synthesis based on morphological approach
 is examined including
 the
 modified version of the approach with multiset estimates of
 design alternatives.
%
 Knapsack-like problems
 with multiset estimates are briefly described as well.
 The assessment approach,
 multiset-estimates, and corresponding combinatorial problems
 are illustrated by numerical examples.
%
%
%
\end{abstract}

\maketitle


\newcounter{cms}
\setlength{\unitlength}{1mm}

\section{Introduction}

 In this article,
 a combinatorial approach to
 ordinal assessment of alternatives is suggested.
 The approach consists in assignment of elements (\(1,2,3,...\))
 into an ordinal scale \([1,2,...,l]\).
%
%
 As a result, a multi-set based estimate is obtained,
 where a basis set involves all levels of the ordinal scale:
 \(\Omega = \{ 1,2,...,l\}\) (the levels are linear ordered:
 \(1 \succ 2 \succ 3 \succ ...\)) and
 the assessment problem (for each alternative)
 consists in selection of a multiset over set \(A\) while taking into
 account two conditions:

 {\it 1.} cardinality of the selected multiset equals a specified
 number of elements \( \eta = 1,2,3,...\)
 (i.e., multisets of cardinality \(\eta \) are considered);

 {\it 2.} ``configuration'' of the multiset is the following:
 the selected elements of \(\Omega\) cover an interval over scale \([1,l]\)
 (i.e., ``interval multiset estimate'').

 Note, fundamentals of multisets can be found
 in
 (\cite{alk11},\cite{knuth98},\cite{syr01},\cite{yager86}).
 Evidently, the assessment case \(\eta = 1\)
 corresponds to traditional ordinal assessment.
 Thus, an estimate \(e\) for an alternative \(A\) is
 (scale \([1,l]\), position-based form or position form):
 \(e(\A) = (\eta_{1},...,\eta_{\iota},...,\eta_{l})\),
%
%
 where \(\eta_{\iota}\) corresponds to the number of elements at the
 level \(\iota\) (\(\iota = \overline{1,l}\)).
 Here, the conditions above are:

 {\it Condition 1}:
 \(\sum_{\iota=1}^{l} \eta_{\iota} = \eta\)
 (or \(\left|  e(A)  \right|  = \eta\)).

 {\it Condition 2}:
 if \(\eta_{\iota} > 0\) and \(\eta_{\iota+2} > 0\)  then
 \(\eta_{\iota+1} > 0\)  (\(\iota = \overline{1,l-2}\)).

 On the other hand, the multiset estimate is:
 \[e(A) = \{ \overbrace{1,...,1}^{\eta_{1}},\overbrace{2,...2}^{\eta_{2}},
 \overbrace{3,...,3}^{\eta_{3}},...,\overbrace{l,...,l}^{\eta_{l}} \}.\]
%
%
%
%
%
 The number of multisets of cardinality \(\eta\),
 with elements taken from a finite set of cardinality \(l\),
 is called the
 ``multiset coefficient'' or ``multiset number''
  (\cite{knuth98},\cite{syr01},\cite{yager86}):
 \[ \mu^{l,\eta} = \left( \left(  \begin{matrix}  l \\
   \eta \end{matrix} \right) \right) = \frac{l(l+1)(l+2)... (l+\eta-1) } {\eta!} =
   \left(  \begin{matrix}  l+\eta-1 \\  \eta \end{matrix} \right) .  \]
%
%
%
%
 This number corresponds to possible estimates
 (without taking into account interval condition 2).
 In the case of condition 2,
 the number of estimates is decreased.
%
%
%
%

~~

  {\bf Example 1.} The ordinal assessment
  is the following:
  (a) the basic ordinal scale (basic set) is: \(\Omega = \{1,2,3,...,l\}\),
  (b) the number of elements or cardinality of multiset (estimate) is:
  \(\eta = 1\).
  Estimates are:

  \(\widehat{e}^{o}_{1} = \{1\}\) and position-based form \(\widehat{e}^{o}_{1} = (1,0,0,...,0)\),

  \(\widehat{e}^{o}_{2} = \{2\}\) and position-based form \(\widehat{e}^{o}_{2} = (0,1,0,...,0)\),

  \(\widehat{e}^{o}_{3} = \{3\}\) and position-based form \(\widehat{e}^{o}_{3} = (0,0,1,...,0)\),

~~~~~~~~~~~~~~  ... ,

  \(\widehat{e}^{o}_{l} = \{l\}\) and position-based form  \(\widehat{e}^{o}_{l} = (0,0,0,...,1)\).

~~

 {\bf Example 2.}
 The basic ordinal scale (basic set) is: \(\Omega = \{1,2,3,4\}\),
 the number of elements or cardinality of multiset (estimate) is:
  \(\eta = 3\).

 {\it Case 1.} Estimates corresponding to basic ordinal assessment are:
 \(e^{o}_{1} = \{1,1,1\}\) or in position-based form
 \(e^{o}_{1} = (3,0,0,0)\);
 \(e^{o}_{2} = \{2,2,2\}\) or in position-based form
 \(e^{o}_{2} = (0,3,0,0)\);
 \(e^{o}_{3} = \{3,3,3\}\) or in position-based form
 \(e^{o}_{3} = (0,0,3,0)\);
 \(e^{o}_{4} = \{4,4,4\}\) or in position-based form
 \(e^{o}_{4} = (0,0,0,3)\).

 {\it Case 2.} Examples of
 correct estimates are:
 \(e'_{1} = \{1,2,3\}\) or in position-based form
 \(e'_{1} = (1,1,1,0)\);
 \(e'_{2} = \{2,2,3\}\) or in position-based form
 \(e'_{2} = (0,2,1,0)\);
 \(e'_{3} = \{4,4,4\}\) or in position-based form
 \(e'_{3} = (0,0,0,3)\).

 {\it Case 3.}  Examples of incorrect estimates are:
 \(e''_{1} = \{1,1,3\}\) or in position-based form
 \(e''_{1} = (2,0,1,0)\);
 \(e''_{2} = \{1,3,4\}\) or in position-based form
 \(e''_{2} = (1,0,1,1)\);
 \(e''_{3} = \{2,4,4\}\) or in position-based form
 \(e''_{3} = (0,1,0,2)\).

~~

 Basic versions of the assessment problems are formulated
 as ~\( P^{l,\eta} \)~:
 (i) traditional assessment based on ordinal scale \([1,2,3]\):~
 \(P^{3,1}\);
 (ii) traditional assessment based on ordinal scale \([1,2,3,4]\):~
 \(P^{4,1}\);
 (iii)  assessment over ordinal scale \([1,2,3]\)
 based on assignment of two elements:~ \(P^{3,2}\);
 (iv)  assessment over ordinal scale \([1,2,3]\)
 based on assignment of three elements:~ \(P^{3,2}\);
 and
 (v) assessment over ordinal scale \([1,2,3]\)
 based on assignment of four elements:~ \(P^{3,4}\).
%
%
%
 In the article, the obtained scale-posets
  are presented and corresponding
 alternative evaluation and composition problems are described.
%
%
%
 In the case  \(\eta \geq 2\),
 the suggested assessment approach can be considered as a very
 simplified discrete version of fuzzy set based assessment
(e.g., \cite{zadeh65}, \cite{zim87}).
 Fig. 1 depicts a framework of our assessment approach.

 Further, operations over multiset estimates
%
 are examined:
 (a) integration of multiset estimates,
 (b) proximity for multiset estimates,
 (c) comparison of multiset estimates,
 (d) aggregation of multiset estimates
 (e.g., searching for a median estimate),
 (e) alignment of multiset estimates
 (and corresponding assessment problems).
 Combinatorial synthesis based on morphological approach
 is examined including
 the suggested modified version of the approach with multiset estimates of
 design alternatives.
 In addition, some knapsack-like problems
 with multiset estimates are briefly described as well.
%
%
%
%
%
%
 The assessment approach, multiset-estimates and the problems above
 are illustrated by numerical examples.

\begin{center}
\begin{picture}(95.5,57)
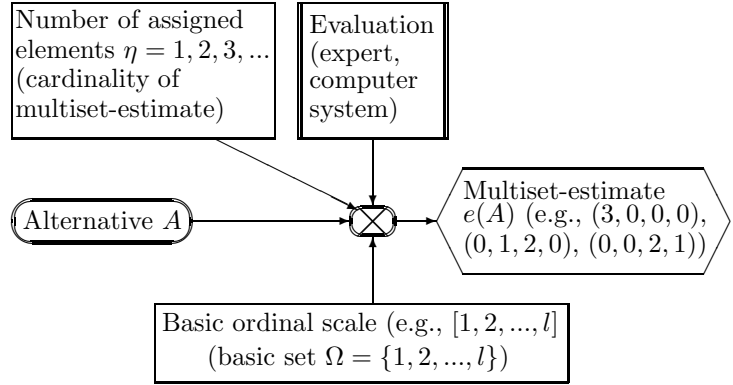
\put(13.4,00){\makebox(0,0)[bl]{Fig. 1. Framework of
   assessment approach}}


\put(12,27){\oval(24,6)} \put(12,27){\oval(23.3,5.3)}

\put(01.5,26){\makebox(0,0)[bl]{Alternative \(A\)}}


\put(24,27){\vector(1,0){021}}


\put(00,38){\line(1,0){35}} \put(00,56){\line(1,0){35}}
\put(00,38){\line(0,1){18}} \put(35,38){\line(0,1){18}}


\put(0.5,52){\makebox(0,0)[bl]{Number of assigned }}
\put(0.5,48){\makebox(0,0)[bl]{elements \(\eta = 1,2,3,...\)}}
\put(0.5,44){\makebox(0,0)[bl]{(cardinality of}}
\put(0.5,40){\makebox(0,0)[bl]{multiset-estimate)}}

\put(27,38){\vector(2,-1){19}}


\put(38,38){\line(1,0){20}} \put(38,56){\line(1,0){20}}
\put(38,38){\line(0,1){18}} \put(58,38){\line(0,1){18}}

\put(38.5,38){\line(0,1){18}} \put(57.5,38){\line(0,1){18}}

\put(39.6,52){\makebox(0,0)[bl]{Evaluation}}
\put(39.6,47.5){\makebox(0,0)[bl]{(expert,}}
\put(39.6,44){\makebox(0,0)[bl]{computer}}
\put(39.6,40){\makebox(0,0)[bl]{system)}}

\put(48,38){\vector(0,-1){09}}



\put(60,20){\line(1,0){32}} \put(60,34){\line(1,0){32}}


\put(56.5,27){\line(1,2){3.5}} \put(56.5,27){\line(1,-2){3.5}}
\put(95.5,27){\line(-1,2){3.5}} \put(95.5,27){\line(-1,-2){3.5}}

\put(60,30){\makebox(0,0)[bl]{Multiset-estimate }}
\put(60,26){\makebox(0,0)[bl]{\(e(A)\) (e.g., \((3,0,0,0)\),}}
\put(60,22){\makebox(0,0)[bl]{\((0,1,2,0)\), \((0,0,2,1)\))}}


\put(48,27){\oval(06,4)} \put(48,27){\oval(5.5,3.5)}


\put(51,27){\vector(1,0){05.5}}

\put(46,25){\line(1,1){4}}

\put(50,25){\line(-1,1){4}}


\put(48,16){\vector(0,1){9}}

\put(19,05.5){\line(1,0){55}} \put(19,16){\line(1,0){55}}
\put(19,05.5){\line(0,1){10.5}} \put(74,05.5){\line(0,1){10.5}}

\put(20,11.5){\makebox(0,0)[bl]{Basic ordinal scale (e.g.,
\([1,2,...,l]\)}}

 \put(26,07){\makebox(0,0)[bl]{(basic set \(\Omega = \{1,2,...,l\}\))}}

\end{picture}
\end{center}

\section{Basic Assessment Problems}

 In this section, several basic assessment problems are considered
 (\(P^{3,1}\), \(P^{4,1}\), \(P^{3,2}\), \(P^{3,3}\), \(P^{3,4}\))
 (Table 1):
 assessment scale,
 order over the scale components.
%

\begin{center}
\begin{picture}(121,52)
\put(033,47){\makebox(0,0)[bl] {Table 1. Basic assessment
problems}}


\put(00,00){\line(1,0){121}} \put(00,23){\line(1,0){121}}
\put(00,45){\line(1,0){121}}

\put(00,00){\line(0,1){45}} \put(05,00){\line(0,1){45}}
\put(20,00){\line(0,1){45}} \put(39,00){\line(0,1){45}}
\put(53,00){\line(0,1){45}} \put(72,00){\line(0,1){45}}
\put(87,00){\line(0,1){45}} \put(102,00){\line(0,1){45}}

\put(121,00){\line(0,1){45}}

\put(02,31){\makebox(0,0)[bl]{ }}

\put(06,41){\makebox(0,0)[bl]{Assess-}}
\put(06,37){\makebox(0,0)[bl]{ment}}
\put(06,33){\makebox(0,0)[bl]{problem}}

\put(20.5,41){\makebox(0,0)[bl]{Number of}}
\put(20.5,37){\makebox(0,0)[bl]{elements}}
\put(20.5,32.7){\makebox(0,0)[bl]{(cardinality}}
\put(20.5,29){\makebox(0,0)[bl]{of multiset)}}
\put(28.5,25){\makebox(0,0)[bl]{\(\eta\) }}

\put(40,41){\makebox(0,0)[bl]{Number }}
\put(40,37){\makebox(0,0)[bl]{of levels }}
\put(40,33){\makebox(0,0)[bl]{of basic }}
\put(40,29){\makebox(0,0)[bl]{ordinal}}
\put(40,25){\makebox(0,0)[bl]{scale \( l\) }}

\put(54,40.5){\makebox(0,0)[bl]{Type of }}
\put(54,37){\makebox(0,0)[bl]{scale}}

\put(73,40.5){\makebox(0,0)[bl]{Type of }}
\put(73,37){\makebox(0,0)[bl]{estimate}}

\put(88,41){\makebox(0,0)[bl]{Multiset}}
\put(88,37){\makebox(0,0)[bl]{coeffici-}}
\put(88,33){\makebox(0,0)[bl]{ent}}
\put(88,28.5){\makebox(0,0)[bl]{(or \(l\))}}

\put(103,41){\makebox(0,0)[bl]{Number of}}
\put(103,37){\makebox(0,0)[bl]{multiset}}
\put(103,33){\makebox(0,0)[bl]{estimates}}
\put(103,28.7){\makebox(0,0)[bl]{(under con-}}
\put(103,25){\makebox(0,0)[bl]{dition 2)}}


\put(01.5,18){\makebox(0,0)[bl]{\(1\)}}

\put(09.7,18){\makebox(0,0)[bl]{\(P^{31}\)}}

\put(28.5,18){\makebox(0,0)[bl]{\(1\)}}

\put(45,18){\makebox(0,0)[bl]{\(3\)}}

\put(53.5,18){\makebox(0,0)[bl]{linear order}}

\put(73,18){\makebox(0,0)[bl]{ordinal}}

\put(93.5,18){\makebox(0,0)[bl]{\(3\)}}

\put(110,18){\makebox(0,0)[bl]{\(3\)}}


\put(01.5,14){\makebox(0,0)[bl]{\(2\)}}

\put(09.7,14){\makebox(0,0)[bl]{\(P^{41}\)}}

\put(28.5,14){\makebox(0,0)[bl]{\(1\)}}

\put(45,14){\makebox(0,0)[bl]{\(4\)}}

\put(53.5,14){\makebox(0,0)[bl]{linear order}}

\put(73,14){\makebox(0,0)[bl]{ordinal}}

\put(93.5,14){\makebox(0,0)[bl]{\(4\)}}

\put(110,14){\makebox(0,0)[bl]{\(4\)}}


\put(01.5,10){\makebox(0,0)[bl]{\(3\)}}

\put(09.7,10){\makebox(0,0)[bl]{\(P^{32}\)}}

\put(28.5,10){\makebox(0,0)[bl]{\(2\)}}

\put(45,10){\makebox(0,0)[bl]{\(3\)}}

\put(53.5,10){\makebox(0,0)[bl]{linear order}}

\put(73,10){\makebox(0,0)[bl]{multiset}}

\put(93.5,10){\makebox(0,0)[bl]{\(6\)}}

\put(110,10){\makebox(0,0)[bl]{\(5\)}}


\put(01.5,06){\makebox(0,0)[bl]{\(4\)}}

\put(09.7,06){\makebox(0,0)[bl]{\(P^{33}\)}}

\put(28.5,06){\makebox(0,0)[bl]{\(3\)}}

\put(45,06){\makebox(0,0)[bl]{\(3\)}}

\put(53.5,06){\makebox(0,0)[bl]{poset}}

\put(73,06){\makebox(0,0)[bl]{multiset}}

\put(92.5,06){\makebox(0,0)[bl]{\(10\)}}

\put(110,06){\makebox(0,0)[bl]{\(8\)}}


\put(01.5,02){\makebox(0,0)[bl]{\(5\)}}

\put(09.7,02){\makebox(0,0)[bl]{\(P^{34}\)}}

\put(28.5,02){\makebox(0,0)[bl]{\(4\)}}

\put(45,02){\makebox(0,0)[bl]{\(3\)}}

\put(53.5,02){\makebox(0,0)[bl]{poset}}

\put(73,02){\makebox(0,0)[bl]{multiset}}

\put(92.5,02){\makebox(0,0)[bl]{\(15\)}}

\put(109,02){\makebox(0,0)[bl]{\(12\)}}

\end{picture}
\end{center}

%

 Fig. 2 illustrates
 the scale and estimates for problem
 \(P^{3,1}\) (ordinal assessment, scale \([1,3]\)).
 In the case of scale \([1,2,3]\),
 the following semantic levels are often considered:
 excellent (\(1\)),
 good (\(2\)), and
 sufficient (\(3\)).
%
%
 Analogically,
 Fig. 3 illustrates
  the scale and estimates for problem
 \(P^{4,1}\) (ordinal assessment, scale \([1,4]\)).

\begin{center}
\begin{picture}(57,42)
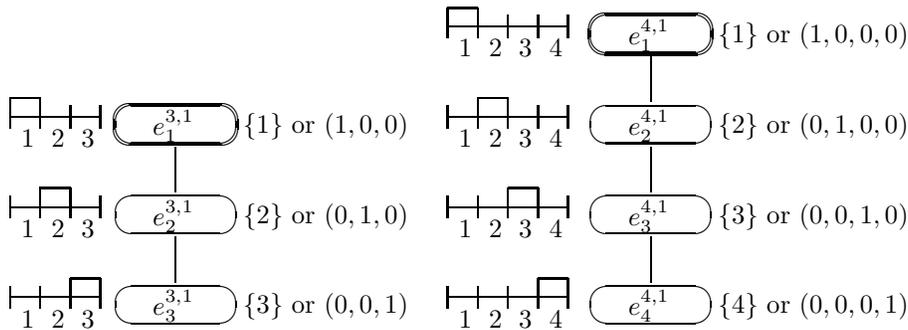
\put(03,00){\makebox(0,0)[bl] {Fig. 2. Scale, estimates
 (\(P^{3,1}\))}}



\put(31,31){\makebox(0,0)[bl]{\(\{1\}\) or \((1,0,0)\) }}

\put(19,30.7){\makebox(0,0)[bl]{\(e^{3,1}_{1}\) }}

\put(22,33){\oval(16,5)} \put(22,33){\oval(16.5,5.5)}


\put(00,34){\line(0,1){02.5}}\put(04,34){\line(0,1){02.5}}
\put(00,36.5){\line(1,0){4}}

\put(00,34){\line(1,0){12}}

\put(00,32.5){\line(0,1){3}} \put(04,32.5){\line(0,1){3}}
\put(08,32.5){\line(0,1){3}} \put(12,32.5){\line(0,1){3}}

\put(01.5,30){\makebox(0,0)[bl]{\(1\)}}
\put(05.5,30){\makebox(0,0)[bl]{\(2\)}}
\put(09.5,30){\makebox(0,0)[bl]{\(3\)}}


\put(22,24){\line(0,1){6}}

\put(31,19){\makebox(0,0)[bl]{\(\{2\}\) or \((0,1,0)\) }}

\put(19,18.7){\makebox(0,0)[bl]{\(e^{3,1}_{2}\) }}

\put(22,21){\oval(16,5)}


\put(04,22){\line(0,1){02.5}}\put(08,22){\line(0,1){02.5}}
\put(04,24.5){\line(1,0){4}}

\put(00,22){\line(1,0){12}}

\put(00,20.5){\line(0,1){3}} \put(04,20.5){\line(0,1){3}}
\put(08,20.5){\line(0,1){3}} \put(12,20.5){\line(0,1){3}}

\put(01.5,18){\makebox(0,0)[bl]{\(1\)}}
\put(05.5,18){\makebox(0,0)[bl]{\(2\)}}
\put(09.5,18){\makebox(0,0)[bl]{\(3\)}}


\put(22,12){\line(0,1){6}}


\put(31,7){\makebox(0,0)[bl]{\(\{3\}\) or \((0,0,1)\) }}

\put(19,06.7){\makebox(0,0)[bl]{\(e^{3,1}_{3}\) }}

\put(22,09){\oval(16,5)}


\put(08,10){\line(0,1){02.5}}\put(12,10){\line(0,1){02.5}}
\put(08,12.5){\line(1,0){4}}

\put(00,10){\line(1,0){12}}

\put(00,08.5){\line(0,1){3}} \put(04,08.5){\line(0,1){3}}
\put(08,08.5){\line(0,1){3}} \put(12,08.5){\line(0,1){3}}

\put(01.5,06){\makebox(0,0)[bl]{\(1\)}}
\put(05.5,06){\makebox(0,0)[bl]{\(2\)}}
\put(09.5,06){\makebox(0,0)[bl]{\(3\)}}

\end{picture}
%
%
\begin{picture}(62,54)
\put(06,00){\makebox(0,0)[bl] {Fig. 3. Scale, estimates
 (\(P^{4,1}\))}}





\put(24,42.7){\makebox(0,0)[bl]{\(e^{4,1}_{1}\) }}

\put(27,45){\oval(16,5)} \put(27,45){\oval(16.5,5.5)}


\put(36,43){\makebox(0,0)[bl]{\(\{1\}\) or \((1,0,0,0)\) }}


\put(0,46){\line(0,1){02.5}}\put(04,46){\line(0,1){02.5}}
\put(0,48.5){\line(1,0){4}}

\put(00,46){\line(1,0){16}}

\put(00,44.5){\line(0,1){3}} \put(04,44.5){\line(0,1){3}}
\put(08,44.5){\line(0,1){3}} \put(12,44.5){\line(0,1){3}}
\put(16,44.5){\line(0,1){3}}

\put(01.5,42){\makebox(0,0)[bl]{\(1\)}}
\put(05.5,42){\makebox(0,0)[bl]{\(2\)}}
\put(09.5,42){\makebox(0,0)[bl]{\(3\)}}
\put(13.5,42){\makebox(0,0)[bl]{\(4\)}}



\put(27,36){\line(0,1){6}}

\put(24,30.7){\makebox(0,0)[bl]{\(e^{4,1}_{2}\) }}

\put(27,33){\oval(16,5)}


\put(36,31){\makebox(0,0)[bl]{\(\{2\}\) or \((0,1,0,0)\) }}


\put(04,34){\line(0,1){02.5}}\put(08,34){\line(0,1){02.5}}
\put(04,36.5){\line(1,0){4}}

\put(00,34){\line(1,0){16}}

\put(00,32.5){\line(0,1){3}} \put(04,32.5){\line(0,1){3}}
\put(08,32.5){\line(0,1){3}} \put(12,32.5){\line(0,1){3}}
\put(16,32.5){\line(0,1){3}}

\put(01.5,30){\makebox(0,0)[bl]{\(1\)}}
\put(05.5,30){\makebox(0,0)[bl]{\(2\)}}
\put(09.5,30){\makebox(0,0)[bl]{\(3\)}}
\put(13.5,30){\makebox(0,0)[bl]{\(4\)}}




\put(27,24){\line(0,1){6}}

\put(24,18.7){\makebox(0,0)[bl]{\(e^{4,1}_{3}\) }}

\put(27,21){\oval(16,5)}


\put(36,19){\makebox(0,0)[bl]{\(\{3\}\) or \((0,0,1,0)\) }}


\put(08,22){\line(0,1){02.5}}\put(12,22){\line(0,1){02.5}}
\put(08,24.5){\line(1,0){4}}

\put(00,22){\line(1,0){16}}

\put(00,20.5){\line(0,1){3}} \put(04,20.5){\line(0,1){3}}
\put(08,20.5){\line(0,1){3}} \put(12,20.5){\line(0,1){3}}
\put(16,20.5){\line(0,1){3}}

\put(01.5,18){\makebox(0,0)[bl]{\(1\)}}
\put(05.5,18){\makebox(0,0)[bl]{\(2\)}}
\put(09.5,18){\makebox(0,0)[bl]{\(3\)}}
\put(13.5,18){\makebox(0,0)[bl]{\(4\)}}



\put(27,12){\line(0,1){6}}


\put(24,06.7){\makebox(0,0)[bl]{\(e^{4,1}_{4}\) }}

\put(27,09){\oval(16,5)}


\put(36,7){\makebox(0,0)[bl]{\(\{4\}\) or \((0,0,0,1)\) }}


\put(12,10){\line(0,1){02.5}}\put(16,10){\line(0,1){02.5}}
\put(12,12.5){\line(1,0){4}}

\put(00,10){\line(1,0){16}}

\put(00,08.5){\line(0,1){3}} \put(04,08.5){\line(0,1){3}}
\put(08,08.5){\line(0,1){3}} \put(12,08.5){\line(0,1){3}}
\put(16,08.5){\line(0,1){3}}

\put(01.5,06){\makebox(0,0)[bl]{\(1\)}}
\put(05.5,06){\makebox(0,0)[bl]{\(2\)}}
\put(09.5,06){\makebox(0,0)[bl]{\(3\)}}
\put(13.5,06){\makebox(0,0)[bl]{\(4\)}}


\end{picture}
\end{center}


 Fig. 4 illustrates
 the scale-poset and estimates for problem
 \(P^{3,2}\) (assessment over scale \([1,3]\) with two elements;
 estimate  \((1,0,1)\) is not used).

\begin{center}
\begin{picture}(66,66)
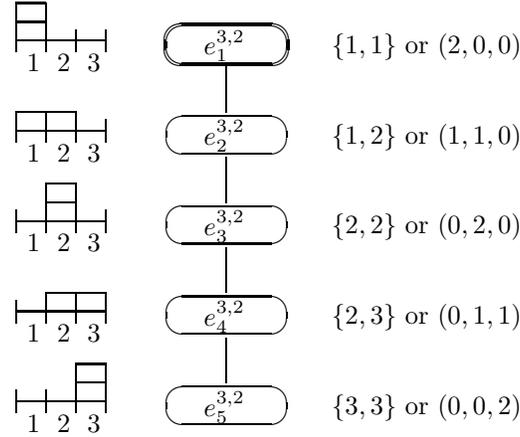
\put(09,00){\makebox(0,0)[bl] {Fig. 4. Scale, estimates
  (\(P^{3,2}\))}}




\put(25,54.75){\makebox(0,0)[bl]{\(e^{3,2}_{1}\) }}

\put(28,57){\oval(16,5)}  \put(28,57){\oval(16.5,5.5)}


\put(42,55){\makebox(0,0)[bl]{\(\{1,1\}\) or \((2,0,0)\) }}

\put(0,57.5){\line(0,1){05}} \put(04,57.5){\line(0,1){05}}
\put(0,60){\line(1,0){04}} \put(0,62.5){\line(1,0){4}}

\put(00,57.5){\line(1,0){12}}

\put(00,56){\line(0,1){3}} \put(04,56){\line(0,1){3}}
\put(08,56){\line(0,1){3}} \put(12,56){\line(0,1){3}}

\put(01.5,53.5){\makebox(0,0)[bl]{\(1\)}}
\put(05.5,53.5){\makebox(0,0)[bl]{\(2\)}}
\put(09.5,53.5){\makebox(0,0)[bl]{\(3\)}}


\put(28,48){\line(0,1){6}}

\put(25,42.7){\makebox(0,0)[bl]{\(e^{3,2}_{2}\) }}

\put(28,45){\oval(16,5)}


\put(42,43){\makebox(0,0)[bl]{\(\{1,2\}\) or \((1,1,0)\) }}

\put(0,45.5){\line(0,1){02.5}} \put(04,45.5){\line(0,1){02.5}}
\put(08,45.5){\line(0,1){02.5}}

\put(00,48){\line(1,0){8}}

\put(00,45.5){\line(1,0){12}}

\put(00,44){\line(0,1){3}} \put(04,44){\line(0,1){3}}
\put(08,44){\line(0,1){3}} \put(12,44){\line(0,1){3}}

\put(01.5,41.5){\makebox(0,0)[bl]{\(1\)}}
\put(05.5,41.5){\makebox(0,0)[bl]{\(2\)}}
\put(09.5,41.5){\makebox(0,0)[bl]{\(3\)}}


\put(28,36){\line(0,1){6}}

\put(25,30.7){\makebox(0,0)[bl]{\(e^{3,2}_{3}\) }}

\put(28,33){\oval(16,5)}


\put(42,31){\makebox(0,0)[bl]{\(\{2,2\}\) or \((0,2,0)\) }}

\put(04,33.5){\line(0,1){05}} \put(08,33.5){\line(0,1){05}}
\put(04,36){\line(1,0){04}} \put(04,38.5){\line(1,0){4}}

\put(00,33.5){\line(1,0){12}}

\put(00,32){\line(0,1){3}} \put(04,32){\line(0,1){3}}
\put(08,32){\line(0,1){3}} \put(12,32){\line(0,1){3}}

\put(01.5,29.5){\makebox(0,0)[bl]{\(1\)}}
\put(05.5,29.5){\makebox(0,0)[bl]{\(2\)}}
\put(09.5,29.5){\makebox(0,0)[bl]{\(3\)}}


\put(28,24){\line(0,1){6}}

\put(25,18.7){\makebox(0,0)[bl]{\(e^{3,2}_{4}\) }}

\put(28,21){\oval(16,5)}


\put(42,19){\makebox(0,0)[bl]{\(\{2,3\}\) or \((0,1,1)\) }}

\put(04,21.5){\line(0,1){02.5}} \put(08,21.5){\line(0,1){02.5}}
\put(12,21.5){\line(0,1){02.5}}

\put(04,24){\line(1,0){8}}

\put(00,21.5){\line(1,0){12}}

\put(00,20){\line(0,1){3}} \put(04,20){\line(0,1){3}}
\put(08,20){\line(0,1){3}} \put(12,20){\line(0,1){3}}

\put(01.5,17.5){\makebox(0,0)[bl]{\(1\)}}
\put(05.5,17.5){\makebox(0,0)[bl]{\(2\)}}
\put(09.5,17.5){\makebox(0,0)[bl]{\(3\)}}


\put(28,12){\line(0,1){6}}


\put(25,06.7){\makebox(0,0)[bl]{\(e^{3,2}_{5}\) }}

\put(28,09){\oval(16,5)}

\put(42,7){\makebox(0,0)[bl]{\(\{3,3\}\) or \((0,0,2)\) }}

\put(08,9.5){\line(0,1){05}} \put(12,9.5){\line(0,1){05}}
\put(08,12){\line(1,0){4}} \put(08,14.5){\line(1,0){4}}

\put(00,9.5){\line(1,0){12}}

\put(00,08){\line(0,1){3}} \put(04,08){\line(0,1){3}}
\put(08,08){\line(0,1){3}} \put(12,08){\line(0,1){3}}

\put(01.5,5.5){\makebox(0,0)[bl]{\(1\)}}
\put(05.5,5.5){\makebox(0,0)[bl]{\(2\)}}
\put(09.5,5.5){\makebox(0,0)[bl]{\(3\)}}


\end{picture}
\end{center}


 Fig. 5 illustrates
  the scale-poset and estimates for problem
 \(P^{3,3}\) (assessment over scale \([1,3]\) with three elements;
 estimates  \((2,0,1)\) and \((1,0,2)\) are not used).

\begin{center}
\begin{picture}(80,91)
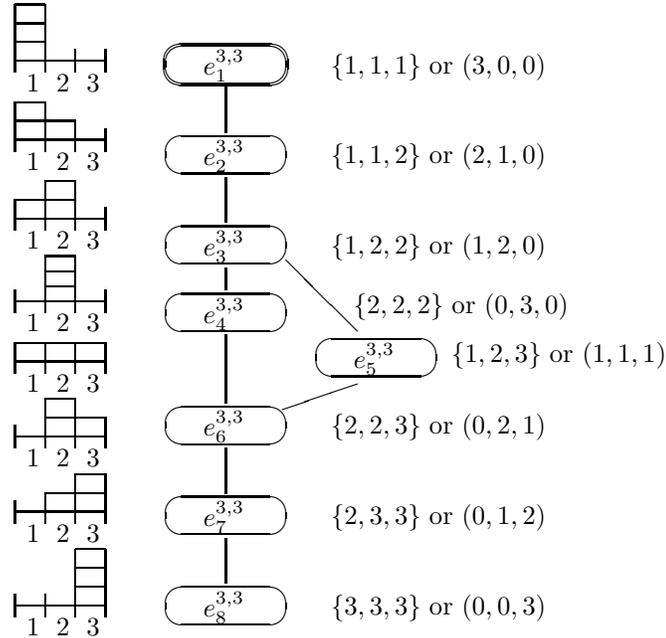
\put(17,00){\makebox(0,0)[bl] {Fig. 5. Scale, estimates
 (\(P^{3,3}\))}}


\put(25,78.7){\makebox(0,0)[bl]{\(e^{3,3}_{1}\) }}

\put(28,81){\oval(16,5)} \put(28,81){\oval(16.5,5.5)}


\put(42,79){\makebox(0,0)[bl]{\(\{1,1,1\}\) or \((3,0,0)\) }}

\put(00,81.5){\line(0,1){07.5}} \put(04,81.5){\line(0,1){07.5}}

\put(00,84){\line(1,0){04}} \put(00,86.5){\line(1,0){04}}
\put(00,89){\line(1,0){4}}

\put(00,81.5){\line(1,0){12}}

\put(00,80){\line(0,1){3}} \put(04,80){\line(0,1){3}}
\put(08,80){\line(0,1){3}} \put(12,80){\line(0,1){3}}

\put(01.5,77.5){\makebox(0,0)[bl]{\(1\)}}
\put(05.5,77.5){\makebox(0,0)[bl]{\(2\)}}
\put(09.5,77.5){\makebox(0,0)[bl]{\(3\)}}


\put(28,72){\line(0,1){6}}


\put(25,66.7){\makebox(0,0)[bl]{\(e^{3,3}_{2}\) }}

\put(28,69){\oval(16,5)}


\put(42,67){\makebox(0,0)[bl]{\(\{1,1,2\}\) or \((2,1,0)\) }}

\put(00,71){\line(0,1){05}} \put(04,71){\line(0,1){05}}
\put(8,71){\line(0,1){02.5}}

\put(00,73.5){\line(1,0){8}} \put(00,76){\line(1,0){4}}

\put(00,71){\line(1,0){12}}

\put(00,69.5){\line(0,1){3}} \put(04,69.5){\line(0,1){3}}
\put(08,69.5){\line(0,1){3}} \put(12,69.5){\line(0,1){3}}

\put(01.5,67){\makebox(0,0)[bl]{\(1\)}}
\put(05.5,67){\makebox(0,0)[bl]{\(2\)}}
\put(09.5,67){\makebox(0,0)[bl]{\(3\)}}


\put(28,60){\line(0,1){6}}

\put(25,54.7){\makebox(0,0)[bl]{\(e^{3,3}_{3}\) }}

\put(28,57){\oval(16,5)}


\put(42,55){\makebox(0,0)[bl]{\(\{1,2,2\}\) or \((1,2,0)\) }}

\put(00,60.5){\line(0,1){02.5}} \put(04,60.5){\line(0,1){05}}
\put(8,60.5){\line(0,1){05}}

\put(00,63){\line(1,0){8}} \put(04,65.5){\line(1,0){4}}

\put(00,60.5){\line(1,0){12}}

\put(00,59){\line(0,1){3}} \put(04,59){\line(0,1){3}}
\put(08,59){\line(0,1){3}} \put(12,59){\line(0,1){3}}

\put(01.5,56.5){\makebox(0,0)[bl]{\(1\)}}
\put(05.5,56.5){\makebox(0,0)[bl]{\(2\)}}
\put(09.5,56.5){\makebox(0,0)[bl]{\(3\)}}


\put(28,51){\line(0,1){3}}

\put(25,45.7){\makebox(0,0)[bl]{\(e^{3,3}_{4}\) }}

\put(28,48){\oval(16,5)}


\put(45,47){\makebox(0,0)[bl]{\(\{2,2,2\}\) or \((0,3,0)\) }}

\put(04,49.5){\line(0,1){06}} \put(08,49.5){\line(0,1){06}}

\put(04,51.5){\line(1,0){04}} \put(04,53.5){\line(1,0){04}}
\put(04,55.5){\line(1,0){04}}

\put(00,49.5){\line(1,0){12}}

\put(00,48){\line(0,1){3}} \put(04,48){\line(0,1){3}}
\put(08,48){\line(0,1){3}} \put(12,48){\line(0,1){3}}

\put(01.5,45.5){\makebox(0,0)[bl]{\(1\)}}
\put(05.5,45.5){\makebox(0,0)[bl]{\(2\)}}
\put(09.5,45.5){\makebox(0,0)[bl]{\(3\)}}


\put(28,36){\line(0,1){9}}

\put(25,30.7){\makebox(0,0)[bl]{\(e^{3,3}_{6}\) }}

\put(28,33){\oval(16,5)}


\put(42,31){\makebox(0,0)[bl]{\(\{2,2,3\}\) or \((0,2,1)\) }}

\put(04,31.5){\line(0,1){05}} \put(08,31.5){\line(0,1){05}}
\put(12,31.5){\line(0,1){02.5}}

\put(04,34){\line(1,0){8}} \put(04,36.5){\line(1,0){4}}

\put(00,31.5){\line(1,0){12}}

\put(00,30){\line(0,1){3}} \put(04,30){\line(0,1){3}}
\put(08,30){\line(0,1){3}} \put(12,30){\line(0,1){3}}

\put(01.5,27.5){\makebox(0,0)[bl]{\(1\)}}
\put(05.5,27.5){\makebox(0,0)[bl]{\(2\)}}
\put(09.5,27.5){\makebox(0,0)[bl]{\(3\)}}


\put(28,24){\line(0,1){6}}

\put(25,18.7){\makebox(0,0)[bl]{\(e^{3,3}_{7}\) }}

\put(28,21){\oval(16,5)}


\put(42,19){\makebox(0,0)[bl]{\(\{2,3,3\}\) or \((0,1,2)\) }}

\put(04,21.5){\line(0,1){02.5}} \put(08,21.5){\line(0,1){05}}
\put(12,21.5){\line(0,1){05}}

\put(04,24){\line(1,0){8}} \put(08,26.5){\line(1,0){4}}

\put(00,21.5){\line(1,0){12}}

\put(00,20){\line(0,1){3}} \put(04,20){\line(0,1){3}}
\put(08,20){\line(0,1){3}} \put(12,20){\line(0,1){3}}

\put(01.5,17.5){\makebox(0,0)[bl]{\(1\)}}
\put(05.5,17.5){\makebox(0,0)[bl]{\(2\)}}
\put(09.5,17.5){\makebox(0,0)[bl]{\(3\)}}


\put(28,12){\line(0,1){6}}


\put(25,06.7){\makebox(0,0)[bl]{\(e^{3,3}_{8}\) }}

\put(28,09){\oval(16,5)}


\put(42,7){\makebox(0,0)[bl]{\(\{3,3,3\}\) or \((0,0,3)\) }}

\put(08,9){\line(0,1){07.5}} \put(12,9){\line(0,1){07.5}}
\put(08,11.5){\line(1,0){4}} \put(08,14){\line(1,0){4}}
\put(08,16.5){\line(1,0){4}}

\put(00,9){\line(1,0){12}}

\put(00,07.5){\line(0,1){3}} \put(04,07.5){\line(0,1){3}}
\put(08,07.5){\line(0,1){3}} \put(12,07.5){\line(0,1){3}}

\put(01.5,5){\makebox(0,0)[bl]{\(1\)}}
\put(05.5,5){\makebox(0,0)[bl]{\(2\)}}
\put(09.5,5){\makebox(0,0)[bl]{\(3\)}}


\put(45.5,45.5){\line(-1,1){09.5}}

\put(45.5,38.5){\line(-3,-1){10}}


\put(45,39.7){\makebox(0,0)[bl]{\(e^{3,3}_{5}\) }}

\put(48,42){\oval(16,5)}


\put(58,40.5){\makebox(0,0)[bl]{\(\{1,2,3\}\) or \((1,1,1)\) }}

\put(00,41.5){\line(0,1){02.5}} \put(04,41.5){\line(0,1){02.5}}
\put(8,41.5){\line(0,1){02.5}} \put(12,41.5){\line(0,1){02.5}}

\put(00,44){\line(1,0){12}}

\put(00,41.5){\line(1,0){12}}

\put(00,40){\line(0,1){3}} \put(04,40){\line(0,1){3}}
\put(08,40){\line(0,1){3}} \put(12,40){\line(0,1){3}}

\put(01.5,37.5){\makebox(0,0)[bl]{\(1\)}}
\put(05.5,37.5){\makebox(0,0)[bl]{\(2\)}}
\put(09.5,37.5){\makebox(0,0)[bl]{\(3\)}}


\end{picture}
\end{center}


 Fig. 6 illustrates
  the scale-poset and estimates for problem
 \(P^{3,4}\) (assessment over scale \([1,3]\) with four elements;
 estimates  \((2,0,2)\), \((3,0,1)\), and \((1,0,3)\) are not used).

\begin{center}
\begin{picture}(82,138)
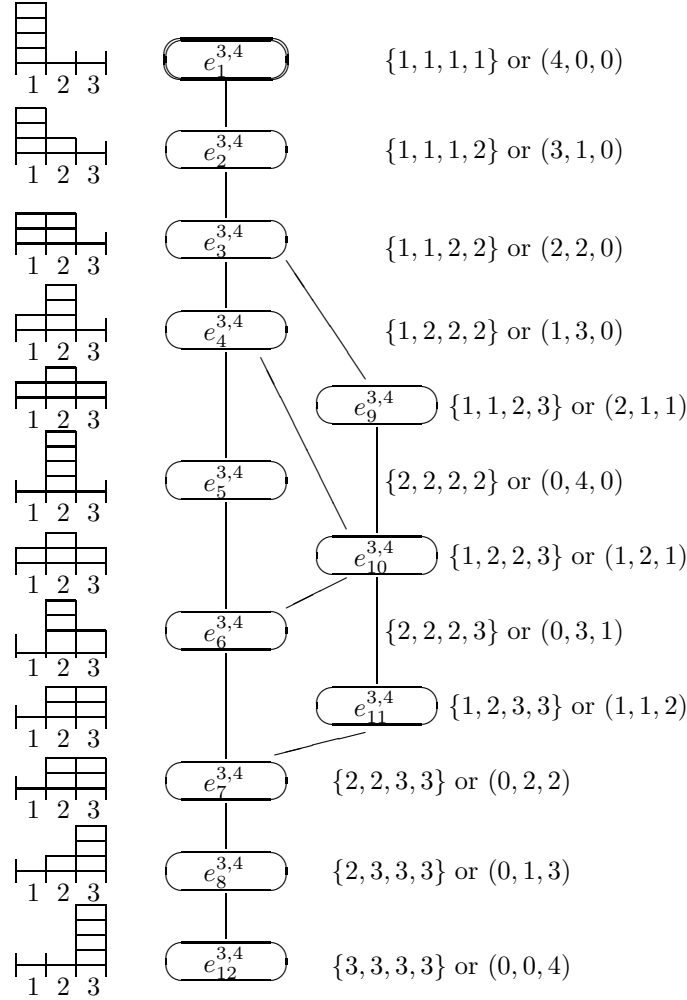
\put(019,00){\makebox(0,0)[bl] {Fig. 6. Scale, estimates
 (\(P^{3,4}\))}}




\put(25,126.7){\makebox(0,0)[bl]{\(e^{3,4}_{1}\) }}

\put(28,129){\oval(16,5)} \put(28,129){\oval(16.5,5.5)}


\put(49,127){\makebox(0,0)[bl]{\(\{1,1,1,1\}\) or \((4,0,0)\) }}

\put(00,128.5){\line(0,1){08}} \put(04,128.5){\line(0,1){08}}

\put(00,130.5){\line(1,0){4}} \put(00,132.5){\line(1,0){4}}
\put(00,134.5){\line(1,0){4}} \put(00,136.5){\line(1,0){4}}

\put(00,128.5){\line(1,0){12}}

\put(00,127){\line(0,1){3}} \put(04,127){\line(0,1){3}}
\put(08,127){\line(0,1){3}} \put(12,127){\line(0,1){3}}

\put(01.5,124.5){\makebox(0,0)[bl]{\(1\)}}
\put(05.5,124.5){\makebox(0,0)[bl]{\(2\)}}
\put(09.5,124.5){\makebox(0,0)[bl]{\(3\)}}


\put(28,120){\line(0,1){6}}


\put(25,114.7){\makebox(0,0)[bl]{\(e^{3,4}_{2}\) }}

\put(28,117){\oval(16,5)}


\put(49,115){\makebox(0,0)[bl]{\(\{1,1,1,2\}\) or \((3,1,0)\) }}

\put(00,116.5){\line(0,1){06}} \put(04,116.5){\line(0,1){06}}
\put(08,116.5){\line(0,1){02}}

\put(00,118.5){\line(1,0){8}} \put(00,120.5){\line(1,0){4}}
\put(00,122.5){\line(1,0){4}}

\put(00,116.5){\line(1,0){12}}

\put(00,115){\line(0,1){3}} \put(04,115){\line(0,1){3}}
\put(08,115){\line(0,1){3}} \put(12,115){\line(0,1){3}}

\put(01.5,112.5){\makebox(0,0)[bl]{\(1\)}}
\put(05.5,112.5){\makebox(0,0)[bl]{\(2\)}}
\put(09.5,112.5){\makebox(0,0)[bl]{\(3\)}}


\put(28,108){\line(0,1){6}}


\put(25,102.7){\makebox(0,0)[bl]{\(e^{3,4}_{3}\) }}

\put(28,105){\oval(16,5)}


\put(49,102){\makebox(0,0)[bl]{\(\{1,1,2,2\}\) or \((2,2,0)\) }}

\put(00,104.5){\line(0,1){04}}

\put(04,104.5){\line(0,1){04}} \put(08,104.5){\line(0,1){04}}
\put(00,106.5){\line(1,0){8}} \put(00,108.5){\line(1,0){8}}

\put(00,104.5){\line(1,0){12}}

\put(00,103){\line(0,1){3}} \put(04,103){\line(0,1){3}}
\put(08,103){\line(0,1){3}} \put(12,103){\line(0,1){3}}

\put(01.5,100.5){\makebox(0,0)[bl]{\(1\)}}
\put(05.5,100.5){\makebox(0,0)[bl]{\(2\)}}
\put(09.5,100.5){\makebox(0,0)[bl]{\(3\)}}


\put(28,96){\line(0,1){6}}


\put(25,90.7){\makebox(0,0)[bl]{\(e^{3,4}_{4}\) }}

\put(28,93){\oval(16,5)}


\put(49,91){\makebox(0,0)[bl]{\(\{1,2,2,2\}\) or \((1,3,0)\) }}

\put(00,93){\line(0,1){02}}

\put(04,93){\line(0,1){06}} \put(08,93){\line(0,1){06}}

\put(00,95){\line(1,0){8}} \put(04,97){\line(1,0){4}}
\put(04,99){\line(1,0){4}}

\put(00,93){\line(1,0){12}}

\put(00,91.5){\line(0,1){3}} \put(04,91.5){\line(0,1){3}}
\put(08,91.5){\line(0,1){3}} \put(12,91.5){\line(0,1){3}}

\put(01.5,89){\makebox(0,0)[bl]{\(1\)}}
\put(05.5,89){\makebox(0,0)[bl]{\(2\)}}
\put(09.5,89){\makebox(0,0)[bl]{\(3\)}}


\put(28,76){\line(0,1){14}}

\put(25,70.7){\makebox(0,0)[bl]{\(e^{3,4}_{5}\) }}

\put(28,73){\oval(16,5)}


\put(49,71){\makebox(0,0)[bl]{\(\{2,2,2,2\}\) or \((0,4,0)\) }}

\put(04,71.5){\line(0,1){08}} \put(08,71.5){\line(0,1){08}}

\put(04,73.5){\line(1,0){4}} \put(04,75.5){\line(1,0){4}}
\put(04,77.5){\line(1,0){4}} \put(04,79.5){\line(1,0){4}}

\put(00,71.5){\line(1,0){12}}

\put(00,70){\line(0,1){3}} \put(04,70){\line(0,1){3}}
\put(08,70){\line(0,1){3}} \put(12,70){\line(0,1){3}}

\put(01.5,67.5){\makebox(0,0)[bl]{\(1\)}}
\put(05.5,67.5){\makebox(0,0)[bl]{\(2\)}}
\put(09.5,67.5){\makebox(0,0)[bl]{\(3\)}}


\put(28,56){\line(0,1){14}}

\put(25,50.7){\makebox(0,0)[bl]{\(e^{3,4}_{6}\) }}

\put(28,53){\oval(16,5)}


\put(49,51){\makebox(0,0)[bl]{\(\{2,2,2,3\}\) or \((0,3,1)\) }}

\put(04,51){\line(0,1){06}} \put(08,51){\line(0,1){06}}
 \put(12,51){\line(0,1){02}}

\put(04,53){\line(1,0){8}} \put(04,55){\line(1,0){4}}
\put(04,57){\line(1,0){4}}

\put(00,50){\line(1,0){12}}

\put(00,49.5){\line(0,1){3}} \put(04,49.5){\line(0,1){3}}
\put(08,49.5){\line(0,1){3}} \put(12,49.5){\line(0,1){3}}

\put(01.5,47){\makebox(0,0)[bl]{\(1\)}}
\put(05.5,47){\makebox(0,0)[bl]{\(2\)}}
\put(09.5,47){\makebox(0,0)[bl]{\(3\)}}


\put(28,36){\line(0,1){14}}

\put(25,30.7){\makebox(0,0)[bl]{\(e^{3,4}_{7}\) }}

\put(28,33){\oval(16,5)}


\put(42,31){\makebox(0,0)[bl]{\(\{2,2,3,3\}\) or \((0,2,2)\) }}

\put(04,32){\line(0,1){04}}

\put(08,32){\line(0,1){04}} \put(12,32){\line(0,1){04}}
\put(04,34){\line(1,0){8}} \put(04,36){\line(1,0){8}}

\put(00,32){\line(1,0){12}}

\put(00,30.5){\line(0,1){3}} \put(04,30.5){\line(0,1){3}}
\put(08,30.5){\line(0,1){3}} \put(12,30.5){\line(0,1){3}}

\put(01.5,28){\makebox(0,0)[bl]{\(1\)}}
\put(05.5,28){\makebox(0,0)[bl]{\(2\)}}
\put(09.5,28){\makebox(0,0)[bl]{\(3\)}}


\put(28,24){\line(0,1){6}}

\put(25,18.7){\makebox(0,0)[bl]{\(e^{3,4}_{8}\) }}

\put(28,21){\oval(16,5)}


\put(42,19){\makebox(0,0)[bl]{\(\{2,3,3,3\}\) or \((0,1,3)\) }}

\put(04,21){\line(0,1){02}}

\put(08,21){\line(0,1){06}} \put(12,21){\line(0,1){06}}

\put(04,23){\line(1,0){8}} \put(08,25){\line(1,0){4}}
\put(08,27){\line(1,0){4}}

\put(00,21){\line(1,0){12}}

\put(00,19.5){\line(0,1){3}} \put(04,19.5){\line(0,1){3}}
\put(08,19.5){\line(0,1){3}} \put(12,19.5){\line(0,1){3}}

\put(01.5,17){\makebox(0,0)[bl]{\(1\)}}
\put(05.5,17){\makebox(0,0)[bl]{\(2\)}}
\put(09.5,17){\makebox(0,0)[bl]{\(3\)}}


\put(28,12){\line(0,1){6}}


\put(25,06.7){\makebox(0,0)[bl]{\(e^{3,4}_{12}\) }}

\put(28,09){\oval(16,5)}


\put(42,6.5){\makebox(0,0)[bl]{\(\{3,3,3,3\}\) or \((0,0,4)\) }}

\put(08,8.5){\line(0,1){08}} \put(12,8.5){\line(0,1){08}}
\put(08,10.5){\line(1,0){4}} \put(08,12.5){\line(1,0){4}}
\put(08,14.5){\line(1,0){4}} \put(08,16.5){\line(1,0){4}}

\put(00,8.5){\line(1,0){12}}

\put(00,07){\line(0,1){3}} \put(04,07){\line(0,1){3}}
\put(08,07){\line(0,1){3}} \put(12,07){\line(0,1){3}}

\put(01.5,4.5){\makebox(0,0)[bl]{\(1\)}}
\put(05.5,4.5){\makebox(0,0)[bl]{\(2\)}}
\put(09.5,4.5){\makebox(0,0)[bl]{\(3\)}}


\put(46.5,86.5){\line(-2,3){10.5}}




\put(45,80.7){\makebox(0,0)[bl]{\(e^{3,4}_{9}\) }}

\put(48,83){\oval(16,5)}


\put(57.5,81){\makebox(0,0)[bl]{\(\{1,1,2,3\}\) or \((2,1,1)\) }}

\put(00,84){\line(0,1){02}} \put(04,84){\line(0,1){04}}
\put(08,84){\line(0,1){04}} \put(12,84){\line(0,1){02}}

\put(00,86){\line(1,0){12}} \put(04,88){\line(1,0){4}}

\put(00,84){\line(1,0){12}}

\put(00,82.5){\line(0,1){3}} \put(04,82.5){\line(0,1){3}}
\put(08,82.5){\line(0,1){3}} \put(12,82.5){\line(0,1){3}}

\put(01.5,80){\makebox(0,0)[bl]{\(1\)}}
\put(05.5,80){\makebox(0,0)[bl]{\(2\)}}
\put(09.5,80){\makebox(0,0)[bl]{\(3\)}}


\put(44,66.5){\line(-1,2){11.4}}


\put(44,60){\line(-2,-1){8}}

\put(48,66){\line(0,1){14}}


\put(45,60.7){\makebox(0,0)[bl]{\(e^{3,4}_{10}\) }}

\put(48,63){\oval(16,5)}


\put(57.5,61){\makebox(0,0)[bl]{\(\{1,2,2,3\}\) or \((1,2,1)\) }}

\put(00,62){\line(0,1){02}} \put(04,62){\line(0,1){04}}
\put(08,62){\line(0,1){04}} \put(12,62){\line(0,1){02}}

\put(00,64){\line(1,0){12}} \put(04,66){\line(1,0){4}}

\put(00,62){\line(1,0){12}}

\put(00,60.5){\line(0,1){3}} \put(04,60.5){\line(0,1){3}}
\put(08,60.5){\line(0,1){3}} \put(12,60.5){\line(0,1){3}}

\put(01.5,58){\makebox(0,0)[bl]{\(1\)}}
\put(05.5,58){\makebox(0,0)[bl]{\(2\)}}
\put(09.5,58){\makebox(0,0)[bl]{\(3\)}}



\put(46.5,39.5){\line(-4,-1){12}}

\put(48,46){\line(0,1){14}}


\put(45,40.7){\makebox(0,0)[bl]{\(e^{3,4}_{11}\) }}

\put(48,43){\oval(16,5)}


\put(57.5,41){\makebox(0,0)[bl]{\(\{1,2,3,3\}\) or \((1,1,2)\) }}

\put(04,41.5){\line(0,1){04}}

\put(08,41.5){\line(0,1){04}} \put(12,41.5){\line(0,1){04}}
\put(04,43.5){\line(1,0){8}} \put(04,45.5){\line(1,0){8}}

\put(00,41.5){\line(1,0){12}}

\put(00,40){\line(0,1){3}} \put(04,40){\line(0,1){3}}
\put(08,40){\line(0,1){3}} \put(12,40){\line(0,1){3}}

\put(01.5,37.5){\makebox(0,0)[bl]{\(1\)}}
\put(05.5,37.5){\makebox(0,0)[bl]{\(2\)}}
\put(09.5,37.5){\makebox(0,0)[bl]{\(3\)}}


\end{picture}
\end{center}

\section{Operations over Multiset Estimates}

 The following operations are considered for the multiset estimates
 (or corresponding alternatives):
 (a) integration of several estimates (e.g., for composite systems),
 (b) proximity between estimates and comparison of the estimates,
 (c) comparison, ordering, selection of Pareto-efficient
 estimates (alternatives), and
 (d) aggregation (e.g., searching for a median estimate for the
 specified set of initial estimates).

\subsection{Integrated Estimates}

 Integration of estimates (mainly, for composite systems)
 is based on summarization of the estimates by components (i.e.,
 positions).
 Let us consider \(n\) estimates (position form):~~
 estimate \(e^{1} = (\eta^{1}_{1},...,\eta^{1}_{\iota},...,\eta^{1}_{l})
 \),
  {\bf . ~. ~.},
 estimate \(e^{\kappa} = (\eta^{\kappa}_{1},...,\eta^{\kappa}_{\iota},...,\eta^{\kappa}_{l})
 \),
  {\bf . ~. ~.},
 estimate \(e^{n} = (\eta^{n}_{1},...,\eta^{n}_{\iota},...,\eta^{n}_{l})
 \).
 Then, the integrated estimate is:~
 estimate \(e^{I} = (\eta^{I}_{1},...,\eta^{I}_{\iota},...,\eta^{I}_{l})
 \),
 where
 \(\eta^{I}_{\iota} = \sum_{\kappa=1}^{n} \eta^{\kappa}_{\iota} ~~ \forall
 \iota = \overline{1,l}\).
 In fact, the operation \(\biguplus\) is used for multiset estimates:
 \(e^{I} = e^{1} \biguplus ... \biguplus e^{\kappa} \biguplus ... \biguplus e^{n}\).

 Further, some examples for integration of multiset-estimates are presented.

~~

 {\bf Example 3.}
 \(S = X \star Y \star Z\).
 Assessment problem \(P^{3,1}\).
 Estimates of system parts are:
 (a) \(X^{i}\):  \(e(X^{i})=\{1\}\) or \(e(X^{i})=(1,0,0)\),
 (b) \(Y^{i}\):  \(e(Y^{i})=\{2\}\) or \(e(Y^{i})=(0,1,0)\),
 (c) \(Z^{i}\):  \(e(Z^{i})=\{1\}\) or \(e(Z^{i})=(1,1,0)\).
 As a result,
 the integrated estimate of composite system is
  (component-based summarization):

  \(S^{i} = X^{i} \star Y^{i} \star Z^{i}\):~
 \(e(S^{i})=\{1,1,2\}\) or \(e(S^{i})=(2,1,0)\).

~~

 {\bf Example 4.}
 \(S = X \star Y \star Z \star V\).
 Assessment problem \(P^{3,1}\).
 Estimates of system parts are:
 (a) \(X^{ii}\):  \(e(X^{ii})=\{1\}\) or \(e(X^{ii})=(1,0,0)\),
 (b) \(Y^{ii}\):  \(e(Y^{ii})=\{2\}\) or \(e(Y^{ii})=(0,1,0)\),
 (c) \(Z^{ii}\):  \(e(Z^{ii})=\{1\}\) or \(e(Z^{ii})=(0,1,0)\),
 (d) \(V^{ii}\):  \(e(V^{ii})=\{3\}\) or \(e(V^{ii})=(0,0,1)\).
 As a result,
 the integrated estimate of composite system is
  (component-based summarization):

  \(S^{ii} = X^{ii} \star Y^{ii} \star Z^{ii} \star V^{ii}\):~
 \(e(S^{ii})=\{1,2,2,3\}\) or \(e(S^{ii})=(1,2,1)\).

 ~~

  {\bf Example 5.}
 \(S = X \star Y \star Z\).
 Assessment problem \(P^{3,2}\).
 Estimates of system parts are:
 (a) \(X^{iii}\):  \(e(X^{iii})=\{1\}\) or \(e(X^{iii})=(1,1,0)\),
 (b) \(Y^{iii}\):  \(e(Y^{iii})=\{2\}\) or \(e(Y^{iii})=(0,1,0)\),
 (c) \(Z^{iii}\):  \(e(Z^{iii})=\{1\}\) or \(e(Z^{iii})=(0,2,0)\).
 As a result,
 the integrated estimate of composite system is
  (component-based summarization):

  \(S^{iii} = X^{iii} \star Y^{iii} \star Z^{iii}\):~
 \(e(S^{iii})=\{1,2,2,2,2\}\) or \(e(S^{iii})=(1,4,0)\).

~~

 {\bf Example 6.}
 \(S = X \star Y \star Z\).
 Assessment problem \(P^{3,3}\).
 Estimates of system parts are:
 (a) \(X^{iv}\):  \(\{1\}\) or \((1,1,1)\),
 (b) \(Y^{iv}\):  \(\{2\}\) or \((1,2,0)\),
 (c) \(Z^{iv}\):  \(\{1\}\) or \((0,2,1)\).
 As a result,
 the integrated estimate of composite system is
  (component-based summarization:
  \(S^{iv} = X^{iv} \star Y^{iv} \star Z^{iv}\):~
 \(\{1,1,2,2,2,2,3,3\}\) or \((2,5,2)\).

~~

 {\bf Example 7.}
 \(S = X \star Y \star Z\).
 Assessment problem \(P^{3,4}\).
 Estimates of system parts are:
 (a) \(X^{v}\):  \(e(X^{v})=\{1\}\) or \(e(X^{v})=(1,2,1)\),
 (b) \(Y^{v}\):  \(e(Y^{v})=\{2\}\) or \(e(Y^{v})=(2,2,0)\),
 (c) \(Z^{v}\):  \(e(Z^{v})=\{1\}\) or \(e(Z^{v})=(1,1,2)\).
 As a result,
 the integrated estimate of composite system is
  (component-based summarization):

  \(S^{v} = X^{v} \star Y^{v} \star Z^{v}\):~
 \(e(S^{v})=\{1,1,1,1,2,2,2,2,2,3,3,3\}\) or \(e(S^{v})=(4,5,3)\).

~~

 Generally, the integrated multiset estimate for multi-part system
  is based on assessment problem \(P^{l,\eta \times m} \)
  (case of complete poset, i.e., without condition 2 on ``interval''),
 where
 \( P^{l,\eta}\) is assessment problem for system parts,
 \(m\) is the number of subsystems (parts).

 ~~

 {\bf Example 8.}
 \(S = X \star Y \star Z\).
 Assessment problem \(P^{3,4}\).

 The integrated estimate for \(S\) is based on the
 assessment problem \(P^{3,12}\) (case of complete poset).

~~

 {\bf Example 9.}
 \(S = X \star Y \star Z\).
 Assessment problem \(P^{4,3}\).

 The integrated estimate for \(S\) is based on the
 assessment problem \(P^{4,9}\) (case of complete poset).
%
%
  Estimates of system parts are:
 (a) \(X^{vii}\):  \(e(X^{vii})=\{1\}\) or \(e(X^{vii})=(3,0,0,0)\),
 (b) \(Y^{vii}\):  \(e(Y^{vii})=\{2\}\) or \(e(Y^{vii})=(0,3,0,0)\),
 (c) \(Z^{vii}\):  \(e(Z^{vii})=\{1\}\) or \(e(Z^{vii})=(0,0,0,3)\).
 As a result,
 the integrated estimate of composite system is
  (component-based summarization):

  \(S^{vii} = X^{vii} \star Y^{vii} \star Z^{vii}\):~
 \(e(S^{vii})=\{1,1,1,2,2,2,4,4,4\}\) or \(e(S^{v})=(3,3,0,3)\).

 This example is the reason to use
 the complete poset for the integrated multiset estimate.

\subsection{Vector-like Proximity}

 Consider estimates of two  alternatives \(e(A_{1})\), \(e(A_{2})\) and the following vector-like proximity:
 \[\delta ( e(A_{1}), e(A_{2})) = (\delta^{-}(A_{1},A_{2}),\delta^{+}(A_{1},A_{2})),\]
 where vector components are:
%
 (i) \(\delta^{-}\) is the number of one-step changes:
 element of quality \(\iota + 1\) into element of quality \(\iota\) (\(\iota = \overline{1,l-1}\))
 (this corresponds to ``improvement'');
 (ii) \(\delta^{+}\) is the number of one-step changes:
 element of quality \(\iota\) into element of quality  \(\iota+1\) (\(\iota = \overline{1,l-1}\))
 (this corresponds to ``degradation'').

 This definition corresponds to change (edition) of \(A_{1}\) into
 \(A_{2}\).
%
 Vector-like proximity \(\delta ( e(A_{1}), e(A_{2}))\)
 is similar to vector-like proximity for rankings that was
 suggested in
 (\cite{lev98a},\cite{lev98},\cite{lev11agg}).
%
%
 Evidently, the following axioms are satisfied:

 {\bf 1.}  \( \forall A_{1},A_{2}\):~~
 \(\delta ( e(A_{1}), e(A_{2}))  \succeq (0,0)\) (nonnegativity).

 {\bf 2.}
 \(\delta ( e(A_{1}), e(A_{1}))  = (0,0)\) (identity).


 {\bf 3.}  \( \forall A_{1},A_{2} \):~~
 \(\delta ( e(A_{1}), e(A_{2})) =
 (\delta^{-} ( e(A_{1}),e(A_{2})),\delta^{+} (e(A_{1}), e(A_{2})))\),

 \(\delta ( e(A_{2}),e(A_{1}) ) = \
 ( \delta^{-} ( e(A_{2}),e(A_{1})),\delta^{+} (e(A_{2}), e(A_{1})) )=\)

  \(( \delta^{+} ( e(A_{1}),e(A_{2})),\delta^{-}
  (e(A_{1}), e(A_{2})) )\).
%

 {\bf 4.}  \( \forall A_{1},A_{2},A_{3} \):~~
 \(\delta ( e(A_{1}), e(A_{2})) +  \delta ( e(A_{2}), e(A_{3})) \succeq \delta ( e(A_{1}), e(A_{3}))\),
 here operation \('+'\) corresponds to summarization by
 components (triangle inequality).

 In addition, the following is defined:
 \( | \delta (e_{1},e_{2}) | =  | \delta^{-}(e_{1},e_{2})| + | \delta^{+}(e_{1},e_{2})
 |\).
 Evidently, \( | \delta (e_{1},e_{2}) | \leq \eta  \times (l-1) \).
 Fig. 7 depicts the domain  of the proximity.

\begin{center}
\begin{picture}(71,53)
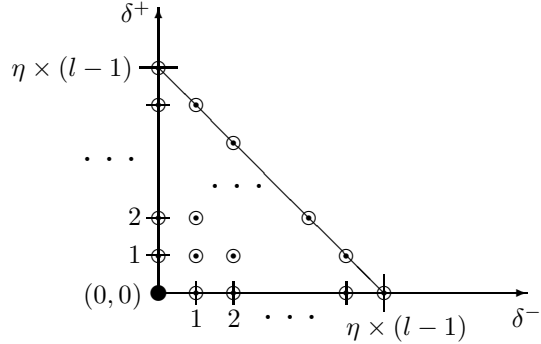
\put(06,00){\makebox(0,0)[bl]{Fig. 7. Discrete domain of
proximity}}

\put(10,08.8){\makebox(0,0)[bl]{\((0,0)\)}}

\put(20,11){\circle*{2}}


\put(50,11){\circle{1.6}} \put(50,11){\circle*{0.8}}

\put(67,6.5){\makebox(0,0)[bl]{\(\delta^{-}\)}}

\put(20,11){\vector(1,0){49}}

\put(24.2,6.5){\makebox(0,0)[bl]{\(1\)}}
\put(29.2,6.5){\makebox(0,0)[bl]{\(2\)}}

\put(45,4.5){\makebox(0,0)[bl]{\(\eta \times (l-1)\)}}


\put(20,11){\vector(0,1){38}}
\put(15,47){\makebox(0,0)[bl]{\(\delta^{+} \)}}

\put(0.5,39){\makebox(0,0)[bl]{\(\eta \times (l-1)\)}}


\put(16,15){\makebox(0,0)[bl]{\(1\)}}
\put(16,20){\makebox(0,0)[bl]{\(2\)}}

\put(18.5,16){\line(1,0){3}} \put(18.5,21){\line(1,0){3}}

\put(10,28.5){\makebox(0,0)[bl]{{\bf . . .} }}

\put(18.5,36){\line(1,0){3}} \put(17.5,41){\line(1,0){5}}

\put(20,41){\circle{1.6}} \put(20,41){\circle*{0.8}}

\put(20,41){\line(1,-1){30}}


\put(25,9.5){\line(0,1){3}} \put(30,9.5){\line(0,1){3}}

\put(34,7.5){\makebox(0,0)[bl]{{\bf . . .} }}

\put(45,9.5){\line(0,1){3}} \put(50,8.5){\line(0,1){5}}


\put(20,36){\circle{1.6}} \put(20,36){\circle*{0.8}}
\put(25,36){\circle{1.6}} \put(25,36){\circle*{0.8}}


\put(30,31){\circle{1.6}} \put(30,31){\circle*{0.8}}


\put(27,25){\makebox(0,0)[bl]{{\bf . . .} }}


\put(20,21){\circle{1.6}} \put(20,21){\circle*{0.8}}
\put(25,21){\circle{1.6}} \put(25,21){\circle*{0.8}}
\put(40,21){\circle{1.6}} \put(40,21){\circle*{0.8}}


\put(20,16){\circle{1.6}} \put(20,16){\circle*{0.8}}
\put(25,16){\circle{1.6}} \put(25,16){\circle*{0.8}}
\put(30,16){\circle{1.6}} \put(30,16){\circle*{0.8}}
\put(45,16){\circle{1.6}} \put(45,16){\circle*{0.8}}


\put(25,11){\circle{1.6}} \put(25,11){\circle*{0.8}}
\put(30,11){\circle{1.6}} \put(30,11){\circle*{0.8}}
\put(45,11){\circle{1.6}} \put(45,11){\circle*{0.8}}


\end{picture}
\end{center}

 Now, let us consider numerical examples of the proximity.

~~

 {\bf Example 10.} Assessment problem \(P^{3,1}\).

 {\it Case 1.}
 \(e_{1} = (0,0,1)\),
  \(e_{2} = (0,1,0)\),
  \(\delta (e_{1},e_{2}) = (1,0)\),
 \(\delta (e_{2},e_{1}) = (0,1)\).

{\it Case 2.}
 \(e_{1} = (0,0,1)\),
  \(e_{2} = (1,0,0)\),
  \(\delta (e_{1},e_{2}) = (2,0)\),
 \(\delta (e_{2},e_{1}) = (0,2)\).

~~

 {\bf Example 11.} Assessment problem \(P^{4,1}\).

  {\it Case 1.}
 \(e_{1} = (0,0,0,1)\),
  \(e_{2} = (0,1,0,0)\),
  \(\delta (e_{1},e_{2}) = (2,0)\),
 \(\delta (e_{2},e_{1}) = (0,2)\).

{\it Case 2.}
 \(e_{1} = (0,0,0,1)\),
  \(e_{2} = (1,0,0,0)\),
  \(\delta (e_{1},e_{2}) = (3,0)\),
 \(\delta (e_{2},e_{1}) = (0,3)\).

~~

{\bf Example 12.} Assessment problem \(P^{3,2}\).

 {\it Case 1.}
 \(e_{1} = (0,0,2)\),
  \(e_{2} = (0,1,1)\),
  \(\delta (e_{1},e_{2}) = (1,0)\),
 \(\delta (e_{2},e_{1}) = (0,1)\).

 {\it Case 2.}
 \(e_{1} = (0,0,2)\),
  \(e_{2} = (0,2,0)\),
  \(\delta (e_{1},e_{2}) = (2,0)\),
 \(\delta (e_{2},e_{1}) = (0,2)\).

{\it Case 3.}
 \(e_{1} = (0,0,2)\),
  \(e_{2} = (2,0,0)\),
  \(\delta (e_{1},e_{2}) = (4,0)\),
 \(\delta (e_{2},e_{1}) = (0,4)\).

~~

 {\bf Example 13.} Assessment problem \(P^{3,3}\).

 {\it Case 1.}
 \(e_{1} = (0,0,3)\),
  \(e_{2} = (0,1,2)\),
  \(\delta (e_{1},e_{2}) = (1,0)\),
 \(\delta (e_{2},e_{1}) = (0,1)\).

 {\it Case 2.}
 \(e_{1} = (0,0,3)\),
  \(e_{2} = (0,2,1)\),
  \(\delta (e_{1},e_{2}) = (2,0)\),
 \(\delta (e_{2},e_{1}) = (0,2)\).

{\it Case 3.}
 \(e_{1} = (0,0,3)\),
  \(e_{2} = (3,0,0)\),
  \(\delta (e_{1},e_{2}) = (6,0)\),
 \(\delta (e_{2},e_{1}) = (0,6)\).

{\it Case 4.}
 \(e_{1} = (0,3,0)\),
  \(e_{2} = (1,1,1)\),
  \(\delta (e_{1},e_{2}) = (1,1)\),
 \(\delta (e_{2},e_{1}) = (1,1)\).

~~

 {\bf Example 14.} Assessment problem \(P^{3,4}\).

 {\it Case 1.}
 \(e_{1} = (0,0,4)\),
  \(e_{2} = (0,1,3)\),
  \(\delta (e_{1},e_{2}) = (1,0)\),
 \(\delta (e_{2},e_{1}) = (0,1)\).

 {\it Case 2.}
 \(e_{1} = (0,0,4)\),
  \(e_{2} = (0,3,1)\),
  \(\delta (e_{1},e_{2}) = (3,0)\),
 \(\delta (e_{2},e_{1}) = (0,3)\).

 {\it Case 3.}
 \(e_{1} = (0,0,4)\),
  \(e_{2} = (4,0,0)\),
  \(\delta (e_{1},e_{2}) = (8,0)\),
 \(\delta (e_{2},e_{1}) = (0,8)\).

 {\it Case 4.}
 \(e_{1} = (0,4,0)\),
  \(e_{2} = (1,2,1)\),
  \(\delta (e_{1},e_{2}) = (1,1)\),
 \(\delta (e_{2},e_{1}) = (1,1)\).

 {\it Case 5.}
 \(e_{1} = (1,1,2)\),
  \(e_{2} = (0,4,0)\),
  \(\delta (e_{1},e_{2}) = (2,1)\),
 \(\delta (e_{2},e_{1}) = (1,2)\).

~~

 Further, for two alternative \(A_{1}\) and \(A_{2}\)
 their proximity is:
 \(\delta (e_{1}(A_{1}),e_{2}(A_{2})) = (\delta^{-},\delta^{+})\).
 Generally, the following is satisfied:

 {\bf 1.} \(\delta^{-} = 0\) and \(\delta^{+} > 0\)
 \(\Longleftrightarrow\)
  \(A_{1} \succ A_{2}\).

 {\bf 2.} \(\delta^{-} > 0\) and \(\delta^{+} = 0\)
 \(\Longleftrightarrow\)
 \(A_{1}  \prec A_{2}\).

 {\bf 3.} \(\delta^{-} > 0\) and \(\delta^{+} > 0\)
 \(\Longleftrightarrow\)
 \(A_{1}\) and \(A_{2}\)
 are incomparable.

\subsection{Comparison of Estimates}

 In our study, the following comparison problems over estimates
 are considered:

 {\it 1}. Comparison of two estimates.
 This operation is based on estimate proximity.

 {\it 2.} Ordering of estimates from
  a specified set of estimates.
 This operation corresponds to a well-known linear ordering algorithm
 while taking into account incomparable estimates
 (as ``equivalent'' ones).
 The computing complexity of the operation is
 \(O(n \lg n)\) (\(n\) is the number of estimates).

 {\it 3.} Selection of Pareto-efficient estimates for a specified set of
 estimates.
 Here the simple algorithm is the following.
 The proximity matrix can be computed and a Pareto-efficient
 estimate (alternative) has its line without
 matrix element \(\prec\) (or \(\succ\)).
 The computing complexity of the algorithm  is
 \(O(n^{2})\) (\(n\) is the number of estimates).
 The basic linear ordering algorithm leads to the same result
 (complexity: \(O(n \lg n)\)),
 but the described algorithm is more friendly for interactive procedure.

 Evidently, in special cases the similar algorithms can be used to
 search for the  ``maximal'' or the ``minimal'' estimate.

~~

 {\bf Example 15.} Assessment problem \(P^{3,3}\).

 The set of estimates is (Fig. 5):~
  \(e^{3,3}_{4} = (0,3,0)\),
  \(e^{3,3}_{5} = (1,1,1)\),
  \(e^{3,3}_{6} = (0,2,1)\),
  \(e^{3,3}_{7} = (0,1,2)\),
  \(e^{3,3}_{8} = (0,0,4)\).
  Proximities for the estimates are presented
  as proximity matrix in Table 2.
 Pareto-efficient estimates are:~
 \(e^{3,3}_{4}\),
  \(e^{3,3}_{5}\).
 Here the ``minimal'' estimate exists:
 \(e^{3,3}_{8}\).

~~

 {\bf Example 16.} Assessment problem \(P^{3,4}\).

 The set of estimates is (Fig. 6):~
  \(e^{3,4}_{9} = (2,1,1)\),
  \(e^{3,4}_{5} = (0,4,0)\),
  \(e^{3,4}_{10} = (1,2,1)\),
  \(e^{3,4}_{6} = (0,2,2)\),
  \(e^{3,4}_{7} = (0,1,3)\),
  \(e^{3,4}_{8} = (0,0,4)\).
  Proximities for the estimates are presented
  as proximity matrix in Table 3.
 Pareto-efficient estimates are:~
 \(e^{3,4}_{9}\),
  \(e^{3,4}_{5}\).
 Here the ``minimal'' estimate exists:
 \(e^{3,4}_{8}\).

\begin{center}
\begin{picture}(74,43)
\put(11.5,36.5){\makebox(0,0)[bl] {Table 2. Proximities
 \( \delta (e^{3,3}_{\kappa},e^{3,3}_{j} ) \)}}


\put(00,00){\line(1,0){74}} \put(00,28){\line(1,0){74}}
\put(00,35){\line(1,0){74}}

\put(00,00){\line(0,1){35}} \put(24,00){\line(0,1){35}}
\put(74,00){\line(0,1){35}}

\put(00,35){\line(3,-1){20}}

\put(02,30){\makebox(0,0)[bl]{\(e_{\kappa}\)}}
\put(16,31){\makebox(0,0)[bl]{\(e_{j}\)}}


\put(26,29.5){\makebox(0,0)[bl]{\(e^{3,3}_{4}\)}}
\put(36,29.5){\makebox(0,0)[bl]{\(e^{3,3}_{5}\)}}
\put(46,29.5){\makebox(0,0)[bl]{\(e^{3,3}_{6}\)}}
\put(56,29.5){\makebox(0,0)[bl]{\(e^{3,3}_{7}\)}}
\put(66,29.5){\makebox(0,0)[bl]{\(e^{3,3}_{8}\)}}

\put(24,28){\line(0,1){07}} \put(34,28){\line(0,1){07}}
\put(44,28){\line(0,1){07}} \put(54,28){\line(0,1){07}}
\put(64,28){\line(0,1){07}}


\put(01,22){\makebox(0,0)[bl]{\(e^{3,3}_{4}=(0,3,0)\)}}
\put(01,17){\makebox(0,0)[bl]{\(e^{3,3}_{5}=(1,1,1)\)}}
\put(01,12){\makebox(0,0)[bl]{\(e^{3,3}_{6}=(0,2,1)\)}}
\put(01,07){\makebox(0,0)[bl]{\(e^{3,3}_{7}=(0,1,2)\)}}
\put(01,02){\makebox(0,0)[bl]{\(e^{3,3}_{8}=(0,0,3)\)}}



\put(25,22){\makebox(0,0)[bl]{\((0,0)\)}}
\put(35,22){\makebox(0,0)[bl]{\((1,1)\)}}
\put(45,22){\makebox(0,0)[bl]{\((0,1)\)}}
\put(55,22){\makebox(0,0)[bl]{\((0,2)\)}}
\put(65,22){\makebox(0,0)[bl]{\((0,3)\)}}


\put(25,17){\makebox(0,0)[bl]{\((1,1)\)}}
\put(35,17){\makebox(0,0)[bl]{\((0,0)\)}}
\put(45,17){\makebox(0,0)[bl]{\((0,1)\)}}
\put(55,17){\makebox(0,0)[bl]{\((0,2)\)}}
\put(65,17){\makebox(0,0)[bl]{\((0,3)\)}}


\put(25,12){\makebox(0,0)[bl]{\((1,0)\)}}
\put(35,12){\makebox(0,0)[bl]{\((1,0)\)}}
\put(45,12){\makebox(0,0)[bl]{\((0,0)\)}}
\put(55,12){\makebox(0,0)[bl]{\((0,1)\)}}
\put(65,12){\makebox(0,0)[bl]{\((0,2)\)}}


\put(25,07){\makebox(0,0)[bl]{\((2,0)\)}}
\put(35,07){\makebox(0,0)[bl]{\((2,0)\)}}
\put(45,07){\makebox(0,0)[bl]{\((1,0)\)}}
\put(55,07){\makebox(0,0)[bl]{\((0,0)\)}}
\put(65,07){\makebox(0,0)[bl]{\((0,1)\)}}


\put(25,02){\makebox(0,0)[bl]{\((3,0)\)}}
\put(35,02){\makebox(0,0)[bl]{\((3,0)\)}}
\put(45,02){\makebox(0,0)[bl]{\((2,0)\)}}
\put(55,02){\makebox(0,0)[bl]{\((1,0)\)}}
\put(65,02){\makebox(0,0)[bl]{\((0,0)\)}}

\end{picture}
\end{center}

\begin{center}
\begin{picture}(84,48)
\put(16.5,41.5){\makebox(0,0)[bl] {Table 3. Proximities
 \( \delta (e^{3,4}_{\kappa},e^{3,4}_{j} ) \)}}


\put(00,00){\line(1,0){84}} \put(00,33){\line(1,0){84}}
\put(00,40){\line(1,0){84}}

\put(00,00){\line(0,1){40}} \put(24,00){\line(0,1){40}}
\put(84,00){\line(0,1){40}}

\put(00,40){\line(3,-1){20}}

\put(02,35){\makebox(0,0)[bl]{\(e_{\kappa}\)}}
\put(16,36){\makebox(0,0)[bl]{\(e_{j}\)}}


\put(26,34.5){\makebox(0,0)[bl]{\(e^{3,4}_{9}\)}}
\put(36,34.5){\makebox(0,0)[bl]{\(e^{3,4}_{5}\)}}
\put(46,34.5){\makebox(0,0)[bl]{\(e^{3,4}_{10}\)}}
\put(56,34.5){\makebox(0,0)[bl]{\(e^{3,4}_{6}\)}}
\put(66,34.5){\makebox(0,0)[bl]{\(e^{3,4}_{7}\)}}
\put(76,34.5){\makebox(0,0)[bl]{\(e^{3,4}_{8}\)}}

\put(24,33){\line(0,1){07}} \put(34,33){\line(0,1){07}}
\put(44,33){\line(0,1){07}} \put(54,33){\line(0,1){07}}
\put(64,33){\line(0,1){07}} \put(74,33){\line(0,1){07}}


\put(01,27){\makebox(0,0)[bl]{\(e^{3,4}_{9}=(2,1,1)\)}}
\put(01,22){\makebox(0,0)[bl]{\(e^{3,4}_{5}=(0,4,0)\)}}
\put(01,17){\makebox(0,0)[bl]{\(e^{3,4}_{10}=(1,2,1)\)}}
\put(01,12){\makebox(0,0)[bl]{\(e^{3,4}_{6}=(0,3,1)\)}}
\put(01,07){\makebox(0,0)[bl]{\(e^{3,4}_{7}=(0,2,2)\)}}
\put(01,02){\makebox(0,0)[bl]{\(e^{3,4}_{8}=(0,1,3)\)}}



\put(25,27){\makebox(0,0)[bl]{\((0,0)\)}}
\put(35,27){\makebox(0,0)[bl]{\((1,2)\)}}
\put(45,27){\makebox(0,0)[bl]{\((0,1)\)}}
\put(55,27){\makebox(0,0)[bl]{\((0,2)\)}}
\put(65,27){\makebox(0,0)[bl]{\((0,3)\)}}
\put(75,27){\makebox(0,0)[bl]{\((0,4)\)}}


\put(25,22){\makebox(0,0)[bl]{\((2,1)\)}}
\put(35,22){\makebox(0,0)[bl]{\((0,0)\)}}
\put(45,22){\makebox(0,0)[bl]{\((1,1)\)}}
\put(55,22){\makebox(0,0)[bl]{\((0,1)\)}}
\put(65,22){\makebox(0,0)[bl]{\((0,2)\)}}
\put(75,22){\makebox(0,0)[bl]{\((0,3)\)}}


\put(25,17){\makebox(0,0)[bl]{\((1,0)\)}}
\put(35,17){\makebox(0,0)[bl]{\((1,1)\)}}
\put(45,17){\makebox(0,0)[bl]{\((0,0)\)}}
\put(55,17){\makebox(0,0)[bl]{\((0,1)\)}}
\put(65,17){\makebox(0,0)[bl]{\((0,2)\)}}
\put(75,17){\makebox(0,0)[bl]{\((0,3)\)}}


\put(25,12){\makebox(0,0)[bl]{\((2,0)\)}}
\put(35,12){\makebox(0,0)[bl]{\((1,0)\)}}
\put(45,12){\makebox(0,0)[bl]{\((1,0)\)}}
\put(55,12){\makebox(0,0)[bl]{\((0,0)\)}}
\put(65,12){\makebox(0,0)[bl]{\((0,1)\)}}
\put(75,12){\makebox(0,0)[bl]{\((0,2)\)}}


\put(25,07){\makebox(0,0)[bl]{\((3,0)\)}}
\put(35,07){\makebox(0,0)[bl]{\((2,0)\)}}
\put(45,07){\makebox(0,0)[bl]{\((2,0)\)}}
\put(55,07){\makebox(0,0)[bl]{\((1,0)\)}}
\put(65,07){\makebox(0,0)[bl]{\((0,0)\)}}
\put(75,07){\makebox(0,0)[bl]{\((0,1)\)}}


\put(25,02){\makebox(0,0)[bl]{\((4,0)\)}}
\put(35,02){\makebox(0,0)[bl]{\((3,0)\)}}
\put(45,02){\makebox(0,0)[bl]{\((3,0)\)}}
\put(55,02){\makebox(0,0)[bl]{\((2,0)\)}}
\put(65,02){\makebox(0,0)[bl]{\((1,0)\)}}
\put(75,02){\makebox(0,0)[bl]{\((0,0)\)}}

\end{picture}
\end{center}

\subsection{Aggregation of Estimates}

 Here searching for a median estimate for the
 specified set of initial estimates is considered.
 Let \(E = \{ e_{1},...,e_{\kappa},...,e_{n}\}\)
 be the set of specified estimates
 (or a corresponding set of specified alternatives),
 let \(D \)
 be the set of all possible estimates
 (or a corresponding set of possible alternatives)
 (\( E  \subseteq D \)).
%
%
  Thus, the median estimate  is
 (e.g.,  \cite{lev11agg}, \cite{solnon07}):

 (a) ``generalized median'':~
%
%
%
 \[M^{g} =   \underset{M \in D}{\operatorname{argmin}}~~
 | \biguplus_{\kappa=1}^{n} ~ \delta (M, e_{\kappa}) ~|; \]
%

 (b) simplified case of the median (approximation)
  as ``set median'' over set \(E\):~
%
%
\[M^{s} =   \underset{M \in E}{\operatorname{argmin}}~~
 | \biguplus_{\kappa=1}^{n} ~ \delta (M, e_{\kappa}) ~|. \]
%
%
%
%
%
%
%
 The problem of searching for the ``generalized median'' is usually NP-hard,
 complexity of searching for ``set median'' is a polynomial problem
 (\(O(n^{2})\)).
 In our study,
 simple problems are considered where
 the set of all multiset estimates is very limited
 (i.e., ``multiset number'')
 and simple enumerative solving schemes can be used for ``generalized median''.

~~

 {\bf Example 17.} Assessment problem \(P^{3,4}\).

 The initial set of estimates  \(E\) is (Fig. 6):
 \(e^{3,4}_{2} = (3,1,0)\),
 \(e^{3,4}_{4} = (1,3,0)\),
  \(e^{3,4}_{5} = (0,4,0)\),
  \(e^{3,4}_{9} = (2,1,1)\),
  \(e^{3,4}_{10} = (1,2,1)\),
  \(e^{3,4}_{7} = (0,2,2)\),
  \(e^{3,4}_{8} = (0,1,3)\),
  \(e^{3,4}_{12} = (0,0,4)\).
 Table 4 contains proximities and the integrated estimate:~
 \(
  \Upsilon_{\kappa} = \biguplus_{j \in E} ~  \delta (e_{\kappa}, e_{j})
  \).
 The median estimates are:
 (a) ``generalized median''
 \(M^{g} = e^{3,4}_{6}=(0,3,1)\) (\(e^{3,4}_{6} \overline{\in}  E\));
 (b) ``set median'' \(M^{s} = e^{3,4}_{10} = (1,2,1)\)
 (\(e^{3,4}_{10} \in  E\)).

\begin{center}
\begin{picture}(115,57)
\put(32,51.5){\makebox(0,0)[bl] {Table 4. Proximities
 \( \delta (e^{3,4}_{\kappa},e^{3,4}_{j} ) \)}}


\put(00,00){\line(1,0){115}} \put(00,43){\line(1,0){115}}
\put(00,50){\line(1,0){115}}

\put(00,00){\line(0,1){50}} \put(24,00){\line(0,1){50}}
\put(104,00){\line(0,1){50}} \put(115,00){\line(0,1){50}}

\put(00,50){\line(3,-1){20}}

\put(02,45){\makebox(0,0)[bl]{\(e_{\kappa}\)}}
\put(16,46){\makebox(0,0)[bl]{\(e_{j}\)}}


\put(26,44.5){\makebox(0,0)[bl]{\(e^{3,4}_{2}\)}}
\put(36,44.5){\makebox(0,0)[bl]{\(e^{3,4}_{4}\)}}
\put(46,44.5){\makebox(0,0)[bl]{\(e^{3,4}_{5}\)}}
\put(56,44.5){\makebox(0,0)[bl]{\(e^{3,4}_{9}\)}}
\put(66,44.5){\makebox(0,0)[bl]{\(e^{3,4}_{10}\)}}
\put(76,44.5){\makebox(0,0)[bl]{\(e^{3,4}_{7}\)}}
\put(86,44.5){\makebox(0,0)[bl]{\(e^{3,4}_{8}\)}}
\put(96,44.5){\makebox(0,0)[bl]{\(e^{3,4}_{12}\)}}

\put(107.5,45){\makebox(0,0)[bl]{\(\Upsilon_{\kappa}
%
  \)}}

\put(24,43){\line(0,1){07}} \put(34,43){\line(0,1){07}}
\put(44,43){\line(0,1){07}} \put(54,43){\line(0,1){07}}
\put(64,43){\line(0,1){07}} \put(74,43){\line(0,1){07}}
\put(84,43){\line(0,1){07}} \put(94,43){\line(0,1){07}}


\put(01,37){\makebox(0,0)[bl]{\(e^{3,4}_{2}=(3,1,0)\)}}
\put(01,32){\makebox(0,0)[bl]{\(e^{3,4}_{4}=(1,3,0)\)}}
\put(01,27){\makebox(0,0)[bl]{\(e^{3,4}_{5}=(0,4,0)\)}}
\put(01,22){\makebox(0,0)[bl]{\(e^{3,4}_{9}=(2,1,1)\)}}
\put(01,17){\makebox(0,0)[bl]{\(e^{3,4}_{10}=(1,2,1)\)}}
\put(01,12){\makebox(0,0)[bl]{\(e^{3,4}_{7}=(0,2,2)\)}}
\put(01,07){\makebox(0,0)[bl]{\(e^{3,4}_{8}=(0,1,3)\)}}
\put(01,02){\makebox(0,0)[bl]{\(e^{3,4}_{12}=(0,0,4)\)}}



\put(25,37){\makebox(0,0)[bl]{\((0,0)\)}}
\put(35,37){\makebox(0,0)[bl]{\((0,2)\)}}
\put(45,37){\makebox(0,0)[bl]{\((0,3)\)}}
\put(55,37){\makebox(0,0)[bl]{\((0,2)\)}}
\put(65,37){\makebox(0,0)[bl]{\((0,3)\)}}
\put(75,37){\makebox(0,0)[bl]{\((0,5)\)}}
\put(85,37){\makebox(0,0)[bl]{\((0,6)\)}}
\put(95,37){\makebox(0,0)[bl]{\((0,7)\)}}

\put(105,37){\makebox(0,0)[bl]{\((0,28)\)}}


\put(25,32){\makebox(0,0)[bl]{\((2,0)\)}}
\put(35,32){\makebox(0,0)[bl]{\((0,0)\)}}
\put(45,32){\makebox(0,0)[bl]{\((0,1)\)}}
\put(55,32){\makebox(0,0)[bl]{\((1,1)\)}}
\put(65,32){\makebox(0,0)[bl]{\((0,1)\)}}
\put(75,32){\makebox(0,0)[bl]{\((0,3)\)}}
\put(85,32){\makebox(0,0)[bl]{\((0,4)\)}}
\put(95,32){\makebox(0,0)[bl]{\((0,5)\)}}

\put(105,32){\makebox(0,0)[bl]{\((3,15)\)}}


\put(25,27){\makebox(0,0)[bl]{\((3,0)\)}}
\put(35,27){\makebox(0,0)[bl]{\((1,0)\)}}
\put(45,27){\makebox(0,0)[bl]{\((0,0)\)}}
\put(55,27){\makebox(0,0)[bl]{\((2,1)\)}}
\put(65,27){\makebox(0,0)[bl]{\((1,1)\)}}
\put(75,27){\makebox(0,0)[bl]{\((0,2)\)}}
\put(85,27){\makebox(0,0)[bl]{\((0,3)\)}}
\put(95,27){\makebox(0,0)[bl]{\((0,4)\)}}

\put(105,27){\makebox(0,0)[bl]{\((7,11)\)}}


\put(25,22){\makebox(0,0)[bl]{\((2,0)\)}}
\put(35,22){\makebox(0,0)[bl]{\((1,1)\)}}
\put(45,22){\makebox(0,0)[bl]{\((1,2)\)}}
\put(55,22){\makebox(0,0)[bl]{\((0,0)\)}}
\put(65,22){\makebox(0,0)[bl]{\((0,1)\)}}
\put(75,22){\makebox(0,0)[bl]{\((0,3)\)}}
\put(85,22){\makebox(0,0)[bl]{\((0,4)\)}}
\put(95,22){\makebox(0,0)[bl]{\((0,5)\)}}

\put(105,22){\makebox(0,0)[bl]{\((4,16)\)}}


\put(25,17){\makebox(0,0)[bl]{\((3,0)\)}}
\put(35,17){\makebox(0,0)[bl]{\((1,0)\)}}
\put(45,17){\makebox(0,0)[bl]{\((1,1)\)}}
\put(55,17){\makebox(0,0)[bl]{\((1,0)\)}}
\put(65,17){\makebox(0,0)[bl]{\((0,0)\)}}
\put(75,17){\makebox(0,0)[bl]{\((0,2)\)}}
\put(85,17){\makebox(0,0)[bl]{\((0,3)\)}}
\put(95,17){\makebox(0,0)[bl]{\((0,4)\)}}

\put(105,17){\makebox(0,0)[bl]{\((6,10)\)}}


\put(25,12){\makebox(0,0)[bl]{\((5,0)\)}}
\put(35,12){\makebox(0,0)[bl]{\((3,0)\)}}
\put(45,12){\makebox(0,0)[bl]{\((2,0)\)}}
\put(55,12){\makebox(0,0)[bl]{\((3,0)\)}}
\put(65,12){\makebox(0,0)[bl]{\((2,0)\)}}
\put(75,12){\makebox(0,0)[bl]{\((0,0)\)}}
\put(85,12){\makebox(0,0)[bl]{\((0,1)\)}}
\put(95,12){\makebox(0,0)[bl]{\((0,2)\)}}

\put(105,12){\makebox(0,0)[bl]{\((15,3)\)}}


\put(25,07){\makebox(0,0)[bl]{\((6,0)\)}}
\put(35,07){\makebox(0,0)[bl]{\((4,0)\)}}
\put(45,07){\makebox(0,0)[bl]{\((3,0)\)}}
\put(55,07){\makebox(0,0)[bl]{\((4,0)\)}}
\put(65,07){\makebox(0,0)[bl]{\((3,0)\)}}
\put(75,07){\makebox(0,0)[bl]{\((1,0)\)}}
\put(85,07){\makebox(0,0)[bl]{\((0,0)\)}}
\put(95,07){\makebox(0,0)[bl]{\((0,1)\)}}

\put(105,07){\makebox(0,0)[bl]{\((21,1)\)}}


\put(25,02){\makebox(0,0)[bl]{\((7,0)\)}}
\put(35,02){\makebox(0,0)[bl]{\((5,0)\)}}
\put(45,02){\makebox(0,0)[bl]{\((4,0)\)}}
\put(55,02){\makebox(0,0)[bl]{\((5,0)\)}}
\put(65,02){\makebox(0,0)[bl]{\((4,0)\)}}
\put(75,02){\makebox(0,0)[bl]{\((2,0)\)}}
\put(85,02){\makebox(0,0)[bl]{\((1,0)\)}}
\put(95,02){\makebox(0,0)[bl]{\((0,0)\)}}

\put(105,02){\makebox(0,0)[bl]{\((28,0)\)}}

\end{picture}
\end{center}

 Now, let us define  the deviation
  of the median estimate (\(M^{g}\) or \(M^{s}\)):
 \[ \Delta(M) =  (\Delta^{-} (M),\Delta^{+} (M)),
  \Delta^{-}(M)=\delta( M, \min_{\kappa=\overline{1,n}} e_{\kappa} ),
  \Delta^{+}(M)=\delta( \max_{\kappa=\overline{1,n}} e_{\kappa} , M ).\]
 In addition,~~
%
  \( | \Delta(M) | =  \max \{ | \Delta^{-}(M) | , |\Delta^{+}(M) | \} \).

~~

 {\bf Example 18.} (Assessment problem \(P^{3,4}\), from the previous section).
  The initial set of estimates  \(E\) is (Fig. 6):
 \(e^{3,4}_{2}\),
 \(e^{3,4}_{4}\),
  \(e^{3,4}_{5}\),
  \(e^{3,4}_{9} \),
  \(e^{3,4}_{10}\),
  \(e^{3,4}_{7}\),
  \(e^{3,4}_{8}\),
  \(e^{3,4}_{12}\).
%
%
%
%
 The median estimates are:
 (a) ``generalized median''
 \(M^{g} = e^{3,4}_{6}=(0,3,1)\);
%
 (b) ``set median'' \(M^{s} = e^{3,4}_{10} = (1,2,1)\).
%
%
 The dispersions are:
%
%

%
  \(\Delta^{-} (M^{g}) =  \delta ( (0,3,1), (0,0,4) ) = (0,3)\),
   \( \Delta^{+} (M^{g}) = \delta ( (3,1,0),(0,3,1) ) = (0,4)\);
 \( | \Delta(M^{g}) | =  4 \).
%
%

%
  \(\Delta^{-} (M^{s}) =  \delta ( (1,2,1), (0,0,4) ) = (0,4)\),
  \( \Delta^{+} (M^{s}) = \delta ( (3,1,0),(1,2,1) ) = (0,3)\);
 \( | \Delta(M^{s}) | =  4 \).

\subsection{Alignment of Estimates}

 Let
 \(\{e^{l_{1},\eta_{1}}_{1},...,e^{l_{i},\eta_{i}}_{i},..., e^{l_{n},\eta_{n}}_{n}\}\)
 be an initial set of \(n\) multiset estimates.
 The alignment problem is:
 \[\{e^{l_{1},\eta_{1}}_{1},...,
 e^{l_{i},\eta_{i}}_{i},...,e^{l_{n},\eta_{n}}_{n}\}~~
 \Longrightarrow
 ~~e^{l,\eta}; ~~
%
%
 \{P^{l_{1},\eta_{1}},...,P^{l_{i},\eta_{i}},...,P^{l_{n},\eta_{n}}\}
 \Longrightarrow
 P^{l,\eta}.\]
 In general, solving approach to this problem has to be based
 on special applied analysis by the domain expert(s).
 Let us consider a simplified approach to alignment of
 multiset alternatives as follows (Fig. 8):
 \[  P^{l_{i},\eta_{i}}
 \Longrightarrow
 P^{l,\eta},  ~~\forall i=\overline{1,n},
%
%
%
 ~~ l = \max_{i=\overline{1,n}} \{l_{i}\},
  ~~\eta = \max_{i=\overline{1,n}} \{\eta_{i}\}.\]
%

\begin{center}
\begin{picture}(115,35)

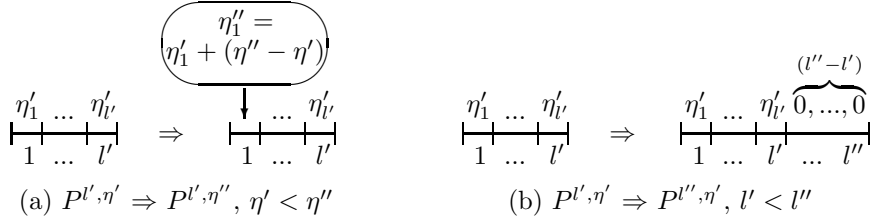
\put(24,00){\makebox(0,0)[bl] {Fig. 8. Alignment of multiset
estimates}}


\put(01,06){\makebox(0,0)[bl] {(a) \(P^{l',\eta'}\)
\(\Rightarrow\) \(P^{l',\eta''} \), \(\eta' < \eta''\)}}


\put(00,17){\line(1,0){14}}

\put(00,15.5){\line(0,1){3}} \put(04,15.5){\line(0,1){3}}
\put(10,15.5){\line(0,1){3}} \put(14,15.5){\line(0,1){3}}

\put(19.5,16){\makebox(0,0)[bl]{\(\Rightarrow\)}}

\put(01.5,13){\makebox(0,0)[bl]{\(1\)}}
\put(05.5,13){\makebox(0,0)[bl]{\({\bf ...}\)}}
\put(11.5,13){\makebox(0,0)[bl]{\(l'\)}}

\put(00.5,19){\makebox(0,0)[bl]{\(\eta'_{1}\)}}
\put(05.5,19){\makebox(0,0)[bl]{\({\bf ...}\)}}
\put(10.5,19){\makebox(0,0)[bl]{\(\eta'_{l'}\)}}


\put(29,17){\line(1,0){14}}

\put(29,15.5){\line(0,1){3}} \put(33,15.5){\line(0,1){3}}
\put(39,15.5){\line(0,1){3}} \put(43,15.5){\line(0,1){3}}

\put(30.5,13){\makebox(0,0)[bl]{\(1\)}}
\put(34.5,13){\makebox(0,0)[bl]{\({\bf ...}\)}}
\put(40.5,13){\makebox(0,0)[bl]{\(l'\)}}

\put(27.5,30){\makebox(0,0)[bl]{\(\eta''_{1}=\)}}
\put(20.5,26){\makebox(0,0)[bl]{\(\eta'_{1} + (\eta'' - \eta')\)}}

\put(31,29){\oval(22,11)}

\put(31,23){\vector(0,-1){4}}


\put(34.5,19){\makebox(0,0)[bl]{\({\bf ...}\)}}
\put(39.5,19){\makebox(0,0)[bl]{\(\eta'_{l'}\)}}


\put(66,06){\makebox(0,0)[bl] {(b) \(P^{l',\eta'}\)
\(\Rightarrow\) \(P^{l'',\eta'} \), \(l' < l''\)}}


\put(60,17){\line(1,0){14}}

\put(60,15.5){\line(0,1){3}} \put(64,15.5){\line(0,1){3}}
\put(70,15.5){\line(0,1){3}} \put(74,15.5){\line(0,1){3}}

\put(79.5,16){\makebox(0,0)[bl]{\(\Rightarrow\)}}

\put(61.5,13){\makebox(0,0)[bl]{\(1\)}}
\put(65.5,13){\makebox(0,0)[bl]{\({\bf ...}\)}}
\put(71.5,13){\makebox(0,0)[bl]{\(l'\)}}

\put(60.5,19){\makebox(0,0)[bl]{\(\eta'_{1}\)}}
\put(65.5,19){\makebox(0,0)[bl]{\({\bf ...}\)}}
\put(70.5,19){\makebox(0,0)[bl]{\(\eta'_{l'}\)}}


\put(89,17){\line(1,0){25}}

\put(89,15.5){\line(0,1){3}} \put(93,15.5){\line(0,1){3}}
\put(99,15.5){\line(0,1){3}} \put(103,15.5){\line(0,1){3}}


\put(114,15.5){\line(0,1){3}}



%
%

\put(104,19){\makebox(0,0)[bl]{\( \overbrace{0,...,0}^{(l''-l')}
\)}}



\put(90.5,13){\makebox(0,0)[bl]{\(1\)}}
\put(94.5,13){\makebox(0,0)[bl]{\({\bf ...}\)}}
\put(100.5,13){\makebox(0,0)[bl]{\(l'\)}}

\put(105,13){\makebox(0,0)[bl]{\({\bf ...}\)}}

\put(110.5,13){\makebox(0,0)[bl]{\(l''\)}}



\put(89.5,19){\makebox(0,0)[bl]{\(\eta'_{1}\)}}

\put(94.5,19){\makebox(0,0)[bl]{\({\bf ...}\)}}
\put(99.5,19){\makebox(0,0)[bl]{\(\eta'_{l'}\)}}


\end{picture}
\end{center}



%
 {\bf Example 19.}
%
 The initial set of multiset estimates is:
 \(e^{3,2}_{1} = (1,1,0)\),
 \(e^{3,3}_{2} = (1,1,1)\),
 \(e^{2,3}_{3} = (2,1)\),
 \(e^{4,4}_{4} = (0,2,1,1)\),
 \(e^{3,4}_{5} = (1,2,1)\),
 Fig. 9 depicts
  alignment.

\begin{center}
\begin{picture}(102,62)
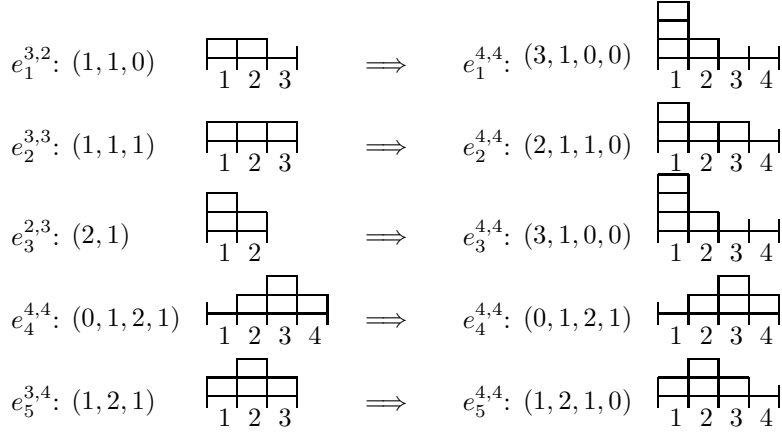
\put(019,00){\makebox(0,0)[bl] {Fig. 9. Example of estimate
alignment}}


\put(00,51.7){\makebox(0,0)[bl]{\(e^{3,2}_{1}\): }}

\put(08,52){\makebox(0,0)[bl]{\((1,1,0)\) }}

\put(26,54.5){\line(0,1){02.5}} \put(30,54.5){\line(0,1){02.5}}
\put(34,54.5){\line(0,1){02.5}}

\put(26,57){\line(1,0){08}}

\put(26,54.5){\line(1,0){12}}

\put(26,53){\line(0,1){3}} \put(30,53){\line(0,1){3}}
\put(34,53){\line(0,1){3}} \put(38,53){\line(0,1){3}}

\put(27.5,50.5){\makebox(0,0)[bl]{\(1\)}}
\put(31.5,50.5){\makebox(0,0)[bl]{\(2\)}}
\put(35.5,50.5){\makebox(0,0)[bl]{\(3\)}}


\put(00,40.7){\makebox(0,0)[bl]{\(e^{3,3}_{2}\): }}

\put(08,41){\makebox(0,0)[bl]{\((1,1,1)\) }}

\put(26,43.5){\line(0,1){02.5}} \put(30,43.5){\line(0,1){02.5}}
\put(34,43.5){\line(0,1){02.5}} \put(38,43.5){\line(0,1){02.5}}

\put(26,46){\line(1,0){12}}

\put(26,43.5){\line(1,0){12}}

\put(26,42){\line(0,1){3}} \put(30,42){\line(0,1){3}}
\put(34,42){\line(0,1){3}} \put(38,42){\line(0,1){3}}

\put(27.5,39.5){\makebox(0,0)[bl]{\(1\)}}
\put(31.5,39.5){\makebox(0,0)[bl]{\(2\)}}
\put(35.5,39.5){\makebox(0,0)[bl]{\(3\)}}


\put(00,28.7){\makebox(0,0)[bl]{\(e^{2,3}_{3}\): }}

\put(08,29){\makebox(0,0)[bl]{\((2,1)\) }}

\put(26,31.5){\line(0,1){05}} \put(30,31.5){\line(0,1){05}}
\put(34,31.5){\line(0,1){02.5}}

\put(26,34){\line(1,0){08}} \put(26,36.5){\line(1,0){4}}

\put(26,31.5){\line(1,0){08}}

\put(26,30){\line(0,1){3}} \put(30,30){\line(0,1){3}}
\put(34,30){\line(0,1){3}}


\put(27.5,27.5){\makebox(0,0)[bl]{\(1\)}}
\put(31.5,27.5){\makebox(0,0)[bl]{\(2\)}}


\put(00,17.7){\makebox(0,0)[bl]{\(e^{4,4}_{4}\): }}

\put(08,18){\makebox(0,0)[bl]{\((0,1,2,1)\) }}

\put(30,20.5){\line(0,1){02.5}} \put(34,20.5){\line(0,1){05}}
\put(38,20.5){\line(0,1){05}} \put(42,20.5){\line(0,1){02.5}}

\put(30,23){\line(1,0){12}} \put(34,25.5){\line(1,0){4}}

\put(26,20.5){\line(1,0){16}}

\put(26,19){\line(0,1){3}} \put(30,19){\line(0,1){3}}
\put(34,19){\line(0,1){3}} \put(38,19){\line(0,1){3}}
\put(42,19){\line(0,1){3}}

\put(27.5,16.5){\makebox(0,0)[bl]{\(1\)}}
\put(31.5,16.5){\makebox(0,0)[bl]{\(2\)}}
\put(35.5,16.5){\makebox(0,0)[bl]{\(3\)}}
\put(39.5,16.5){\makebox(0,0)[bl]{\(4\)}}


\put(00,06.7){\makebox(0,0)[bl]{\(e^{3,4}_{5}\): }}

\put(08,7){\makebox(0,0)[bl]{\((1,2,1)\) }}

\put(26,9.5){\line(0,1){02.5}} \put(30,9.5){\line(0,1){05}}
\put(34,9.5){\line(0,1){05}} \put(38,9.5){\line(0,1){02.5}}

\put(26,12){\line(1,0){12}} \put(30,14.5){\line(1,0){4}}

\put(26,9.5){\line(1,0){12}}

\put(26,08){\line(0,1){3}} \put(30,08){\line(0,1){3}}
\put(34,08){\line(0,1){3}} \put(38,08){\line(0,1){3}}

\put(27.5,5.5){\makebox(0,0)[bl]{\(1\)}}
\put(31.5,5.5){\makebox(0,0)[bl]{\(2\)}}
\put(35.5,5.5){\makebox(0,0)[bl]{\(3\)}}



\put(47,52.7){\makebox(0,0)[bl]{\(\Longrightarrow\)}}

\put(60,51.7){\makebox(0,0)[bl]{\(e^{4,4}_{1}\): }}

\put(68,53){\makebox(0,0)[bl]{\((3,1,0,0)\) }}

\put(86,54.5){\line(0,1){07.5}} \put(90,54.5){\line(0,1){07.5}}
\put(94,54.5){\line(0,1){02.5}}

\put(86,59.5){\line(1,0){04}} \put(86,62){\line(1,0){04}}

\put(86,57){\line(1,0){08}}

\put(86,54.5){\line(1,0){16}}

\put(86,53){\line(0,1){3}} \put(90,53){\line(0,1){3}}
\put(94,53){\line(0,1){3}} \put(98,53){\line(0,1){3}}
\put(102,53){\line(0,1){3}}

\put(87.5,50.5){\makebox(0,0)[bl]{\(1\)}}
\put(91.5,50.5){\makebox(0,0)[bl]{\(2\)}}
\put(95.5,50.5){\makebox(0,0)[bl]{\(3\)}}
\put(99.5,50.5){\makebox(0,0)[bl]{\(4\)}}


\put(47,41.7){\makebox(0,0)[bl]{\(\Longrightarrow\)}}

\put(60,40.7){\makebox(0,0)[bl]{\(e^{4,4}_{2}\): }}

\put(68,41){\makebox(0,0)[bl]{\((2,1,1,0)\) }}

\put(86,43.5){\line(0,1){05}} \put(90,43.5){\line(0,1){05}}
\put(94,43.5){\line(0,1){02.5}} \put(98,43.5){\line(0,1){02.5}}

\put(86,46){\line(1,0){12}} \put(86,48.5){\line(1,0){4}}

\put(86,43.5){\line(1,0){16}}

\put(86,42){\line(0,1){3}} \put(90,42){\line(0,1){3}}
\put(94,42){\line(0,1){3}} \put(98,42){\line(0,1){3}}
\put(102,42){\line(0,1){3}}

\put(87.5,39.5){\makebox(0,0)[bl]{\(1\)}}
\put(91.5,39.5){\makebox(0,0)[bl]{\(2\)}}
\put(95.5,39.5){\makebox(0,0)[bl]{\(3\)}}
\put(99.5,39.5){\makebox(0,0)[bl]{\(4\)}}


\put(47,29.7){\makebox(0,0)[bl]{\(\Longrightarrow\)}}

\put(60,28.7){\makebox(0,0)[bl]{\(e^{4,4}_{3}\): }}

\put(68,29){\makebox(0,0)[bl]{\((3,1,0,0)\) }}

\put(86,31.5){\line(0,1){07.5}} \put(90,31.5){\line(0,1){07.5}}
\put(94,31.5){\line(0,1){02.5}}

\put(86,34){\line(1,0){08}} \put(86,36.5){\line(1,0){4}}
\put(86,39){\line(1,0){4}}

\put(86,31.5){\line(1,0){16}}

\put(86,30){\line(0,1){3}} \put(90,30){\line(0,1){3}}
\put(94,30){\line(0,1){3}} \put(98,30){\line(0,1){3}}
\put(102,30){\line(0,1){3}}


\put(87.5,27.5){\makebox(0,0)[bl]{\(1\)}}
\put(91.5,27.5){\makebox(0,0)[bl]{\(2\)}}
\put(95.5,27.5){\makebox(0,0)[bl]{\(3\)}}
\put(99.5,27.5){\makebox(0,0)[bl]{\(4\)}}


\put(47,18.7){\makebox(0,0)[bl]{\(\Longrightarrow\)}}

\put(60,17.7){\makebox(0,0)[bl]{\(e^{4,4}_{4}\): }}

\put(68,18){\makebox(0,0)[bl]{\((0,1,2,1)\) }}

\put(90,20.5){\line(0,1){02.5}} \put(94,20.5){\line(0,1){05}}
\put(98,20.5){\line(0,1){05}} \put(102,20.5){\line(0,1){02.5}}

\put(90,23){\line(1,0){12}} \put(94,25.5){\line(1,0){4}}

\put(86,20.5){\line(1,0){16}}

\put(86,19){\line(0,1){3}} \put(90,19){\line(0,1){3}}
\put(94,19){\line(0,1){3}} \put(98,19){\line(0,1){3}}
\put(102,19){\line(0,1){3}}

\put(87.5,16.5){\makebox(0,0)[bl]{\(1\)}}
\put(91.5,16.5){\makebox(0,0)[bl]{\(2\)}}
\put(95.5,16.5){\makebox(0,0)[bl]{\(3\)}}
\put(99.5,16.5){\makebox(0,0)[bl]{\(4\)}}


\put(47,7.7){\makebox(0,0)[bl]{\(\Longrightarrow\)}}

\put(60,06.7){\makebox(0,0)[bl]{\(e^{4,4}_{5}\): }}

\put(68,7){\makebox(0,0)[bl]{\((1,2,1,0)\) }}

\put(86,9.5){\line(0,1){02.5}} \put(90,9.5){\line(0,1){05}}
\put(94,9.5){\line(0,1){05}} \put(98,9.5){\line(0,1){02.5}}

\put(86,12){\line(1,0){12}} \put(90,14.5){\line(1,0){4}}

\put(86,9.5){\line(1,0){16}}

\put(86,08){\line(0,1){3}} \put(90,08){\line(0,1){3}}
\put(94,08){\line(0,1){3}} \put(98,08){\line(0,1){3}}
\put(102,08){\line(0,1){3}}

\put(87.5,5.5){\makebox(0,0)[bl]{\(1\)}}
\put(91.5,5.5){\makebox(0,0)[bl]{\(2\)}}
\put(95.5,5.5){\makebox(0,0)[bl]{\(3\)}}
\put(99.5,5.5){\makebox(0,0)[bl]{\(4\)}}


\end{picture}
\end{center}


\section{Combinatorial Synthesis (morphological approach)}

 Here system composition as combinatorial synthesis
 of design alternatives (for system components)
 which are evaluated via the suggested assessment approach is examined.
 An additional problem consist in integration of design
 alternatives estimates into the total estimate of the composed system.
 This estimate integration is based on summarization of
 component estimates by estimate elements (i.e, position).
 As a result,
 system estimates are based on some analogical poset-like scales.
 This case corresponds to
 combinatorial
 synthesis in
  Hierarchical Morphological Multicriteria Design (HMMD) method.
 Descriptions of HMMD are presented in accessible literature:
 (i) two books  (\cite{lev98},\cite{lev06});
 (ii) journal articles
 (\cite{lev96},\cite{lev01},\cite{lev02},\cite{lev05},\cite{lev09});
%
 (iii) electronic preprint \cite{lev12morph}.
 The method is based on morphological clique problem.
 In basic HMMD method, ordinal assessment for design alternatives
 is used (e.g., problems \(P^{3,1}\), \(P^{4,1}\)).
 The section contains the following:

 (1) a brief description of basic HMMD method
 (combinatorial synthesis with ordinal assessment of design
 alternatives);

 (2) two numerical examples for basic HMMD method
 (three-component system and four-component system);
 and

 (3) modified  version of HMMD method that is based
 on the suggested  multiset estimates
  for evaluation of design alternatives
 (numerical examples:
 (i) three-component system and assessment problem \(P^{3,3}\),
 (ii) four-component system and assessment problem \(P^{3,4}\),
 and
 (iii) three-layer hierarchical system.

 Table 5 contains a list of the described
  combinatorial synthesis problems.

\begin{center}
\begin{picture}(115,39)
\put(023,35){\makebox(0,0)[bl] {Table 5. Problem of combinatorial
synthesis}}


\put(00,00){\line(1,0){115}} \put(00,23){\line(1,0){115}}
\put(00,33){\line(1,0){115}}

\put(00,00){\line(0,1){33}} \put(05,00){\line(0,1){33}}
\put(28,00){\line(0,1){33}} \put(50,00){\line(0,1){33}}
\put(70,00){\line(0,1){33}} \put(98,00){\line(0,1){33}}
\put(115,00){\line(0,1){33}}

\put(02,29){\makebox(0,0)[bl]{ }}

\put(6,29){\makebox(0,0)[bl]{Number of}}
\put(6,25){\makebox(0,0)[bl]{system layers}}

\put(29,29){\makebox(0,0)[bl]{Number of}}
\put(29,25){\makebox(0,0)[bl]{system parts}}

\put(51,29){\makebox(0,0)[bl]{Assessment}}
\put(51,25){\makebox(0,0)[bl]{problem}}

\put(71,28.5){\makebox(0,0)[bl]{Type of }}
\put(71,25){\makebox(0,0)[bl]{estimate for DAs}}

\put(99,29){\makebox(0,0)[bl]{Version of}}
\put(99,25){\makebox(0,0)[bl]{HMMD}}


\put(01.5,18){\makebox(0,0)[bl]{\(1\)}}

\put(16,18){\makebox(0,0)[bl]{\(2\)}}

\put(38,18){\makebox(0,0)[bl]{\(3\)}}

\put(58,18){\makebox(0,0)[bl]{\(P^{31}\)}}

\put(71,18){\makebox(0,0)[bl]{ordinal}}

\put(99,18){\makebox(0,0)[bl]{basic}}


\put(01.5,14){\makebox(0,0)[bl]{\(2\)}}

\put(16,14){\makebox(0,0)[bl]{\(2\)}}

\put(38,14){\makebox(0,0)[bl]{\(4\)}}

\put(58,14){\makebox(0,0)[bl]{\(P^{41}\)}}

\put(71,14){\makebox(0,0)[bl]{ordinal}}

\put(99,14){\makebox(0,0)[bl]{basic}}


\put(01.5,10){\makebox(0,0)[bl]{\(3\)}}

\put(16,10){\makebox(0,0)[bl]{\(2\)}}

\put(38,10){\makebox(0,0)[bl]{\(3\)}}

\put(58,10){\makebox(0,0)[bl]{\(P^{33}\)}}

\put(71,10){\makebox(0,0)[bl]{multiset}}

\put(99,10){\makebox(0,0)[bl]{modified}}


\put(01.5,06){\makebox(0,0)[bl]{\(4\)}}

\put(16,06){\makebox(0,0)[bl]{\(2\)}}

\put(38,06){\makebox(0,0)[bl]{\(4\)}}

\put(58,06){\makebox(0,0)[bl]{\(P^{34}\)}}

\put(71,06){\makebox(0,0)[bl]{multiset}}

\put(99,06){\makebox(0,0)[bl]{modified}}


\put(01.5,02){\makebox(0,0)[bl]{\(5\)}}

\put(16,02){\makebox(0,0)[bl]{\(3\)}}

\put(38,02){\makebox(0,0)[bl]{\(3\)}}

\put(58,02){\makebox(0,0)[bl]{\(P^{33}\)}}

\put(71,02){\makebox(0,0)[bl]{multiset}}

\put(99,02){\makebox(0,0)[bl]{modified}}

\end{picture}
\end{center}

\subsection{Basic Combinatorial Synthesis (ordinal assessment of alternatives)}

%
 An examined composite
 (modular, decomposable) system consists
 of components and their interconnection or compatibility (IC).
 Basic assumptions of HMMD are the following:
 ~{\it (a)} a tree-like structure of the system;
 ~{\it (b)} a composite estimate for system quality
     that integrates components (subsystems, parts) qualities and
    qualities of IC (compatibility) across subsystems;
 ~{\it (c)} monotonic criteria for the system and its components;
 ~{\it (d)} quality of system components and IC are evaluated on the basis
    of coordinated ordinal scales.
 The designations are:
  ~(1) design alternatives (DAs) for leaf nodes of the model;
  ~(2) priorities of DAs (\(\iota = \overline{1,l}\);
      \(1\) corresponds to the best one);
  ~(3) ordinal compatibility for each pair of DAs
  (\(w=\overline{1,\nu}\); \(\nu\) corresponds to the best one).
 The basic phases of HMMD are:

  ~{\it Phase 1.} design of the tree-like system model;

  ~{\it Phase 2.} generation of DAs for leaf nodes of the model;

  ~{\it Phase 3.} hierarchical selection and composing of DAs into composite
    DAs for the corresponding higher level of the system hierarchy;

  ~{\it Phase 4.} analysis and improvement of composite DAs (decisions).

%

 Let \(S\) be a system consisting of \(m\) parts (components):
 \(R(1),...,R(i),...,R(m)\).
 A set of design alternatives
 is generated for each system part above.
 The problem is:

~~

 {\it Find a composite design alternative}
 ~~ \(S=S(1)\star ...\star S(i)\star ...\star S(m)\)~~
 {\it of DAs (one representative design alternative}
 ~\(S(i)\)
 {\it for each system component/part}
  ~\(R(i)\), \(i=\overline{1,m}\)
  {\it )}
 {\it with non-zero compatibility}
 {\it between design alternatives.}

~~

 A discrete space of the system excellence on the basis of the
 following vector is used:
 ~~\(N(S)=(w(S);e(S))\),
 ~where \(w(S)\) is the minimum of pairwise compatibility
 between DAs which correspond to different system components
 (i.e.,
 \(~\forall ~R_{j_{1}}\) and \( R_{j_{2}}\),
 \(1 \leq j_{1} \neq j_{2} \leq m\))
 in \(S\),
 ~\(e(S)=(\eta_{1},...,\eta_{\iota},...,\eta_{l})\),
 ~where \(\eta_{\iota}\) is the number of DAs of the \(\iota\)th quality in \(S\).
 Thus, the modified problem is (two objectives, one constraint):
%
%
%
%
 \[ \max~ e(S)
%
 ,\]
 \[ \max~~ w(S), \]
%
%
 \[s.t.
%
  ~~~w(S) \geq 1  .\]
%

 As a result,
 we search for composite decisions which are nondominated by \(N(S)\)
 (i.e., Pareto-efficient solutions).
 ``Maximization''  of \(e(S)\) is based on the corresponding poset.
%
%
%
 The considered combinatorial problem is NP-hard
 and an enumerative solving scheme is used.

\subsection{Example: basic HMMD, three-component system}

 Here
 three component system and assessment problem \(P^{3,1}\)
 are examined.
%
%
%
 Fig. 10 illustrate the composition problem by a numerical example,
 Table 6 contains compatibility estimates.
%
 In the example, Pareto-efficient composite DAs are:

 (a) ~\(S_{1}=X_{1}\star Y_{1}\star Z_{2}\), ~\(N(S_{1})=(3;1,1,1)\);

 (b) ~\(S_{2}=X_{2}\star Y_{1}\star Z_{2}\), ~\(N(S_{2})=(2;2,0,1)\);
 and

 (c) ~\(S_{3}=X_{3}\star Y_{2}\star Z_{1}\), ~\(N(S_{3})=(1;3,0,0)\).

 Fig. 11 depicts the poset for integrated system estimate by
 components, Fig. 12 depicts the general poset of system quality.

\begin{center}
\begin{picture}(60,48)

\put(00,00){\makebox(0,0)[bl] {Fig. 10. Example of composition}}

\put(4,11){\makebox(0,8)[bl]{\(X_{3}(1)\)}}
\put(4,15){\makebox(0,8)[bl]{\(X_{2}(1)\)}}
\put(4,19){\makebox(0,8)[bl]{\(X_{1}(2)\)}}

\put(19,07){\makebox(0,8)[bl]{\(Y_{4}(3)\)}}
\put(19,11){\makebox(0,8)[bl]{\(Y_{3}(3)\)}}
\put(19,15){\makebox(0,8)[bl]{\(Y_{2}(1)\)}}
\put(19,19){\makebox(0,8)[bl]{\(Y_{1}(3)\)}}

\put(34,11){\makebox(0,8)[bl]{\(Z_{3}(3)\)}}
\put(34,15){\makebox(0,8)[bl]{\(Z_{2}(1)\)}}
\put(34,19){\makebox(0,8)[bl]{\(Z_{1}(1)\)}}
\put(3,24){\circle{2}} \put(18,24){\circle{2}}
\put(33,24){\circle{2}}

\put(0,24){\line(1,0){02}} \put(15,24){\line(1,0){02}}
\put(30,24){\line(1,0){02}}

\put(0,24){\line(0,-1){13}} \put(15,24){\line(0,-1){17}}
\put(30,24){\line(0,-1){13}}

\put(30,19){\line(1,0){01}} \put(30,15){\line(1,0){01}}
\put(30,11){\line(1,0){01}}

\put(32,19){\circle{2}} \put(32,19){\circle*{1}}
\put(32,15){\circle{2}} \put(32,15){\circle*{1}}
\put(32,11){\circle{2}} \put(32,11){\circle*{1}}

\put(15,07){\line(1,0){01}} \put(15,11){\line(1,0){01}}
\put(15,15){\line(1,0){01}} \put(15,19){\line(1,0){01}}

\put(17,15){\circle{2}} \put(17,15){\circle*{1}}
\put(17,11){\circle{2}} \put(17,19){\circle{2}}
\put(17,11){\circle*{1}} \put(17,19){\circle*{1}}
\put(17,07){\circle{2}} \put(17,07){\circle*{1}}

\put(0,11){\line(1,0){01}} \put(0,15){\line(1,0){01}}
\put(0,19){\line(1,0){01}}

\put(2,15){\circle{2}} \put(2,19){\circle{2}}
\put(2,15){\circle*{1}} \put(2,19){\circle*{1}}
\put(2,11){\circle{2}} \put(2,11){\circle*{1}}
\put(3,29){\line(0,-1){04}} \put(18,29){\line(0,-1){04}}
\put(33,29){\line(0,-1){04}}

\put(3,29){\line(1,0){30}} \put(16,29){\line(0,1){15}}

\put(16,44){\circle*{3}}

\put(4,26){\makebox(0,8)[bl]{X}} \put(14,26){\makebox(0,8)[bl]{Y}}
\put(29,26){\makebox(0,8)[bl]{Z}}

\put(21,43){\makebox(0,8)[bl]{\(S = X\star Y\star Z\)}}


\put(18,39){\makebox(0,8)[bl] {\(S_{1}=X_{1}\star Y_{1}\star
Z_{2}\)}}

\put(18,35){\makebox(0,8)[bl] {\(S_{2}=X_{2}\star Y_{1}\star
Z_{2}\)}}

\put(18,31){\makebox(0,8)[bl] {\(S_{3}=X_{3}\star Y_{2}\star
Z_{1}\)}}

\end{picture}
%
\begin{picture}(42,43)

\put(03,38){\makebox(0,0)[bl]{Table 6. Compatibility}}

\put(00,0){\line(1,0){42}} \put(00,30){\line(1,0){42}}
\put(00,36){\line(1,0){42}}

\put(00,0){\line(0,1){36}} \put(07,0){\line(0,1){36}}
\put(42,0){\line(0,1){36}}


\put(01,26){\makebox(0,0)[bl]{\(X_{1}\)}}
\put(01,22){\makebox(0,0)[bl]{\(X_{2}\)}}
\put(01,18){\makebox(0,0)[bl]{\(X_{3}\)}}

\put(01,14){\makebox(0,0)[bl]{\(Y_{1}\)}}
\put(01,10){\makebox(0,0)[bl]{\(Y_{2}\)}}
\put(01,06){\makebox(0,0)[bl]{\(Y_{3}\)}}
\put(01,02){\makebox(0,0)[bl]{\(Y_{4}\)}}


\put(12,30){\line(0,1){6}} \put(17,30){\line(0,1){6}}
\put(22,30){\line(0,1){6}} \put(27,30){\line(0,1){6}}
\put(32,30){\line(0,1){6}} \put(37,30){\line(0,1){6}}

\put(07.4,32){\makebox(0,0)[bl]{\(Y_{1}\)}}
\put(12.4,32){\makebox(0,0)[bl]{\(Y_{2}\)}}
\put(17.4,32){\makebox(0,0)[bl]{\(Y_{3}\)}}
\put(22.4,32){\makebox(0,0)[bl]{\(Y_{4}\)}}

\put(27.4,32){\makebox(0,0)[bl]{\(Z_{1}\)}}
\put(32.4,32){\makebox(0,0)[bl]{\(Z_{2}\)}}
\put(37.4,32){\makebox(0,0)[bl]{\(Z_{3}\)}}



\put(09,26){\makebox(0,0)[bl]{\(3\)}}
\put(14,26){\makebox(0,0)[bl]{\(0\)}}
\put(19,26){\makebox(0,0)[bl]{\(0\)}}
\put(24,26){\makebox(0,0)[bl]{\(0\)}}
\put(29,26){\makebox(0,0)[bl]{\(0\)}}
\put(34,26){\makebox(0,0)[bl]{\(3\)}}
\put(39,26){\makebox(0,0)[bl]{\(0\)}}


\put(09,22){\makebox(0,0)[bl]{\(2\)}}
\put(14,22){\makebox(0,0)[bl]{\(0\)}}
\put(19,22){\makebox(0,0)[bl]{\(1\)}}
\put(24,22){\makebox(0,0)[bl]{\(1\)}}
\put(29,22){\makebox(0,0)[bl]{\(0\)}}
\put(34,22){\makebox(0,0)[bl]{\(2\)}}
\put(39,22){\makebox(0,0)[bl]{\(2\)}}


\put(09,18){\makebox(0,0)[bl]{\(0\)}}
\put(14,18){\makebox(0,0)[bl]{\(1\)}}
\put(19,18){\makebox(0,0)[bl]{\(0\)}}
\put(24,18){\makebox(0,0)[bl]{\(0\)}}
\put(29,18){\makebox(0,0)[bl]{\(1\)}}
\put(34,18){\makebox(0,0)[bl]{\(0\)}}
\put(39,18){\makebox(0,0)[bl]{\(0\)}}


\put(29,14){\makebox(0,0)[bl]{\(0\)}}
\put(34,14){\makebox(0,0)[bl]{\(3\)}}
\put(39,14){\makebox(0,0)[bl]{\(2\)}}


\put(29,10){\makebox(0,0)[bl]{\(1\)}}
\put(34,10){\makebox(0,0)[bl]{\(0\)}}
\put(39,10){\makebox(0,0)[bl]{\(0\)}}


\put(29,06){\makebox(0,0)[bl]{\(1\)}}
\put(34,06){\makebox(0,0)[bl]{\(1\)}}
\put(39,06){\makebox(0,0)[bl]{\(2\)}}


\put(29,02){\makebox(0,0)[bl]{\(0\)}}
\put(34,02){\makebox(0,0)[bl]{\(2\)}}
\put(39,02){\makebox(0,0)[bl]{\(3\)}}

\end{picture}
\end{center}
%

%

\begin{center}
\begin{picture}(62,84)

\put(00,00){\makebox(0,0)[bl] {Fig. 11. Poset
 \(e(S)=(\eta_{1},\eta_{2},\eta_{3})\)}}


\put(28,77){\makebox(0,0)[bl]{The ideal}}
\put(28,74){\makebox(0,0)[bl]{point}}

\put(10,75){\makebox(0,0)[bl]{\(<3,0,0>\) }}
\put(18,77){\oval(16,5)} \put(18,77){\oval(16.5,5.5)}

\put(00,69.1){\makebox(0,0)[bl]{\(e(S_{3})\)}}
\put(05,73.6){\vector(2,1){04}}

\put(18,70){\line(0,1){4}}
\put(10,65){\makebox(0,0)[bl]{\(<2,1,0>\)}}
\put(18,67){\oval(16,5)}

\put(29,65.5){\makebox(0,0)[bl]{\(e(S_{2})\)}}
\put(31,65){\vector(-1,-2){4}}

\put(18,58){\line(0,1){6}}
\put(10,53){\makebox(0,0)[bl]{\(<2,0,1>\) }}
\put(18,55){\oval(16,5)}

\put(00,49.4){\makebox(0,0)[bl]{\(e(S_{1})\)}}
\put(05,49){\vector(1,-1){04}}

\put(18,46){\line(0,1){6}}
\put(10,41){\makebox(0,0)[bl]{\(<1,1,1>\) }}
\put(18,43){\oval(16,5)}

\put(18,34){\line(0,1){6}}
\put(10,29){\makebox(0,0)[bl]{\(<1,0,2>\) }}
\put(18,31){\oval(16,5)}

\put(18,22){\line(0,1){6}}
\put(10,17){\makebox(0,0)[bl]{\(<0,1,2>\) }}
\put(18,19){\oval(16,5)}

\put(18,12){\line(0,1){4}}

\put(10,07){\makebox(0,0)[bl]{\(<0,0,3>\) }}
\put(18,09){\oval(16,5)}

\put(28,11){\makebox(0,0)[bl]{The worst}}
\put(28,08){\makebox(0,0)[bl]{point}}
\put(20.5,63.5){\line(3,-1){15}}

\put(35.5,52){\line(-3,-1){15}}

\put(30,53){\makebox(0,0)[bl]{\(<1,2,0>\) }}
\put(38,55){\oval(16,5)}

\put(38,46){\line(0,1){6}}
\put(30,41){\makebox(0,0)[bl]{\(<0,3,0>\) }}
\put(38,43){\oval(16,5)}
\put(20.5,39.5){\line(3,-1){15}}

\put(38,34){\line(0,1){6}}
\put(30,29){\makebox(0,0)[bl]{\(<0,2,1>\) }}
\put(38,31){\oval(16,5)}
\put(35.5,27.5){\line(-3,-1){15}}

\end{picture}
%
\begin{picture}(60,63)
\put(00,00){\makebox(0,0)[bl] {Fig. 12. Poset of system quality
\(N(S)\)}}

\put(00,010){\line(0,1){40}} \put(00,010){\line(3,4){15}}
\put(00,050){\line(3,-4){15}}

\put(20,015){\line(0,1){40}} \put(20,015){\line(3,4){15}}
\put(20,055){\line(3,-4){15}}

\put(40,020){\line(0,1){40}} \put(40,020){\line(3,4){15}}
\put(40,060){\line(3,-4){15}}

\put(00,50){\circle*{2}}
\put(01.5,50){\makebox(0,0)[bl]{\(N(S_{3})\)}}

\put(20,45){\circle*{2}}
\put(08.5,43){\makebox(0,0)[bl]{\(N(S_{2})\)}}

\put(40,40){\circle*{2}}
\put(40.5,41.6){\makebox(0,0)[bl]{\(N(S_{1})\)}}

\put(40,60){\circle{2}} \put(40,60){\circle*{1}}


\put(43,58.6){\makebox(0,0)[bl]{The ideal}}
\put(45,55.6){\makebox(0,0)[bl]{point}}

\put(00,07){\makebox(0,0)[bl]{\(w=1\)}}
\put(20,12){\makebox(0,0)[bl]{\(w=2\)}}
\put(40,17){\makebox(0,0)[bl]{\(w=3\)}}

\put(00,10){\circle{1.7}}

\put(01.5,14){\makebox(0,0)[bl]{The worst}}
\put(02.5,11){\makebox(0,0)[bl]{point}}

\end{picture}
\end{center}

\subsection{Example: basic HMMD, four-component system}
%
%
 Here the following composition problem is considered:

 (i) the system consists of 4 components,

 (ii) assessment problem \(P^{3,1}\)
 is used for the evaluation of
 DAs for components (Fig. 2), and

 (iii) integrated system quality is based on poset
 (Fig. 13).

%
 Let us describe the corresponding numerical example
 (a corrected version of the example from \cite{lev96}).
 The system structure and DAs are presented in Fig. 14
 (ordinal estimates of DAs are shown in parentheses).
%
%
 In addition, estimates of compatibility between DAs
 (scale \([0,1,2,3,4,5]\)) are presented in Table 7.
 Resultant Pareto-efficient composite DAs are:

 (a)
~\(S_{1}=X_{3}\star Y_{5}\star Z_{3} \star V_{4}\),
%
%
 ~\(N(S_{1})  (4;1,3,0) \);

 (b)
 %
%
~\(S_{2}=X_{3}\star Y_{5}\star Z_{2} \star V_{5}\),
  \(N(S_{2})  = (3;2,2,0) \); and

  (c)
~\(S_{3}=X_{3}\star Y_{3}\star Z_{2} \star V_{4}\),
%
%
  \(N(S_{3})= (2;3,1,0) \).

 Fig. 13 illustrates the poset
  of quality for obtained composite DAs by components
 (i.e., \(e(S)=(\eta_{1},\eta_{2},\eta_{3})\)).
  The general
  poset
  of quality
 (by \(N(S)\)) is depicted in Fig. 15.

\begin{center}
\begin{picture}(82,106)
\put(011,00){\makebox(0,0)[bl] {Fig. 13. Poset   
\(e(S)=(\eta_{1},\eta_{2},\eta_{3})\)}}


\put(01,101){\makebox(0,0)[bl]{The ideal}}
\put(01,98){\makebox(0,0)[bl]{point}}

\put(20,99){\makebox(0,0)[bl]{\(<4,0,0>\) }}
\put(28,101){\oval(16,5)} \put(28,101){\oval(16.5,5.5)}

\put(28,94){\line(0,1){4}}

\put(20,89){\makebox(0,0)[bl]{\(<3,1,0>\)}}
\put(28,91){\oval(16,5)}

\put(28,82){\line(0,1){6}}

\put(20,77){\makebox(0,0)[bl]{\(<3,0,1>\) }}
\put(28,79){\oval(16,5)}



\put(28,70){\line(0,1){6}}
\put(20,65){\makebox(0,0)[bl]{\(<2,1,1>\)}}
\put(28,67){\oval(16,5)}


\put(28,58){\line(0,1){6}}
\put(20,53){\makebox(0,0)[bl]{\(<2,0,2>\) }}
\put(28,55){\oval(16,5)}

%
\put(03,81){\makebox(0,0)[bl]{\(e(S_{3})\)}}
\put(10,85){\vector(2,1){08}}

%
\put(63,86){\makebox(0,0)[bl]{\(e(S_{2})\)}}

\put(65,85){\vector(-2,-1){08}}

%
\put(63,74){\makebox(0,0)[bl]{\(e(S_{1})\)}}

\put(65,73){\vector(-2,-1){08}}


\put(28,46){\line(0,1){6}}
\put(20,41){\makebox(0,0)[bl]{\(<1,1,2>\) }}
\put(28,43){\oval(16,5)}

\put(28,34){\line(0,1){6}}
\put(20,29){\makebox(0,0)[bl]{\(<1,0,3>\) }}
\put(28,31){\oval(16,5)}

\put(28,22){\line(0,1){6}}
\put(20,17){\makebox(0,0)[bl]{\(<0,1,2>\) }}
\put(28,19){\oval(16,5)}

\put(28,12){\line(0,1){4}}

\put(20,07){\makebox(0,0)[bl]{\(<0,0,4>\) }}
\put(28,09){\oval(16,5)}

\put(01,11){\makebox(0,0)[bl]{The worst}}
\put(01,08){\makebox(0,0)[bl]{point}}


\put(45.5,82.5){\line(-3,1){15}}

\put(45.5,75.5){\line(-3,-1){15}}

\put(40,77){\makebox(0,0)[bl]{\(<2,2,0>\) }}
\put(48,79){\oval(16,5)}

\put(48,70){\line(0,1){6}}
\put(40,65){\makebox(0,0)[bl]{\(<1,3,0>\) }}
\put(48,67){\oval(16,5)}

\put(30.5,63.5){\line(3,-1){15}}

\put(48,58){\line(0,1){6}}
\put(40,53){\makebox(0,0)[bl]{\(<1,2,1>\) }}
\put(48,55){\oval(16,5)}
\put(30.5,46.5){\line(3,1){15}}

\put(48,46){\line(0,1){6}}

\put(40,41){\makebox(0,0)[bl]{\(<0,3,1>\) }}
\put(48,43){\oval(16,5)}
\put(30.5,39.5){\line(3,-1){15}}

\put(48,34){\line(0,1){6}}
\put(40,29){\makebox(0,0)[bl]{\(<0,2,2>\) }}
\put(48,31){\oval(16,5)}
\put(45.5,27.5){\line(-3,-1){15}}


\put(65.5,58.5){\line(-3,1){15}} \put(65.5,51.5){\line(-3,-1){15}}

\put(60,53){\makebox(0,0)[bl]{\(<0,4,0>\) }}
\put(68,55){\oval(16,5)}

\end{picture}
\end{center}

\begin{center}
\begin{picture}(64,53)
\put(02,00){\makebox(0,0)[bl] {Fig. 14. Example of composition}}

\put(4,21){\makebox(0,8)[bl]{\(X_{1}(2)\)}}
\put(4,17){\makebox(0,8)[bl]{\(X_{2}(3)\)}}
\put(4,13){\makebox(0,8)[bl]{\(X_{3}(2)\)}}

\put(19,21){\makebox(0,8)[bl]{\(Y_{1}(3)\)}}
\put(19,17){\makebox(0,8)[bl]{\(Y_{2}(2)\)}}
\put(19,13){\makebox(0,8)[bl]{\(Y_{3}(1)\)}}
\put(19,09){\makebox(0,8)[bl]{\(Y_{4}(3)\)}}
\put(19,05){\makebox(0,8)[bl]{\(Y_{5}(2)\)}}

\put(34,21){\makebox(0,8)[bl]{\(Z_{1}(2)\)}}
\put(34,17){\makebox(0,8)[bl]{\(Z_{2}(1)\)}}
\put(34,13){\makebox(0,8)[bl]{\(Z_{3}(2)\)}}

\put(49,21){\makebox(0,8)[bl]{\(V_{1}(3)\)}}
\put(49,17){\makebox(0,8)[bl]{\(V_{2}(2)\)}}
\put(49,13){\makebox(0,8)[bl]{\(V_{3}(2)\)}}
\put(49,09){\makebox(0,8)[bl]{\(V_{4}(1)\)}}
\put(49,05){\makebox(0,8)[bl]{\(V_{5}(2)\)}}

\put(3,28){\circle{2}} \put(18,28){\circle{2}}
\put(33,28){\circle{2}} \put(48,28){\circle{2}}

\put(0,28){\line(1,0){02}} \put(15,28){\line(1,0){02}}
\put(30,28){\line(1,0){02}} \put(45,28){\line(1,0){02}}

\put(0,28){\line(0,-1){13}} \put(15,28){\line(0,-1){21}}
\put(30,28){\line(0,-1){13}} \put(45,28){\line(0,-1){21}}

\put(45,19){\line(1,0){01}} \put(45,19){\line(1,0){01}}
\put(45,15){\line(1,0){01}} \put(45,11){\line(1,0){01}}
\put(45,07){\line(1,0){01}}

\put(47,23){\circle{2}} \put(47,23){\circle*{1}}
\put(47,19){\circle{2}} \put(47,19){\circle*{1}}
\put(47,15){\circle{2}} \put(47,15){\circle*{1}}
\put(47,11){\circle{2}} \put(47,11){\circle*{1}}
\put(47,07){\circle{2}} \put(47,07){\circle*{1}}

\put(30,23){\line(1,0){01}} \put(30,19){\line(1,0){01}}
\put(30,15){\line(1,0){01}}

\put(32,23){\circle{2}} \put(32,23){\circle*{1}}
\put(32,19){\circle{2}} \put(32,19){\circle*{1}}
\put(32,15){\circle{2}} \put(32,15){\circle*{1}}

\put(15,07){\line(1,0){01}} \put(15,11){\line(1,0){01}}
\put(15,15){\line(1,0){01}} \put(15,19){\line(1,0){01}}
\put(15,23){\line(1,0){01}}

\put(17,19){\circle{2}} \put(17,19){\circle*{1}}
\put(17,15){\circle{2}} \put(17,23){\circle{2}}
\put(17,15){\circle*{1}} \put(17,23){\circle*{1}}
\put(17,11){\circle{2}} \put(17,11){\circle*{1}}
\put(17,07){\circle{2}} \put(17,07){\circle*{1}}

\put(0,15){\line(1,0){01}} \put(0,19){\line(1,0){01}}
\put(0,23){\line(1,0){01}}

\put(2,19){\circle{2}} \put(2,23){\circle{2}}
\put(2,19){\circle*{1}} \put(2,23){\circle*{1}}
\put(2,15){\circle{2}} \put(2,15){\circle*{1}}
\put(3,33){\line(0,-1){04}} \put(18,33){\line(0,-1){04}}
\put(33,33){\line(0,-1){04}} \put(48,33){\line(0,-1){04}}

\put(3,33){\line(1,0){45}} \put(17,33){\line(0,1){15}}

\put(17,48){\circle*{3}}

\put(04,30){\makebox(0,8)[bl]{X}}
\put(14,30){\makebox(0,8)[bl]{Y}}
\put(29,30){\makebox(0,8)[bl]{Z}}
\put(44,30){\makebox(0,8)[bl]{V}}

\put(21,47.5){\makebox(0,8)[bl]{\(S = X\star Y\star Z \star V\)}}

\put(19,43){\makebox(0,8)[bl] {\(S_{1}=X_{3}\star Y_{5}\star Z_{3}
 \star V_{4}\)}}


\put(19,39){\makebox(0,8)[bl] {\(S_{2}=X_{3}\star Y_{5}\star Z_{2}
 \star V_{5}\)}}

\put(19,35){\makebox(0,8)[bl] {\(S_{3}=X_{3}\star Y_{3}\star Z_{2}
 \star V_{4}\)}}

\end{picture}
\end{center}

\begin{center}
\begin{picture}(72,60)

\put(09.5,56){\makebox(0,0)[bl]{Table 7. Compatibility estimates}}

\put(00,0){\line(1,0){72}} \put(00,47){\line(1,0){72}}
\put(00,54){\line(1,0){72}}

\put(00,0){\line(0,1){54}} \put(07,0){\line(0,1){54}}
\put(72,0){\line(0,1){54}}

\put(01,42){\makebox(0,0)[bl]{\(X_{3}\)}}
\put(01,38){\makebox(0,0)[bl]{\(X_{1}\)}}
\put(01,34){\makebox(0,0)[bl]{\(X_{2}\)}}

\put(01,30){\makebox(0,0)[bl]{\(Y_{3}\)}}
\put(01,26){\makebox(0,0)[bl]{\(Y_{2}\)}}
\put(01,22){\makebox(0,0)[bl]{\(Y_{5}\)}}
\put(01,18){\makebox(0,0)[bl]{\(Y_{1}\)}}
\put(01,14){\makebox(0,0)[bl]{\(Y_{4}\)}}

\put(01,10){\makebox(0,0)[bl]{\(Z_{2}\)}}
\put(01,06){\makebox(0,0)[bl]{\(Z_{1}\)}}
\put(01,02){\makebox(0,0)[bl]{\(Z_{3}\)}}

\put(12,47){\line(0,1){7}} \put(17,47){\line(0,1){7}}
\put(22,47){\line(0,1){7}} \put(27,47){\line(0,1){7}}
\put(32,47){\line(0,1){7}} \put(37,47){\line(0,1){7}}
\put(42,47){\line(0,1){7}} \put(47,47){\line(0,1){7}}
\put(52,47){\line(0,1){7}} \put(57,47){\line(0,1){7}}
\put(62,47){\line(0,1){7}} \put(67,47){\line(0,1){7}}

\put(07.4,49){\makebox(0,0)[bl]{\(Y_{3}\)}}
\put(12.4,49){\makebox(0,0)[bl]{\(Y_{2}\)}}
\put(17.4,49){\makebox(0,0)[bl]{\(Y_{5}\)}}
\put(22.4,49){\makebox(0,0)[bl]{\(Y_{1}\)}}
\put(27.4,49){\makebox(0,0)[bl]{\(Y_{4}\)}}

\put(32.4,49){\makebox(0,0)[bl]{\(Z_{2}\)}}
\put(37.4,49){\makebox(0,0)[bl]{\(Z_{1}\)}}
\put(42.4,49){\makebox(0,0)[bl]{\(Z_{3}\)}}

\put(47.4,49){\makebox(0,0)[bl]{\(V_{4}\)}}
\put(52.4,49){\makebox(0,0)[bl]{\(V_{5}\)}}
\put(57.4,49){\makebox(0,0)[bl]{\(V_{2}\)}}
\put(62.4,49){\makebox(0,0)[bl]{\(V_{3}\)}}
\put(67.4,49){\makebox(0,0)[bl]{\(V_{1}\)}}



\put(09,42){\makebox(0,0)[bl]{\(5\)}}
\put(14,42){\makebox(0,0)[bl]{\(0\)}}
\put(19,42){\makebox(0,0)[bl]{\(5\)}}
\put(24,42){\makebox(0,0)[bl]{\(0\)}}
\put(29,42){\makebox(0,0)[bl]{\(5\)}}

\put(34,42){\makebox(0,0)[bl]{\(3\)}}
\put(39,42){\makebox(0,0)[bl]{\(1\)}}
\put(44,42){\makebox(0,0)[bl]{\(4\)}}

\put(49,42){\makebox(0,0)[bl]{\(4\)}}
\put(54,42){\makebox(0,0)[bl]{\(4\)}}
\put(59,42){\makebox(0,0)[bl]{\(3\)}}
\put(64,42){\makebox(0,0)[bl]{\(3\)}}
\put(69,42){\makebox(0,0)[bl]{\(0\)}}


\put(09,38){\makebox(0,0)[bl]{\(0\)}}
\put(14,38){\makebox(0,0)[bl]{\(5\)}}
\put(19,38){\makebox(0,0)[bl]{\(1\)}}
\put(24,38){\makebox(0,0)[bl]{\(3\)}}
\put(29,38){\makebox(0,0)[bl]{\(0\)}}

\put(34,38){\makebox(0,0)[bl]{\(2\)}}
\put(39,38){\makebox(0,0)[bl]{\(5\)}}
\put(44,38){\makebox(0,0)[bl]{\(3\)}}

\put(49,38){\makebox(0,0)[bl]{\(1\)}}
\put(54,38){\makebox(0,0)[bl]{\(1\)}}
\put(59,38){\makebox(0,0)[bl]{\(1\)}}
\put(64,38){\makebox(0,0)[bl]{\(1\)}}
\put(69,38){\makebox(0,0)[bl]{\(3\)}}


\put(09,34){\makebox(0,0)[bl]{\(0\)}}
\put(14,34){\makebox(0,0)[bl]{\(3\)}}
\put(19,34){\makebox(0,0)[bl]{\(1\)}}
\put(24,34){\makebox(0,0)[bl]{\(5\)}}
\put(29,34){\makebox(0,0)[bl]{\(0\)}}

\put(34,34){\makebox(0,0)[bl]{\(2\)}}
\put(39,34){\makebox(0,0)[bl]{\(4\)}}
\put(44,34){\makebox(0,0)[bl]{\(3\)}}

\put(49,34){\makebox(0,0)[bl]{\(1\)}}
\put(54,34){\makebox(0,0)[bl]{\(1\)}}
\put(59,34){\makebox(0,0)[bl]{\(1\)}}
\put(64,34){\makebox(0,0)[bl]{\(1\)}}
\put(69,34){\makebox(0,0)[bl]{\(3\)}}



\put(34,30){\makebox(0,0)[bl]{\(2\)}}
\put(39,30){\makebox(0,0)[bl]{\(5\)}}
\put(44,30){\makebox(0,0)[bl]{\(1\)}}

\put(49,30){\makebox(0,0)[bl]{\(4\)}}
\put(54,30){\makebox(0,0)[bl]{\(2\)}}
\put(59,30){\makebox(0,0)[bl]{\(1\)}}
\put(64,30){\makebox(0,0)[bl]{\(1\)}}
\put(69,30){\makebox(0,0)[bl]{\(5\)}}



\put(34,26){\makebox(0,0)[bl]{\(2\)}}
\put(39,26){\makebox(0,0)[bl]{\(3\)}}
\put(44,26){\makebox(0,0)[bl]{\(2\)}}

\put(49,26){\makebox(0,0)[bl]{\(1\)}}
\put(54,26){\makebox(0,0)[bl]{\(1\)}}
\put(59,26){\makebox(0,0)[bl]{\(1\)}}
\put(64,26){\makebox(0,0)[bl]{\(1\)}}
\put(69,26){\makebox(0,0)[bl]{\(5\)}}



\put(34,22){\makebox(0,0)[bl]{\(5\)}}
\put(39,22){\makebox(0,0)[bl]{\(1\)}}
\put(44,22){\makebox(0,0)[bl]{\(5\)}}

\put(49,22){\makebox(0,0)[bl]{\(5\)}}
\put(54,22){\makebox(0,0)[bl]{\(5\)}}
\put(59,22){\makebox(0,0)[bl]{\(5\)}}
\put(64,22){\makebox(0,0)[bl]{\(5\)}}
\put(69,22){\makebox(0,0)[bl]{\(1\)}}



\put(34,18){\makebox(0,0)[bl]{\(2\)}}
\put(39,18){\makebox(0,0)[bl]{\(3\)}}
\put(44,18){\makebox(0,0)[bl]{\(2\)}}

\put(49,18){\makebox(0,0)[bl]{\(1\)}}
\put(54,18){\makebox(0,0)[bl]{\(1\)}}
\put(59,18){\makebox(0,0)[bl]{\(1\)}}
\put(64,18){\makebox(0,0)[bl]{\(1\)}}
\put(69,18){\makebox(0,0)[bl]{\(5\)}}



\put(34,14){\makebox(0,0)[bl]{\(5\)}}
\put(39,14){\makebox(0,0)[bl]{\(3\)}}
\put(44,14){\makebox(0,0)[bl]{\(5\)}}

\put(49,14){\makebox(0,0)[bl]{\(1\)}}
\put(54,14){\makebox(0,0)[bl]{\(1\)}}
\put(59,14){\makebox(0,0)[bl]{\(1\)}}
\put(64,14){\makebox(0,0)[bl]{\(1\)}}
\put(69,14){\makebox(0,0)[bl]{\(5\)}}


\put(49,10){\makebox(0,0)[bl]{\(5\)}}
\put(54,10){\makebox(0,0)[bl]{\(5\)}}
\put(59,10){\makebox(0,0)[bl]{\(5\)}}
\put(64,10){\makebox(0,0)[bl]{\(5\)}}
\put(69,10){\makebox(0,0)[bl]{\(5\)}}


\put(49,06){\makebox(0,0)[bl]{\(5\)}}
\put(54,06){\makebox(0,0)[bl]{\(5\)}}
\put(59,06){\makebox(0,0)[bl]{\(5\)}}
\put(64,06){\makebox(0,0)[bl]{\(5\)}}
\put(69,06){\makebox(0,0)[bl]{\(5\)}}


\put(49,02){\makebox(0,0)[bl]{\(5\)}}
\put(54,02){\makebox(0,0)[bl]{\(5\)}}
\put(59,02){\makebox(0,0)[bl]{\(5\)}}
\put(64,02){\makebox(0,0)[bl]{\(5\)}}
\put(69,02){\makebox(0,0)[bl]{\(5\)}}

\end{picture}
\end{center}
\begin{center}
\begin{picture}(108,70)
\put(20,00){\makebox(0,0)[bl] {Fig. 15. Poset of system quality
\(N(S)\)}}

\put(20,010){\line(0,1){40}} \put(20,010){\line(3,4){15}}
\put(20,050){\line(3,-4){15}}

\put(40,015){\line(0,1){40}} \put(40,015){\line(3,4){15}}
\put(40,055){\line(3,-4){15}}

\put(60,020){\line(0,1){40}} \put(60,020){\line(3,4){15}}
\put(60,060){\line(3,-4){15}}

\put(80,025){\line(0,1){40}} \put(80,025){\line(3,4){15}}
\put(80,065){\line(3,-4){15}}

\put(40,52){\circle*{2}}

\put(28,50){\makebox(0,0)[bl]{\(N(S_{3})\)}}

\put(63,50){\circle*{2}}

\put(57,45){\makebox(0,0)[bl]{\(N(S_{2})\)}}

\put(83,48){\circle*{2}}
\put(81,43){\makebox(0,0)[bl]{\(N(S_{1})\)}}


\put(80,65){\circle{2}} \put(80,65){\circle*{1}}


\put(83,63.6){\makebox(0,0)[bl]{The ideal}}
\put(85,60.6){\makebox(0,0)[bl]{point}}


\put(20,07){\makebox(0,0)[bl]{\(w=1\)}}
\put(40,12){\makebox(0,0)[bl]{\(w=2\)}}
\put(60,17){\makebox(0,0)[bl]{\(w=3\)}}
\put(80,22){\makebox(0,0)[bl]{\(w=4\)}}

\put(20,10){\circle{1.7}}

\put(00,12){\makebox(0,0)[bl]{The worst}}
\put(07,09){\makebox(0,0)[bl]{point}}

\end{picture}
\end{center}

\subsection{Modified version of HMMD method}
 Here combinatorial synthesis is based on usage of multiset
 estimates of design alternatives for system parts.
%
 For the resultant system \(S = S(1) \star ... \star S(i) \star ... \star S(m) \)
 the same multiset estimate is examined as an aggregated estimate
 (``generalized median'')
  of corresponding multiset estimates of its components
 (i.e., selected DAs).
 In addition, a condition quality for the selected DAs is used.
 Thus, \( N(S) = (w(S);e(S))\), where
 \(e(S)\) is the ``generalized median'' of estimates of the solution
 components.
 Evidently, the constraint for the resultant
 compatibility estimate in the obtained solution is used as well.
%
%
 Finally, the modified problem is (two objectives, two constraints):
%
%
%
%
 \[ \max~ e(S) = M^{g} =   \underset{M \in D}{\operatorname{argmin}}~
 | \biguplus_{i=1}^{m} ~ \delta (M, e(S_{i})) ~|,\]
 \[ \max~~ w(S), \]
%
%
 \[s.t. ~~~~ e(S_{i}) \succeq e_{o},~ \forall i=\overline{1,m};\]
  \[~~~w(S) \geq 1  .\]
%
 Note, the usage of the aggregated  multiset estimate
 for composite system \(S\) provides
 the opportunity
 to use the same type of multset estimate
 (e.g., \(P^{3,2}\), \(P^{3,3}\), \(P^{3,4}\))
  computing of the system quality (i.e., \(N(S)\))
 at various hierarchical layers of a multi-layer hierarchical
 system (during the 'Bottom-Up' system design framework).
 The constraint
 \(e(S_{i}) \succeq e_{o},~ \forall i=\overline{1,m}\)
  provides selection of only ``good'' DAs
  (level \(e_{o}\) can be changed, e.g., \(e^{3,3}_{6}=(0,2,1)\)).
  Here estimate \(e_{o}\) is used as a ``reference point''.
 Evidently, it is possible to use several similar constraints
 based on several ``reference points''
 (e.g.,
 \(e^{3,3}_{4}=(0,3,0)\),
 \(e^{3,3}_{5}=(1,1,1)\)).

 It is necessary to point out
 the basic modified version of HMMD method has been  described.
 It is possible to consider two other versions:

~~

 {\it Version 1:} Integrated  estimate for the composed
 system (objective 1):
 \[ \max~ e(S) = \biguplus_{i=1}^{m} ~ e(S_{i}). \]
 This case leads to changing the assessment problem for system:~~
 \(P^{l,\eta}\) \(\Rightarrow \) \(P^{l,\eta \times m}\).

 ~~

 {\it Version 2:} Aggregated  estimate for the composed
 system (objective 1) as the ``set median'':
%
%
%
 \[ \max~ e(S) = M^{s} =   \underset{M \in \{e(S_{1}),...,e(S_{m})\}}{\operatorname{argmin}}~
 | \biguplus_{i=1}^{m} ~ \delta (M, e(S_{i})) ~|.\]
 Evidently, here an approximation is used and the computational complexity
 is decreased (to polynomial case, i.e., selection of a system component estimate).

\subsection{Example:
 three-component system,
 three-element assessment}
%
  Here three component system with assessment problem \(P^{3,3}\)
 is considered.
 Thus, priorities of DAs are evaluated by
 assessment problem
 \(P^{3,3}\) (see Fig. 5).
 The illustrative example is presented in Fig. 16
 (3-position priorities are shown in parentheses),
 compatibility estimates from Table 6 are used.
 This example is close to example from Fig. 10.
  The poset for integrated estimates
 of quality for
 3-component system and assessment problem
 \(P^{3,3}\) is presented in Fig. 17.
 Note, the following points of the basic lattice are impossible:
 \(<8,0,1>\), \(<7,0,2>\), \(<5,0,4>\),
 \(<4,0,5>\), \(<2,0,7>\), and \(<1,0,8>\).

\begin{center}
\begin{picture}(68,50)

\put(05,00){\makebox(0,0)[bl] {Fig. 16. Example of composition}}

\put(4,09){\makebox(0,8)[bl]{\(X_{3}(3,0,0)\)}}
\put(4,13){\makebox(0,8)[bl]{\(X_{2}(2,1,0)\)}}
\put(4,17){\makebox(0,8)[bl]{\(X_{1}(0,3,0)\)}}

\put(26,05){\makebox(0,8)[bl]{\(Y_{4}(0,1,2)\)}}
\put(26,09){\makebox(0,8)[bl]{\(Y_{3}(0,0,3)\)}}
\put(26,13){\makebox(0,8)[bl]{\(Y_{2}(2,1,0)\)}}
\put(26,17){\makebox(0,8)[bl]{\(Y_{1}(0,1,2)\)}}

\put(48,09){\makebox(0,8)[bl]{\(Z_{3}(0,1,2)\)}}
\put(48,13){\makebox(0,8)[bl]{\(Z_{2}(2,1,0)\)}}
\put(48,17){\makebox(0,8)[bl]{\(Z_{1}(3,0,0)\)}}
\put(3,24){\circle{2}} \put(25,24){\circle{2}}
\put(47,24){\circle{2}}

\put(0,24){\line(1,0){02}} \put(22,24){\line(1,0){02}}
\put(44,24){\line(1,0){02}}

\put(0,24){\line(0,-1){13}} \put(22,24){\line(0,-1){17}}
\put(44,24){\line(0,-1){13}}

\put(44,19){\line(1,0){01}} \put(44,15){\line(1,0){01}}
\put(44,11){\line(1,0){01}}

\put(46,19){\circle{2}} \put(46,19){\circle*{1}}
\put(46,15){\circle{2}} \put(46,15){\circle*{1}}
\put(46,11){\circle{2}} \put(46,11){\circle*{1}}

\put(22,07){\line(1,0){01}} \put(22,11){\line(1,0){01}}
\put(22,15){\line(1,0){01}} \put(22,19){\line(1,0){01}}

\put(24,15){\circle{2}} \put(24,15){\circle*{1}}
\put(24,11){\circle{2}} \put(24,19){\circle{2}}
\put(24,11){\circle*{1}} \put(24,19){\circle*{1}}
\put(24,07){\circle{2}} \put(24,07){\circle*{1}}

\put(0,11){\line(1,0){01}} \put(0,15){\line(1,0){01}}
\put(0,19){\line(1,0){01}}

\put(2,15){\circle{2}} \put(2,19){\circle{2}}
\put(2,15){\circle*{1}} \put(2,19){\circle*{1}}
\put(2,11){\circle{2}} \put(2,11){\circle*{1}}
\put(3,29){\line(0,-1){04}} \put(25,29){\line(0,-1){04}}
\put(47,29){\line(0,-1){04}}

\put(3,29){\line(1,0){44}} \put(14,29){\line(0,1){15}}

\put(14,44){\circle*{3}}

\put(4,26){\makebox(0,8)[bl]{X}} \put(21,26){\makebox(0,8)[bl]{Y}}
\put(43,26){\makebox(0,8)[bl]{Z}}

\put(18,44){\makebox(0,8)[bl]{\(S = X\star Y\star Z\)}}

\put(16,39.5){\makebox(0,8)[bl] {\(S^{1}_{1}=X_{1}\star Y_{1}\star
 Z_{2} (3;2,5,2) \)}}

\put(16,35){\makebox(0,8)[bl] {\(S^{1}_{2}=X_{2}\star Y_{1}\star
 Z_{2} (2;4,3,2)\)}}

\put(16,30.5){\makebox(0,8)[bl] {\(S^{1}_{3}=X_{3}\star Y_{2}\star
 Z_{1} (1;8,1,0)\)}}

\end{picture}
\end{center}

\begin{center}
\begin{picture}(116,224)

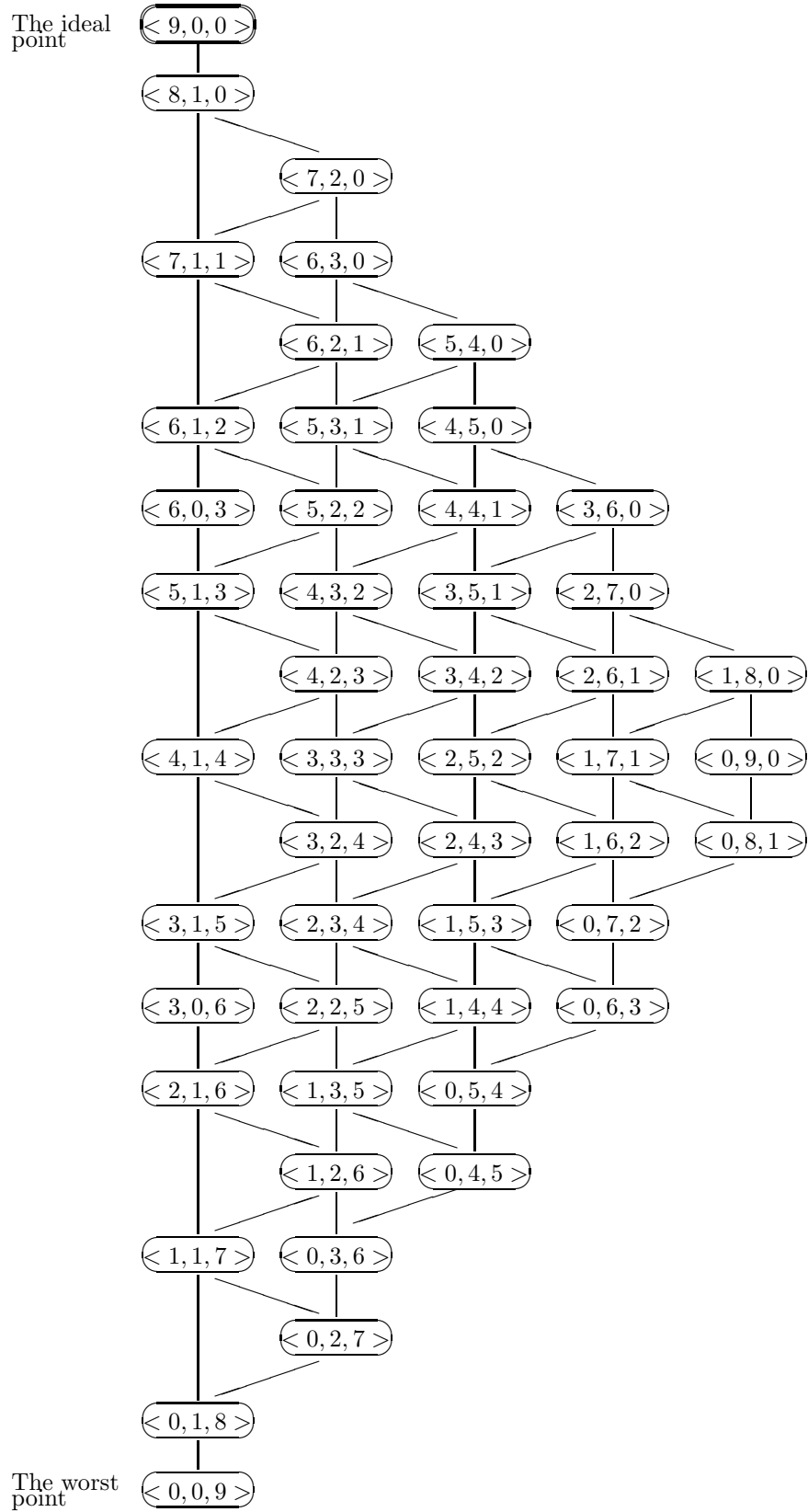
\put(020,00){\makebox(0,0)[bl] {Fig. 17. Poset
 \(e(S)=(\eta_{1},\eta_{2},\eta_{3})\)
%
 (\(P^{3,3}\))}}


\put(01,220){\makebox(0,0)[bl]{The ideal}}
\put(01,217){\makebox(0,0)[bl]{point}}

\put(01,09){\makebox(0,0)[bl]{The worst}}
\put(01,06){\makebox(0,0)[bl]{point}}


\put(20,219){\makebox(0,0)[bl]{\(<9,0,0>\) }}
\put(28,221){\oval(16,5)} \put(28,221){\oval(16.5,5.5)}

\put(28,214){\line(0,1){4}}
\put(20,209){\makebox(0,0)[bl]{\(<8,1,0>\)}}
\put(28,211){\oval(16,5)}


\put(28,190){\line(0,1){18}}
\put(20,185){\makebox(0,0)[bl]{\(<7,1,1>\) }}
\put(28,187){\oval(16,5)}


\put(28,166){\line(0,1){18}}
\put(20,161){\makebox(0,0)[bl]{\(<6,1,2>\) }}
\put(28,163){\oval(16,5)}

\put(28,154){\line(0,1){6}}
\put(20,149){\makebox(0,0)[bl]{\(<6,0,3>\) }}
\put(28,151){\oval(16,5)}


\put(28,142){\line(0,1){6}}
\put(20,137){\makebox(0,0)[bl]{\(<5,1,3>\)}}
\put(28,139){\oval(16,5)}


\put(28,118){\line(0,1){18}}
\put(20,113){\makebox(0,0)[bl]{\(<4,1,4>\) }}
\put(28,115){\oval(16,5)}


\put(28,94){\line(0,1){18}}
\put(20,89){\makebox(0,0)[bl]{\(<3,1,5>\) }}
\put(28,91){\oval(16,5)}


\put(45.5,202.5){\line(-3,1){15}}

\put(45.5,195.5){\line(-3,-1){15}}

\put(40,197){\makebox(0,0)[bl]{\(<7,2,0>\) }}
\put(48,199){\oval(16,5)}


\put(48,190){\line(0,1){6}}

\put(40,185){\makebox(0,0)[bl]{\(<6,3,0>\) }}
\put(48,187){\oval(16,5)}


\put(45.5,178.5){\line(-3,1){15}}

\put(45.5,171.5){\line(-3,-1){15}}

\put(48,178){\line(0,1){6}}
\put(40,173){\makebox(0,0)[bl]{\(<6,2,1>\) }}
\put(48,175){\oval(16,5)}


\put(48,166){\line(0,1){6}}

\put(40,161){\makebox(0,0)[bl]{\(<5,3,1>\) }}
\put(48,163){\oval(16,5)}


\put(45.5,154.5){\line(-3,1){15}}

\put(45.5,147.5){\line(-3,-1){15}}

\put(48,154){\line(0,1){6}}
\put(40,149){\makebox(0,0)[bl]{\(<5,2,2>\) }}
\put(48,151){\oval(16,5)}


\put(48,142){\line(0,1){6}}

\put(40,137){\makebox(0,0)[bl]{\(<4,3,2>\) }}
\put(48,139){\oval(16,5)}


\put(45.5,130.5){\line(-3,1){15}}

\put(45.5,123.5){\line(-3,-1){15}}

\put(48,130){\line(0,1){6}}
\put(40,125){\makebox(0,0)[bl]{\(<4,2,3>\) }}
\put(48,127){\oval(16,5)}

\put(30.5,111.5){\line(3,-1){15}}

\put(48,118){\line(0,1){6}}

\put(40,113){\makebox(0,0)[bl]{\(<3,3,3>\) }}
\put(48,115){\oval(16,5)}

\put(48,106){\line(0,1){6}}
\put(40,101){\makebox(0,0)[bl]{\(<3,2,4>\) }}
\put(48,103){\oval(16,5)}

\put(45.5,99.5){\line(-3,-1){15}}

\put(48,94){\line(0,1){6}}
\put(40,89){\makebox(0,0)[bl]{\(<2,3,4>\) }}
\put(48,91){\oval(16,5)}

\put(45.5,82.5){\line(-3,1){15}} \put(45.5,75.5){\line(-3,-1){15}}

\put(48,82){\line(0,1){6}}
\put(40,77){\makebox(0,0)[bl]{\(<2,2,5>\) }}
\put(48,79){\oval(16,5)}

\put(48,70){\line(0,1){6}}
\put(40,65){\makebox(0,0)[bl]{\(<1,3,5>\) }}
\put(48,67){\oval(16,5)}

\put(65.5,178.5){\line(-3,1){15}}

\put(60,173){\makebox(0,0)[bl]{\(<5,4,0>\) }}
\put(68,175){\oval(16,5)}

\put(65.5,171.5){\line(-3,-1){15}}


\put(68,166){\line(0,1){6}}
\put(60,161){\makebox(0,0)[bl]{\(<4,5,0>\) }}
\put(68,163){\oval(16,5)}

\put(65.5,154.5){\line(-3,1){15}}

\put(68,154){\line(0,1){6}}
\put(60,149){\makebox(0,0)[bl]{\(<4,4,1>\) }}
\put(68,151){\oval(16,5)}

\put(65.5,147.5){\line(-3,-1){15}}


\put(68,142){\line(0,1){6}}
\put(60,137){\makebox(0,0)[bl]{\(<3,5,1>\) }}
\put(68,139){\oval(16,5)}

\put(65.5,130.5){\line(-3,1){15}}

\put(68,130){\line(0,1){6}}
\put(60,125){\makebox(0,0)[bl]{\(<3,4,2>\) }}
\put(68,127){\oval(16,5)}

\put(65.5,123.5){\line(-3,-1){15}}


\put(68,118){\line(0,1){6}}
\put(60,113){\makebox(0,0)[bl]{\(<2,5,2>\) }}
\put(68,115){\oval(16,5)}

\put(65.5,106.5){\line(-3,1){15}}

\put(68,106){\line(0,1){6}}
\put(60,101){\makebox(0,0)[bl]{\(<2,4,3>\) }}
\put(68,103){\oval(16,5)}

\put(65.5,99.5){\line(-3,-1){15}}

\put(68,94){\line(0,1){6}}
\put(60,89){\makebox(0,0)[bl]{\(<1,5,3>\) }}
\put(68,91){\oval(16,5)}

\put(65.5,82.5){\line(-3,1){15}} \put(65.5,75.5){\line(-3,-1){15}}

\put(68,82){\line(0,1){6}}
\put(60,77){\makebox(0,0)[bl]{\(<1,4,4>\) }}
\put(68,79){\oval(16,5)}

\put(68,70){\line(0,1){6}}
\put(60,65){\makebox(0,0)[bl]{\(<0,5,4>\) }}
\put(68,67){\oval(16,5)}

\put(65.5,58.5){\line(-3,1){15}} \put(65.5,52.5){\line(-3,-1){15}}

\put(68,58){\line(0,1){6}}
\put(60,53){\makebox(0,0)[bl]{\(<0,4,5>\) }}
\put(68,55){\oval(16,5)}



\put(28,82){\line(0,1){6}}
\put(20,77){\makebox(0,0)[bl]{\(<3,0,6>\) }}
\put(28,79){\oval(16,5)}


\put(28,70){\line(0,1){6}}
\put(20,65){\makebox(0,0)[bl]{\(<2,1,6>\)}}
\put(28,67){\oval(16,5)}


\put(28,46){\line(0,1){18}}
\put(20,41){\makebox(0,0)[bl]{\(<1,1,7>\) }}
\put(28,43){\oval(16,5)}


\put(28,22){\line(0,1){18}}
\put(20,17){\makebox(0,0)[bl]{\(<0,1,8>\) }}
\put(28,19){\oval(16,5)}

\put(28,12){\line(0,1){4}}

\put(20,07){\makebox(0,0)[bl]{\(<0,0,9>\) }}
\put(28,09){\oval(16,5)}

\put(30.5,63.5){\line(3,-1){15}}

\put(48,58){\line(0,1){6}}
\put(40,53){\makebox(0,0)[bl]{\(<1,2,6>\) }}
\put(48,55){\oval(16,5)}
\put(30.5,46.5){\line(3,1){15}}

\put(48,46){\line(0,1){6}}
\put(40,41){\makebox(0,0)[bl]{\(<0,3,6>\) }}
\put(48,43){\oval(16,5)}
\put(30.5,39.5){\line(3,-1){15}}

\put(48,34){\line(0,1){6}}
\put(40,29){\makebox(0,0)[bl]{\(<0,2,7>\) }}
\put(48,31){\oval(16,5)}
\put(45.5,27.5){\line(-3,-1){15}}

\put(85.5,154.5){\line(-3,1){15}}

\put(80,149){\makebox(0,0)[bl]{\(<3,6,0>\) }}
\put(88,151){\oval(16,5)}

\put(85.5,147.5){\line(-3,-1){15}}


\put(88,142){\line(0,1){6}}
\put(80,137){\makebox(0,0)[bl]{\(<2,7,0>\) }}
\put(88,139){\oval(16,5)}

\put(85.5,130.5){\line(-3,1){15}}

\put(88,130){\line(0,1){6}}
\put(80,125){\makebox(0,0)[bl]{\(<2,6,1>\) }}
\put(88,127){\oval(16,5)}

\put(85.5,123.5){\line(-3,-1){15}}


\put(88,118){\line(0,1){6}}
\put(80,113){\makebox(0,0)[bl]{\(<1,7,1>\) }}
\put(88,115){\oval(16,5)}

\put(85.5,106.5){\line(-3,1){15}}

\put(88,106){\line(0,1){6}}
\put(80,101){\makebox(0,0)[bl]{\(<1,6,2>\) }}
\put(88,103){\oval(16,5)}

\put(85.5,99.5){\line(-3,-1){15}}

\put(88,94){\line(0,1){6}}
\put(80,89){\makebox(0,0)[bl]{\(<0,7,2>\) }}
\put(88,91){\oval(16,5)}

\put(85.5,82.5){\line(-3,1){15}} \put(85.5,75.5){\line(-3,-1){15}}

\put(88,82){\line(0,1){6}}
\put(80,77){\makebox(0,0)[bl]{\(<0,6,3>\) }}
\put(88,79){\oval(16,5)}

\put(105.5,130.5){\line(-3,1){15}}

\put(100,125){\makebox(0,0)[bl]{\(<1,8,0>\) }}
\put(108,127){\oval(16,5)}

\put(105.5,123.5){\line(-3,-1){15}}


\put(108,118){\line(0,1){6}}
\put(100,113){\makebox(0,0)[bl]{\(<0,9,0>\) }}
\put(108,115){\oval(16,5)}

\put(105.5,106.5){\line(-3,1){15}}

\put(108,106){\line(0,1){6}}
\put(100,101){\makebox(0,0)[bl]{\(<0,8,1>\) }}
\put(108,103){\oval(16,5)}

\put(105.5,99.5){\line(-3,-1){15}}

\end{picture}
\end{center}

 Now, let us consider three modified versions of HMMD:
 (i) integrated estimate for system quality (version 1),
 (ii) aggregated estimate for system quality
 (version 2, ``set median''),
 and
 (ii) aggregated estimate for system quality
 (basic version, ``generalized median'').
 The illustration is targeted to system estimates.
 The considered solutions correspond to the solutions
 in Fig. 10.

 In the case 1 (integrated system estimate),
 the following solutions and their estimates are examined:

 (a) \(S^{1}_{1} = X_{1} \star Y_{1} \star Z_{2}\),
 \(N(S^{1}_{1}) = (3; 2,5,2)\),

 (b) \(S^{1}_{2} = X_{2} \star Y_{1} \star Z_{2}\),
  \(N(S^{1}_{2}) = (2; 4,3,2)\),

 (c) \(S^{1}_{3} = X_{3} \star Y_{2} \star Z_{1}\),
 \(N(S^{1}_{3}) = (1; 8,1,0)\).

 Fig. 18 illustrates the general poset of system quality
 (each local poset for \(w=1,2,3\)
 corresponds to the poset from Fig. 17).

\begin{center}
\begin{picture}(60,63)
\put(00,00){\makebox(0,0)[bl] {Fig. 18. Poset of system quality
\(N(S)\)}}

\put(00,010){\line(0,1){40}} \put(00,010){\line(3,4){15}}
\put(00,050){\line(3,-4){15}}

\put(20,015){\line(0,1){40}} \put(20,015){\line(3,4){15}}
\put(20,055){\line(3,-4){15}}

\put(40,020){\line(0,1){40}} \put(40,020){\line(3,4){15}}
\put(40,060){\line(3,-4){15}}

\put(00,47){\circle*{2}}
\put(01.5,48){\makebox(0,0)[bl]{\(N(S^{1}_{3})\)}}

\put(20,42){\circle*{2}}
\put(08.5,40){\makebox(0,0)[bl]{\(N(S^{1}_{2})\)}}

\put(40,40){\circle*{2}}
\put(40.5,41.6){\makebox(0,0)[bl]{\(N(S^{1}_{1})\)}}

\put(40,60){\circle{2}} \put(40,60){\circle*{1}}


\put(43,58.6){\makebox(0,0)[bl]{The ideal}}
\put(45,55.6){\makebox(0,0)[bl]{point}}

\put(00,07){\makebox(0,0)[bl]{\(w=1\)}}
\put(20,12){\makebox(0,0)[bl]{\(w=2\)}}
\put(40,17){\makebox(0,0)[bl]{\(w=3\)}}

\put(00,10){\circle{1.7}}

\put(01.5,14){\makebox(0,0)[bl]{The worst}}
\put(02.5,11){\makebox(0,0)[bl]{point}}

\end{picture}
\end{center}

 In the case 2 (aggregated system estimate as ``set median''),
 the following solutions and their estimates are considered
 (the poset from Fig. 11 is used):

 (a) \(S^{2}_{1} = X_{1} \star Y_{1} \star Z_{2}\),
 \(N(S^{2}_{1}) = (3; 0,3,0)\),

 (b) \(S^{2}_{2} = X_{2} \star Y_{1} \star Z_{2}\),
  \(N(S^{2}_{2}) = (2; 2,1,0)\),

 (c) \(S^{2}_{3} = X_{3} \star Y_{2} \star Z_{1}\),
 \(N(S^{2}_{3}) = (1; 3,0,0)\).

 In the case 3
 (aggregated system estimate as ``generalized median''),
 the following solutions and their estimates are considered
 (the poset from Fig. 11 is used):

 (a) \(S^{3}_{1} = X_{1} \star Y_{1} \star Z_{2}\),
 \(N(S^{3}_{1}) = (3; 0,3,0)\),

 (b) \(S^{3}_{2} = X_{2} \star Y_{1} \star Z_{2}\),
  \(N(S^{3}_{2}) = (2; 2,0,1)\) or
  \(N(S^{3}_{2}) = (2; 1,2,0)\)
  (incomparable estimates, i.e., two ``generalized medians''),

  (c) \(S^{3}_{3} = X_{3} \star Y_{2} \star Z_{1}\),
 \(N(S^{3}_{3}) = (1; 3,0,0)\).

 Note the cases 2 and 3 are more easy
  from the viewpoint of the future usage in multi-layer systems
  at the higher hierarchical layer
  (i.e., assessment problem for composite system
  and the corresponding  multiset system estimate are more easy).

\subsection{Example: four-component system, four elements
  assessment}

 Here three component system with assessment problem \(P^{3,4}\)
 is considered.
 Thus, priorities of DAs are evaluated by assessment problem
 \(P^{3,4}\) (see Fig. 6).
 The illustrative example is presented in Fig. 19
 (4-position priorities are shown in parentheses),
 compatibility estimates from Table 7 are used.

 Three modified versions of HMMD are considered:
 (i) integrated estimate for system quality (version 1),
 (ii) aggregated estimate for system quality
 (version 2, ``set median''),
 and
 (ii) aggregated estimate for system quality
 (basic version, ``generalized median'').
 The illustration is targeted to system estimates.
 The considered solutions correspond to the solutions
 in Fig. 14.

\begin{center}
\begin{picture}(95,49)
\put(24,00){\makebox(0,0)[bl] {Fig. 19. Example of composition}}

\put(4,13){\makebox(0,8)[bl]{\(X_{3}(1,3,0)\)}}
\put(4,17){\makebox(0,8)[bl]{\(X_{2}(0,1,3)\)}}
\put(4,21){\makebox(0,8)[bl]{\(X_{1}(0,4,0)\)}}

\put(29,05){\makebox(0,8)[bl]{\(Y_{5}(1,3,0)\)}}
\put(29,09){\makebox(0,8)[bl]{\(Y_{4}(0,1,3)\)}}
\put(29,13){\makebox(0,8)[bl]{\(Y_{3}(3,1,0)\)}}
\put(29,17){\makebox(0,8)[bl]{\(Y_{2}(1,2,1)\)}}
\put(29,21){\makebox(0,8)[bl]{\(Y_{1}(0,0,4)\)}}

\put(54,13){\makebox(0,8)[bl]{\(Z_{3}(0,3,1)\)}}
\put(54,17){\makebox(0,8)[bl]{\(Z_{2}(4,0,0)\)}}
\put(54,21){\makebox(0,8)[bl]{\(Z_{1}(1,3,0)\)}}

\put(79,05){\makebox(0,8)[bl]{\(V_{5}(0,3,1)\)}}
\put(79,09){\makebox(0,8)[bl]{\(V_{4}(4,0,0)\)}}
\put(79,13){\makebox(0,8)[bl]{\(V_{3}(1,2,1)\)}}
\put(79,17){\makebox(0,8)[bl]{\(V_{2}(1,3,0)\)}}
\put(79,21){\makebox(0,8)[bl]{\(V_{1}(0,1,3)\)}}

\put(3,28){\circle{2}} \put(28,28){\circle{2}}
\put(53,28){\circle{2}} \put(78,28){\circle{2}}

\put(0,28){\line(1,0){02}} \put(25,28){\line(1,0){02}}
\put(50,28){\line(1,0){02}} \put(75,28){\line(1,0){02}}

\put(0,28){\line(0,-1){13}} \put(25,28){\line(0,-1){21}}
\put(50,28){\line(0,-1){13}} \put(75,28){\line(0,-1){21}}

\put(75,07){\line(1,0){01}} \put(75,11){\line(1,0){01}}
\put(75,15){\line(1,0){01}} \put(75,19){\line(1,0){01}}
\put(75,23){\line(1,0){01}}

\put(77,19){\circle{2}} \put(77,19){\circle*{1}}
\put(77,15){\circle{2}} \put(77,23){\circle{2}}
\put(77,15){\circle*{1}} \put(77,23){\circle*{1}}
\put(77,11){\circle{2}} \put(77,11){\circle*{1}}
\put(77,07){\circle{2}} \put(77,07){\circle*{1}}

\put(50,23){\line(1,0){01}} \put(50,19){\line(1,0){01}}
\put(50,15){\line(1,0){01}}

\put(52,23){\circle{2}} \put(52,23){\circle*{1}}
\put(52,19){\circle{2}} \put(52,19){\circle*{1}}
\put(52,15){\circle{2}} \put(52,15){\circle*{1}}

\put(25,07){\line(1,0){01}} \put(25,11){\line(1,0){01}}
\put(25,15){\line(1,0){01}} \put(25,19){\line(1,0){01}}
\put(25,23){\line(1,0){01}}

\put(27,19){\circle{2}} \put(27,19){\circle*{1}}
\put(27,15){\circle{2}} \put(27,23){\circle{2}}
\put(27,15){\circle*{1}} \put(27,23){\circle*{1}}
\put(27,11){\circle{2}} \put(27,11){\circle*{1}}
\put(27,07){\circle{2}} \put(27,07){\circle*{1}}

\put(0,15){\line(1,0){01}} \put(0,19){\line(1,0){01}}
\put(0,23){\line(1,0){01}}

\put(2,19){\circle{2}} \put(2,23){\circle{2}}
\put(2,19){\circle*{1}} \put(2,23){\circle*{1}}
\put(2,15){\circle{2}} \put(2,15){\circle*{1}}
\put(3,33){\line(0,-1){04}} \put(28,33){\line(0,-1){04}}
\put(53,33){\line(0,-1){04}} \put(78,33){\line(0,-1){04}}

\put(3,33){\line(1,0){75}} \put(17,33){\line(0,1){11}}

\put(17,44){\circle*{3}}

\put(4,30){\makebox(0,8)[bl]{X}} \put(24,30){\makebox(0,8)[bl]{Y}}
\put(49,30){\makebox(0,8)[bl]{Z}}
\put(74,30){\makebox(0,8)[bl]{V}}

\put(21,44){\makebox(0,8)[bl]{\(S = X \star Y \star Z \star V\)}}

\put(19,39){\makebox(0,8)[bl] {\(S^{1}_{1}=X_{3}\star Y_{5}\star
Z_{3}
 \star V_{4}(4;6,9,1)\)}}



\put(19,34.6){\makebox(0,8)[bl] {\(S^{1}_{2}=X_{3}\star Y_{3}\star
Z_{2} \star V_{4}(2;12,4,0)\)}}

\end{picture}
\end{center}

 In the case 1 (integrated system estimate),
 the following solutions and their estimates are examined
  (Fig. 20 depicts the general poset of system quality):

 (a) \(S^{1}_{1} = X_{3} \star Y_{5} \star Z_{3} \star V_{4}\),
 \(N(S^{1}_{1}) = (4; 6,9,1)\),

%


 (b) \(S^{1}_{2} = X_{3} \star Y_{3} \star Z_{2} \star V_{4}\),
 \(N(S^{1}_{4}) = (2; 12,4,0)\).

\begin{center}
\begin{picture}(108,70)
\put(20,00){\makebox(0,0)[bl] {Fig. 20. Poset of system quality
\(N(S)\)}}

\put(20,010){\line(0,1){40}} \put(20,010){\line(3,4){15}}
\put(20,050){\line(3,-4){15}}

\put(40,015){\line(0,1){40}} \put(40,015){\line(3,4){15}}
\put(40,055){\line(3,-4){15}}

\put(60,020){\line(0,1){40}} \put(60,020){\line(3,4){15}}
\put(60,060){\line(3,-4){15}}

\put(80,025){\line(0,1){40}} \put(80,025){\line(3,4){15}}
\put(80,065){\line(3,-4){15}}

\put(40,52){\circle*{2}}

\put(28,50){\makebox(0,0)[bl]{\(N(S^{1}_{2})\)}}


\put(83,48){\circle*{2}}
\put(80.6,43){\makebox(0,0)[bl]{\(N(S^{1}_{1})\)}}


\put(80,65){\circle{2}} \put(80,65){\circle*{1}}


\put(83,63.6){\makebox(0,0)[bl]{The ideal}}
\put(85,60.6){\makebox(0,0)[bl]{point}}


\put(20,07){\makebox(0,0)[bl]{\(w=1\)}}
\put(40,12){\makebox(0,0)[bl]{\(w=2\)}}
\put(60,17){\makebox(0,0)[bl]{\(w=3\)}}
\put(80,22){\makebox(0,0)[bl]{\(w=4\)}}

\put(20,10){\circle{1.7}}

\put(00,12){\makebox(0,0)[bl]{The worst}}
\put(07,09){\makebox(0,0)[bl]{point}}

\end{picture}
\end{center}

 In the case 2 (aggregated system estimate as ``set median''),
 the following solutions and their estimates are considered:

 (a) \(S^{2}_{1} = X_{3} \star Y_{5} \star Z_{3} \star V_{4}\),
 \(N(S^{2}_{1}) = (4; 1,3,0)\),


 (b) \(S^{2}_{2} = X_{3} \star Y_{3} \star Z_{2} \star V_{4}\),
 here two ``set medians'' exist:
 \((3,1,0)\) or \((4,0,0)\),
  the best ``set median'' is selected:
  \(N(S^{2}_{2}) = (2; 4,0,0)\).

 In the case 3
 (aggregated system estimate as ``generalized median''),
 the following solutions and their estimates are considered
 (the poset from Fig. 11 is used):

 (a) \(S^{3}_{1} = X_{3} \star Y_{5} \star Z_{3} \star V_{4}\),
 \(N(S^{3}_{1}) = (4; 1,3,0)\),

%


  (b) \(S^{3}_{2} = X_{3} \star Y_{3} \star Z_{2} \star V_{4}\),
 here two ``generalized medians'' exist:
 \((3,1,0)\) or \((4,0,0)\),
  the best ``generalized median'' is selected:
  \(N(S^{3}_{2}) = (2; 4,0,0)\).

\subsection{Example: Three-layer Hierarchical System}
 Here three-layer hierarchical system with assessment problem \(P^{3,3}\)
 is examined (Fig. 21).

\begin{center}

\begin{picture}(114,114)

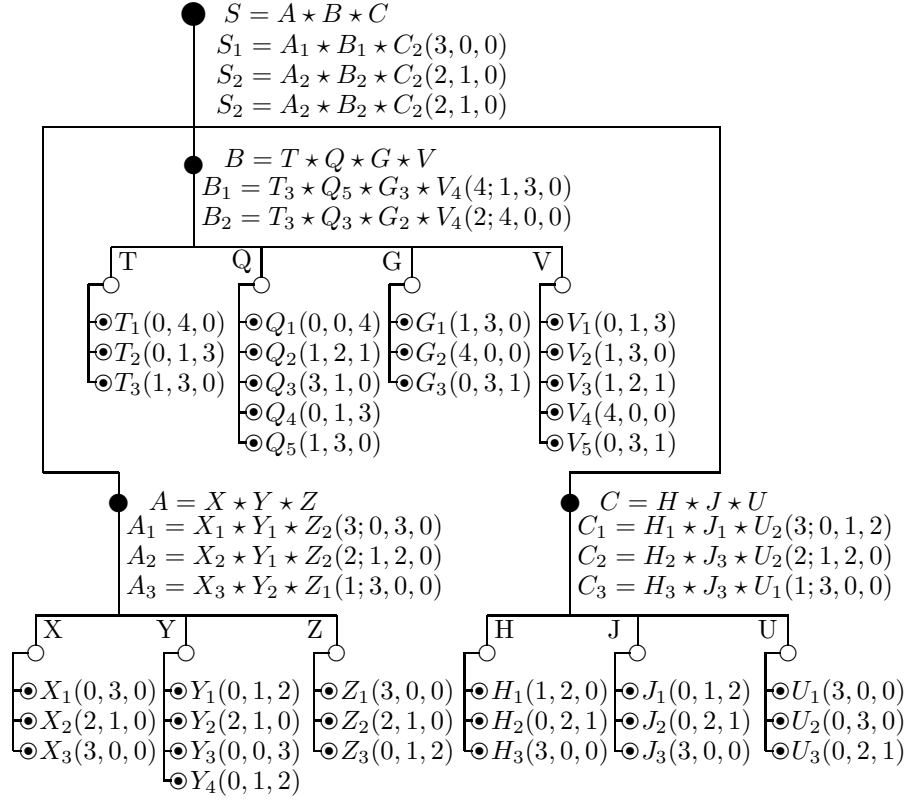
\put(15,00){\makebox(0,0)[bl] {Fig. 21. Example of composition
(three-layer system)}}


\put(13.5,58){\makebox(0,8)[bl]{\(T_{3}(1,3,0)\)}}
\put(13.5,62){\makebox(0,8)[bl]{\(T_{2}(0,1,3)\)}}
\put(13.5,66){\makebox(0,8)[bl]{\(T_{1}(0,4,0)\)}}

\put(33.5,50){\makebox(0,8)[bl]{\(Q_{5}(1,3,0)\)}}
\put(33.5,54){\makebox(0,8)[bl]{\(Q_{4}(0,1,3)\)}}
\put(33.5,58){\makebox(0,8)[bl]{\(Q_{3}(3,1,0)\)}}
\put(33.5,62){\makebox(0,8)[bl]{\(Q_{2}(1,2,1)\)}}
\put(33.5,66){\makebox(0,8)[bl]{\(Q_{1}(0,0,4)\)}}

\put(53.5,58){\makebox(0,8)[bl]{\(G_{3}(0,3,1)\)}}
\put(53.5,62){\makebox(0,8)[bl]{\(G_{2}(4,0,0)\)}}
\put(53.5,66){\makebox(0,8)[bl]{\(G_{1}(1,3,0)\)}}

\put(73.5,50){\makebox(0,8)[bl]{\(V_{5}(0,3,1)\)}}
\put(73.5,54){\makebox(0,8)[bl]{\(V_{4}(4,0,0)\)}}
\put(73.5,58){\makebox(0,8)[bl]{\(V_{3}(1,2,1)\)}}
\put(73.5,62){\makebox(0,8)[bl]{\(V_{2}(1,3,0)\)}}
\put(73.5,66){\makebox(0,8)[bl]{\(V_{1}(0,1,3)\)}}

\put(13,73){\circle{2}} \put(33,73){\circle{2}}
\put(53,73){\circle{2}} \put(73,73){\circle{2}}

\put(10,73){\line(1,0){02}} \put(30,73){\line(1,0){02}}
\put(50,73){\line(1,0){02}} \put(70,73){\line(1,0){02}}

\put(10,73){\line(0,-1){13}} \put(30,73){\line(0,-1){21}}
\put(50,73){\line(0,-1){13}} \put(70,73){\line(0,-1){21}}

\put(50,68){\line(1,0){01}} \put(50,64){\line(1,0){01}}
\put(50,60){\line(1,0){01}}

\put(70,68){\line(1,0){01}} \put(70,64){\line(1,0){01}}
\put(70,60){\line(1,0){01}} \put(70,56){\line(1,0){01}}
\put(70,52){\line(1,0){01}}

\put(52,68){\circle{2}} \put(52,68){\circle*{1}}
\put(52,64){\circle{2}} \put(52,64){\circle*{1}}
\put(52,60){\circle{2}} \put(52,60){\circle*{1}}

\put(72,68){\circle{2}} \put(72,68){\circle*{1}}
\put(72,64){\circle{2}} \put(72,64){\circle*{1}}
\put(72,60){\circle{2}} \put(72,60){\circle*{1}}
\put(72,56){\circle{2}} \put(72,56){\circle*{1}}
\put(72,52){\circle{2}} \put(72,52){\circle*{1}}

\put(30,52){\line(1,0){01}} \put(30,56){\line(1,0){01}}
\put(30,60){\line(1,0){01}} \put(30,64){\line(1,0){01}}
\put(30,68){\line(1,0){01}}

\put(32,52){\circle{2}} \put(32,52){\circle*{1}}
\put(32,56){\circle{2}} \put(32,56){\circle*{1}}
\put(32,60){\circle{2}} \put(32,60){\circle*{1}}
\put(32,64){\circle{2}} \put(32,64){\circle*{1}}
\put(32,68){\circle*{1}} \put(32,68){\circle{2}}


\put(10,60){\line(1,0){01}} \put(10,64){\line(1,0){01}}
\put(10,68){\line(1,0){01}}

\put(12,64){\circle{2}} \put(12,68){\circle{2}}
\put(12,64){\circle*{1}} \put(12,68){\circle*{1}}
\put(12,60){\circle{2}} \put(12,60){\circle*{1}}
\put(13,78){\line(0,-1){04}} \put(33,78){\line(0,-1){04}}
\put(53,78){\line(0,-1){04}} \put(73,78){\line(0,-1){04}}

\put(13,78){\line(1,0){60}} \put(24,78){\line(0,1){11}}

\put(24,89){\circle*{2.5}}

\put(14,75){\makebox(0,8)[bl]{T}}
\put(29,74.5){\makebox(0,8)[bl]{Q}}
\put(49,75){\makebox(0,8)[bl]{G}}
\put(69,75){\makebox(0,8)[bl]{V}}

\put(28,88){\makebox(0,8)[bl]{\(B = T\star Q\star G \star V\)}}

\put(25,84){\makebox(0,8)[bl] {\(B_{1}=T_{3}\star Q_{5}\star G_{3}
 \star V_{4} (4;1,3,0)\)}}

\put(25,80){\makebox(0,8)[bl] {\(B_{2}=T_{3}\star Q_{3}\star G_{2}
  \star V_{4} (2;4,0,0)\)}}



\put(24,89){\line(0,1){20}}

\put(04,94){\line(1,0){90}}

\put(24,109){\circle*{3}}

\put(28,108){\makebox(0,8)[bl]{\(S = A\star B\star C\)}}

\put(27,103){\makebox(0,8)[bl]
 {\(S_{1}=A_{1}\star B_{1}\star C_{2}(3,0,0)\)}}

\put(27,99){\makebox(0,8)[bl]
 {\(S_{2}=A_{2}\star B_{2}\star C_{2}(2,1,0)\)}}

\put(27,95){\makebox(0,8)[bl]
 {\(S_{2}=A_{2}\star B_{2}\star C_{2}(2,1,0)\)}}


\put(3.5,09){\makebox(0,8)[bl]{\(X_{3}(3,0,0)\)}}
\put(3.5,13){\makebox(0,8)[bl]{\(X_{2}(2,1,0)\)}}
\put(3.5,17){\makebox(0,8)[bl]{\(X_{1}(0,3,0)\)}}

\put(23.5,05){\makebox(0,8)[bl]{\(Y_{4}(0,1,2)\)}}
\put(23.5,09){\makebox(0,8)[bl]{\(Y_{3}(0,0,3)\)}}
\put(23.5,13){\makebox(0,8)[bl]{\(Y_{2}(2,1,0)\)}}
\put(23.5,17){\makebox(0,8)[bl]{\(Y_{1}(0,1,2)\)}}

\put(43.5,09){\makebox(0,8)[bl]{\(Z_{3}(0,1,2)\)}}
\put(43.5,13){\makebox(0,8)[bl]{\(Z_{2}(2,1,0)\)}}
\put(43.5,17){\makebox(0,8)[bl]{\(Z_{1}(3,0,0)\)}}
\put(3,24){\circle{2}} \put(23,24){\circle{2}}
\put(43,24){\circle{2}}

\put(0,24){\line(1,0){02}} \put(20,24){\line(1,0){02}}
\put(40,24){\line(1,0){02}}

\put(0,24){\line(0,-1){13}} \put(20,24){\line(0,-1){17}}
\put(40,24){\line(0,-1){13}}

\put(40,19){\line(1,0){01}} \put(40,15){\line(1,0){01}}
\put(40,11){\line(1,0){01}}

\put(42,19){\circle{2}} \put(42,19){\circle*{1}}
\put(42,15){\circle{2}} \put(42,15){\circle*{1}}
\put(42,11){\circle{2}} \put(42,11){\circle*{1}}

\put(20,07){\line(1,0){01}} \put(20,11){\line(1,0){01}}
\put(20,15){\line(1,0){01}} \put(20,19){\line(1,0){01}}

\put(22,15){\circle{2}} \put(22,15){\circle*{1}}
\put(22,11){\circle{2}} \put(22,19){\circle{2}}
\put(22,11){\circle*{1}} \put(22,19){\circle*{1}}
\put(22,07){\circle{2}} \put(22,07){\circle*{1}}

\put(0,11){\line(1,0){01}} \put(0,15){\line(1,0){01}}
\put(0,19){\line(1,0){01}}

\put(2,15){\circle{2}} \put(2,19){\circle{2}}
\put(2,15){\circle*{1}} \put(2,19){\circle*{1}}
\put(2,11){\circle{2}} \put(2,11){\circle*{1}}
\put(3,29){\line(0,-1){04}} \put(23,29){\line(0,-1){04}}
\put(43,29){\line(0,-1){04}}

\put(3,29){\line(1,0){40}} \put(14,29){\line(0,1){15}}

\put(14,44){\circle*{2.5}}

\put(4,26){\makebox(0,8)[bl]{X}} \put(19,26){\makebox(0,8)[bl]{Y}}
\put(39,26){\makebox(0,8)[bl]{Z}}

\put(18,43){\makebox(0,8)[bl]{\(A = X\star Y\star Z\)}}

\put(15,39){\makebox(0,8)[bl] {\(A_{1}=X_{1}\star Y_{1}\star Z_{2}
(3;0,3,0)\)}}

\put(15,35){\makebox(0,8)[bl] {\(A_{2}=X_{2}\star Y_{1}\star
Z_{2}(2;1,2,0)\)}}

\put(15,31){\makebox(0,8)[bl] {\(A_{3}=X_{3}\star Y_{2}\star Z_{1}
(1;3,0,0)\)}}


\put(4,48){\line(0,1){46}}

\put(14,48){\line(-1,0){10}}

\put(14,44){\line(0,1){4}}


\put(63.5,09){\makebox(0,8)[bl]{\(H_{3}(3,0,0)\)}}
\put(63.5,13){\makebox(0,8)[bl]{\(H_{2}(0,2,1)\)}}
\put(63.5,17){\makebox(0,8)[bl]{\(H_{1}(1,2,0)\)}}

\put(83.5,09){\makebox(0,8)[bl]{\(J_{3}(3,0,0)\)}}
\put(83.5,13){\makebox(0,8)[bl]{\(J_{2}(0,2,1)\)}}
\put(83.5,17){\makebox(0,8)[bl]{\(J_{1}(0,1,2)\)}}

\put(103.5,09){\makebox(0,8)[bl]{\(U_{3}(0,2,1)\)}}
\put(103.5,13){\makebox(0,8)[bl]{\(U_{2}(0,3,0)\)}}
\put(103.5,17){\makebox(0,8)[bl]{\(U_{1}(3,0,0)\)}}
\put(63,24){\circle{2}} \put(83,24){\circle{2}}
\put(103,24){\circle{2}}

\put(60,24){\line(1,0){02}} \put(80,24){\line(1,0){02}}
\put(100,24){\line(1,0){02}}

\put(60,24){\line(0,-1){13}} \put(80,24){\line(0,-1){13}}
\put(100,24){\line(0,-1){13}}

\put(100,19){\line(1,0){01}} \put(100,15){\line(1,0){01}}
\put(100,11){\line(1,0){01}}

\put(102,19){\circle{2}} \put(102,19){\circle*{1}}
\put(102,15){\circle{2}} \put(102,15){\circle*{1}}
\put(102,11){\circle{2}} \put(102,11){\circle*{1}}


\put(80,11){\line(1,0){01}} \put(80,15){\line(1,0){01}}
\put(80,19){\line(1,0){01}}

\put(82,15){\circle{2}} \put(82,15){\circle*{1}}
\put(82,11){\circle{2}} \put(82,19){\circle{2}}
\put(82,11){\circle*{1}} \put(82,19){\circle*{1}}


\put(60,11){\line(1,0){01}} \put(60,15){\line(1,0){01}}
\put(60,19){\line(1,0){01}}

\put(62,15){\circle{2}} \put(62,19){\circle{2}}
\put(62,15){\circle*{1}} \put(62,19){\circle*{1}}
\put(62,11){\circle{2}} \put(62,11){\circle*{1}}
\put(63,29){\line(0,-1){04}} \put(83,29){\line(0,-1){04}}
\put(103,29){\line(0,-1){04}}

\put(63,29){\line(1,0){40}} \put(74,29){\line(0,1){15}}

\put(74,44){\circle*{2.5}}

\put(64,26){\makebox(0,8)[bl]{H}}
\put(79,26){\makebox(0,8)[bl]{J}}
\put(99,26){\makebox(0,8)[bl]{U}}

\put(78,43){\makebox(0,8)[bl]{\(C = H\star J\star U\)}}

\put(75,39){\makebox(0,8)[bl] {\(C_{1}=H_{1}\star J_{1}\star
U_{2}(3;0,1,2)\)}}

\put(75,35){\makebox(0,8)[bl] {\(C_{2}=H_{2}\star J_{3}\star
U_{2}(2;1,2,0)\)}}

\put(75,31){\makebox(0,8)[bl] {\(C_{3}=H_{3}\star J_{3}\star
U_{1}(1;3,0,0)\)}}


\put(94,48){\line(0,1){46}}

\put(74,48){\line(1,0){20}}

\put(74,44){\line(0,1){4}}

\end{picture}
\end{center}

\begin{center}
\begin{picture}(37,39)

\put(01.5,34){\makebox(0,0)[bl]{Table 8. Compatibility}}

\put(00,0){\line(1,0){37}} \put(00,26){\line(1,0){37}}
\put(00,32){\line(1,0){37}}

\put(00,0){\line(0,1){32}} \put(07,0){\line(0,1){32}}
\put(37,0){\line(0,1){32}}


\put(01,22){\makebox(0,0)[bl]{\(H_{1}\)}}
\put(01,18){\makebox(0,0)[bl]{\(H_{2}\)}}
\put(01,14){\makebox(0,0)[bl]{\(H_{3}\)}}

\put(01,10){\makebox(0,0)[bl]{\(J_{1}\)}}
\put(01,06){\makebox(0,0)[bl]{\(J_{2}\)}}
\put(01,02){\makebox(0,0)[bl]{\(J_{3}\)}}


\put(12,26){\line(0,1){6}} \put(17,26){\line(0,1){6}}
\put(22,26){\line(0,1){6}} \put(27,26){\line(0,1){6}}
\put(32,26){\line(0,1){6}}

\put(07.4,28){\makebox(0,0)[bl]{\(J_{1}\)}}
\put(12.4,28){\makebox(0,0)[bl]{\(J_{2}\)}}
\put(17.4,28){\makebox(0,0)[bl]{\(J_{3}\)}}

\put(22.4,28){\makebox(0,0)[bl]{\(U_{1}\)}}
\put(27.4,28){\makebox(0,0)[bl]{\(U_{2}\)}}
\put(32.4,28){\makebox(0,0)[bl]{\(U_{3}\)}}



\put(09,22){\makebox(0,0)[bl]{\(3\)}}
\put(14,22){\makebox(0,0)[bl]{\(1\)}}
\put(19,22){\makebox(0,0)[bl]{\(0\)}}
\put(24,22){\makebox(0,0)[bl]{\(1\)}}
\put(29,22){\makebox(0,0)[bl]{\(3\)}}
\put(34,22){\makebox(0,0)[bl]{\(1\)}}


\put(09,18){\makebox(0,0)[bl]{\(2\)}}
\put(14,18){\makebox(0,0)[bl]{\(1\)}}
\put(19,18){\makebox(0,0)[bl]{\(2\)}}
\put(24,18){\makebox(0,0)[bl]{\(1\)}}
\put(29,18){\makebox(0,0)[bl]{\(2\)}}
\put(34,18){\makebox(0,0)[bl]{\(2\)}}


\put(09,14){\makebox(0,0)[bl]{\(0\)}}
\put(14,14){\makebox(0,0)[bl]{\(1\)}}
\put(19,14){\makebox(0,0)[bl]{\(1\)}}
\put(24,14){\makebox(0,0)[bl]{\(1\)}}
\put(29,14){\makebox(0,0)[bl]{\(1\)}}
\put(34,14){\makebox(0,0)[bl]{\(1\)}}


\put(24,10){\makebox(0,0)[bl]{\(2\)}}
\put(29,10){\makebox(0,0)[bl]{\(3\)}}
\put(34,10){\makebox(0,0)[bl]{\(2\)}}


\put(24,06){\makebox(0,0)[bl]{\(1\)}}
\put(29,06){\makebox(0,0)[bl]{\(1\)}}
\put(34,06){\makebox(0,0)[bl]{\(0\)}}


\put(24,02){\makebox(0,0)[bl]{\(3\)}}
\put(29,02){\makebox(0,0)[bl]{\(2\)}}
\put(34,02){\makebox(0,0)[bl]{\(1\)}}

\end{picture}
\end{center}

 The composition problem for subsystem \(A\)
 corresponds to the example from the previous sections
 (Fig. 16, Table 6).
 Thus, the following composite solutions
 are examined (case of ``generalized median''):

 (a) \(A_{1} =
   X_{1} \star Y_{1} \star Z_{2}\),
 \(N(A_{1})= (3;0,3,0)\),

 (b) \(A_{2} =
  X_{2} \star Y_{1} \star Z_{2}\),
 \(N(A_{2})= (2;1,2,0)\),

 (c) \(A_{3} =
  X_{1} \star Y_{1} \star Z_{2}\),
  \(N(A_{3})= (1;3,0,0)\).

 The composition problem for subsystem \(B\)
 corresponds to the example from the previous sections
 (Fig. 19, Table 7).
 Thus, the following composite solutions
 are examined (case of ``generalized median''):

 (a) \(B_{1}  = T_{1} \star Q_{1} \star V_{2}\),
 \(N(B_{1})= (4;0,3,0)\),

 (b) \(B_{2} = T_{2} \star Q_{1} \star V_{2}\),
 \(N(B_{2})= (2;1,2,0)\).

  The following composite solutions
 are examined for subsystem \(C\)
 (compatibility estimates are presented in Table 8):

 (a) \(C_{1}  = H_{1} \star J_{1} \star U_{2}\),
 \(N(B_{1})= (3;0,2,1)\),

 (b) \(C_{2} = H_{2} \star J_{3} \star U_{2}\),
 \(N(B_{2})= (2;1,2,0)\),

 (c) \(C_{3} = H_{3} \star J_{3} \star U_{1}\),
  \(N(B_{3})= (1;3,0,0)\).

 The synthesis problem at the higher hierarchical
 layer is depicted  in Fig. 22.

 Now, it is reasonable to consider several possible
 solving scheme for the composition:

 {\it Scheme 1.} Combination of all possible composite solutions
 and an additional analysis of the solutions.

 {\it Scheme 2.} Combination of all possible composite solutions
 and the selection of the best solution(s) while taking into account
 their integrated  multiset estimates.

 {\it Scheme 3.}
 Combination of all possible composite solutions
 and the selection of the best solution(s) while taking into account
 their aggregated multiset estimates.

 The resultant composite solutions are (including their
 integrated estimates, assessment problem \(P^{3,10}\),
 alignment of estimates is not used;
  compatibility estimates are considered concurrently:
%
  \(w^{I} = \min \{ w(A), w(B),  w(C) \}\)):

 \(S_{1} = A_{1} \star B_{1} \star C_{1}\),
 \(N(S_{1})= (3;1,8,1)\);
 \(S_{2} = A_{1} \star B_{1} \star C_{2}\),
 \(N(S_{2})= (2;2,8,0)\);

 \(S_{3} = A_{1} \star B_{1} \star C_{3}\),
 \(N(S_{3})=(1;4,6,0)\);
 \(S_{4} = A_{1} \star B_{2} \star C_{1}\),
 \(N(S_{4})= (2;4,5,1)\);

 \(S_{5} = A_{1} \star B_{2} \star C_{2}\),
 \(N(S_{5})= (2;5,5,0)\);
 \(S_{6} = A_{1} \star B_{2} \star C_{3}\),
 \(N(S_{6})= (1;7,3,0)\);

 \(S_{7} = A_{2} \star B_{1} \star C_{1}\),
 \(N(S_{7})= (2;3,6,1)\);
 \(S_{8} = A_{2} \star B_{1} \star C_{2}\),
 \(N(S_{8})= (2;4,6,0)\);

 \(S_{9} = A_{2} \star B_{1} \star C_{3}\),
 \(N(S_{9})= (1;6,4,0)\);
 \(S_{10} = A_{2} \star B_{2} \star C_{1}\),
 \(N(S_{10})= (2;6,3,1)\);

 \(S_{11} = A_{2} \star B_{2} \star C_{2}\),
 \(N(S_{11})= (2;7,3,0)\);
 \(S_{12} = A_{2} \star B_{2} \star C_{3}\),
 \(N(S_{12})= (1;9,1,0)\);

 \(S_{13} = A_{3} \star B_{1} \star C_{1}\),
 \(N(S_{13})= (1;7,2,1)\);
 \(S_{14} = A_{3} \star B_{1} \star C_{2}\),
 \(N(S_{14})= (1;5,5,0)\);

 \(S_{15} = A_{3} \star B_{1} \star C_{3} \),
 \(N(S_{15})= (1;7,3,0)\);
 \(S_{16} = A_{3} \star B_{2} \star C_{1} \),
 \(N(S_{16})= (1;7,2,1)\);

 \(S_{17} = A_{3} \star B_{2} \star C_{2}\),
 \(N(S_{17})= (2;8,2,0)\);
 \(S_{18} = A_{3} \star B_{2} \star C_{3}\),
 \(N(S_{18})= (1;10,0,0)\).

\begin{center}
\begin{picture}(70,45.5)
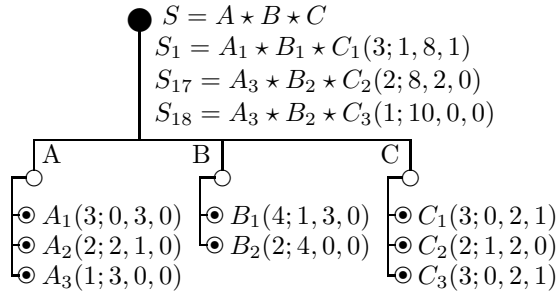
\put(00,00){\makebox(0,0)[bl] {Fig. 22. Composition at higher
hierarchical layer}}

\put(4,05){\makebox(0,8)[bl]{\(A_{3}(1;3,0,0)\)}}
\put(4,09){\makebox(0,8)[bl]{\(A_{2}(2;2,1,0)\)}}
\put(4,13){\makebox(0,8)[bl]{\(A_{1}(3;0,3,0)\)}}

\put(29,09){\makebox(0,8)[bl]{\(B_{2}(2;4,0,0)\)}}
\put(29,13){\makebox(0,8)[bl]{\(B_{1}(4;1,3,0)\)}}

\put(54,05){\makebox(0,8)[bl]{\(C_{3}(3;0,2,1)\)}}
\put(54,09){\makebox(0,8)[bl]{\(C_{2}(2;1,2,0)\)}}
\put(54,13){\makebox(0,8)[bl]{\(C_{1}(3;0,2,1)\)}}

\put(3,20){\circle{2}} \put(28,20){\circle{2}}
\put(53,20){\circle{2}}

\put(0,20){\line(1,0){02}} \put(25,20){\line(1,0){02}}
\put(50,20){\line(1,0){02}}

\put(0,20){\line(0,-1){13}} \put(25,20){\line(0,-1){09}}
\put(50,20){\line(0,-1){13}}

\put(50,15){\line(1,0){01}} \put(50,11){\line(1,0){01}}
\put(50,07){\line(1,0){01}}

\put(52,15){\circle{2}} \put(52,15){\circle*{1}}
\put(52,11){\circle{2}} \put(52,11){\circle*{1}}
\put(52,07){\circle{2}} \put(52,07){\circle*{1}}

\put(25,11){\line(1,0){01}} \put(25,15){\line(1,0){01}}

\put(27,15){\circle{2}} \put(27,15){\circle*{1}}
\put(27,11){\circle{2}} \put(27,11){\circle*{1}}

\put(0,07){\line(1,0){01}} \put(0,11){\line(1,0){01}}
\put(0,15){\line(1,0){01}}

\put(2,11){\circle{2}} \put(2,15){\circle{2}}
\put(2,11){\circle*{1}} \put(2,15){\circle*{1}}
\put(2,07){\circle{2}} \put(2,07){\circle*{1}}
\put(3,25){\line(0,-1){04}} \put(28,25){\line(0,-1){04}}
\put(53,25){\line(0,-1){04}}

\put(3,25){\line(1,0){50}} \put(17,25){\line(0,1){16}}

\put(17,41){\circle*{3}}

\put(4,22){\makebox(0,8)[bl]{A}} \put(24,22){\makebox(0,8)[bl]{B}}
\put(49,22){\makebox(0,8)[bl]{C}}

\put(20,40.5){\makebox(0,8)[bl]{\(S = A \star B \star C \)}}

\put(19,35.5){\makebox(0,8)[bl] {\(S_{1}=A_{1}\star B_{1}\star
 C_{1} (3;1,8,1)\)}}

\put(19,31){\makebox(0,8)[bl] {\(S_{17}=A_{3}\star B_{2}\star
 C_{2} (2;8,2,0)\)}}

\put(19,26.6){\makebox(0,8)[bl] {\(S_{18}=A_{3}\star B_{2}\star
 C_{3} (1;10,0,0)\)}}

\end{picture}
\end{center}

 Evidently, three Pareto-efficient solutions are obtained
 (Fig. 22):

 (a) \(S_{1} = A_{1} \star B_{1} \star C_{1} =
 (X_{1} \star Y_{1} \star Z_{2}) \star
  (T_{3} \star Q_{5} \star G_{3} \star V_{4}) \star
 (H_{1} \star J_{1} \star U_{2})\);
%

 (b) \(S_{17} = A_{3} \star B_{2} \star C_{2}=
 (X_{3} \star Y_{2} \star Z_{1}) \star
  (T_{3} \star Q_{3} \star G_{2} \star V_{4}) \star
 (H_{2} \star J_{3} \star U_{2})\);
%

 (c) \(S_{18} = A_{3} \star B_{2} \star C_{3}=
 (X_{3} \star Y_{2} \star Z_{1}) \star
  (T_{3} \star Q_{3} \star G_{2} \star V_{4}) \star
 (H_{3} \star J_{3} \star U_{1})\).
%

  Fig. 23 illustrates the general poset of system quality
 (each local poset for \(w=1,2,3\) corresponds to
 assessment problem \(P^{3,10}\)).

 Note, the usage of aggregated multiset estimates does not
 increase the dimension of assessment problem at the higher
 hierarchical layer.
 On the other hand,
 it may  be necessary to examine additional compatibility estimates at the higher
 hierarchical layer as well
 (i.e., compatibility between design alternatives for \(A\), \(B\), and \(C\)).

\begin{center}
\begin{picture}(60,63)
\put(00,00){\makebox(0,0)[bl] {Fig. 23. Poset of system quality
\(N(S)\)}}

\put(00,010){\line(0,1){40}} \put(00,010){\line(3,4){15}}
\put(00,050){\line(3,-4){15}}

\put(20,015){\line(0,1){40}} \put(20,015){\line(3,4){15}}
\put(20,055){\line(3,-4){15}}

\put(40,020){\line(0,1){40}} \put(40,020){\line(3,4){15}}
\put(40,060){\line(3,-4){15}}

\put(00,50){\circle*{2}}
\put(01.5,49){\makebox(0,0)[bl]{\(N(S_{18})\)}}

\put(20,44){\circle*{2}}
\put(07.7,42){\makebox(0,0)[bl]{\(N(S_{17})\)}}

\put(40,39){\circle*{2}}
\put(40.5,40.6){\makebox(0,0)[bl]{\(N(S_{1})\)}}

\put(40,60){\circle{2}} \put(40,60){\circle*{1}}


\put(43,58.6){\makebox(0,0)[bl]{The ideal}}
\put(45,55.6){\makebox(0,0)[bl]{point}}

\put(00,07){\makebox(0,0)[bl]{\(w=1\)}}
\put(20,12){\makebox(0,0)[bl]{\(w=2\)}}
\put(40,17){\makebox(0,0)[bl]{\(w=3\)}}

\put(00,10){\circle{1.7}}

\put(01.5,14){\makebox(0,0)[bl]{The worst}}
\put(02.5,11){\makebox(0,0)[bl]{point}}

\end{picture}
\end{center}


%
%
%


\section{Multiset Estimates in Knapsack-like Problems}

 Let us consider basic knapsack-like  problems and their
 modification in the case of multiset estimates.
 The basic knapsack problem is
 (e.g., \cite{gar79}, \cite{kellerer04}):
%
 \[\max\sum_{i=1}^{m} c_{i} x_{i}\]
 \[s.t. ~~ \sum_{i=1}^{m} a_{i} x_{i} \leq b,
 ~~ x_{i} \in\{0,1\}, ~ i=\overline{1,m}\]
%
%
%
 where \(x_{i}=1\) if item \(i\) is selected,
 for \(i\)th item \(c_{i}\) is a value ('utility'), and
 \(a_{i}\) is a weight (i.e., resource requirement).
 Often nonnegative coefficients are assumed.

%
 In the case of  multiple choice problem,
 the items are divided into groups
 and
 it is necessary to select elements  (items)
 or the only one element
 from each group
 while taking into account a total resource constraint (or constraints)
(e.g., \cite{gar79}, \cite{kellerer04}):
%
 \[\max\sum_{i=1}^{m} \sum_{j=1}^{q_{i}} c_{ij} x_{ij}\]
 \[ s.t. ~~ \sum_{i=1}^{m} \sum_{j=1}^{q_{i}} a_{ij} x_{ij} \leq b;
  ~~ \sum_{j=1}^{q_{i}} x_{ij} \leq 1,~ i=\overline{1,m};
 ~~ x_{ij} \in \{0, 1\}.\]
%
%

 The knapsack-like problems above
  are NP-hard
 and can be solved by the following approaches
 (\cite{gar79}, \cite{kellerer04}):
 (i) enumerative methods
 (e.g., Branch-and-Bound, dynamic programming),
 (ii) fully polynomial approximate schemes, and
 (iii) heuristics (e.g., greedy algorithms).
%

%
 In the case of multiset estimates of
 item ``utility'' \(e_{i}, i \in \{1,...,i,...,n\}\)
 (instead of \(c_{i}\)),
 the following aggregated multiset estimate can be used
 for the objective function  (``maximization''):

 (a) an aggregated multiset estimate as the ``generalized median'',

 (b) an aggregated multiset estimate as the ``set median'',
 and

 (c) an integrated  multiset estimate.

 First, let us consider a special case of multiple choice problem
 as follows:

 (1) multiset estimates of item ``utility''
  \(e_{i,j}, i \in \{1,...,i,...,n\}, j = \overline{1,q_{i}}\)
 (instead of \(c_{ij}\)),

 (2) an aggregated multiset estimate as the ``generalized median''
 (or ``set median'')
 is used for the objective function (``maximization'').

 The initial item set is:
 \[\{(1,1),(1,2),(1,q_{1}),...,(i,1),(i,2),(i,q_{i}),...,(n,1),(n,2),(n,q_{n})\}.\]
 Boolean variable \(x_{i,j}\) corresponds to selection of the
 item \((i,j)\).
 The solution  is a subset of the initial item set:
 \( S = \{ (i,j) | x_{i,j}=1 \} \).
  The problem is:
%
%
 \[ \max~ e(S) =   \max~ M = ~ \underset{M \in D}{\operatorname{argmin}}
 ~~
 | \biguplus_{(i,j) \in S=\{(i,j)| x_{i,j}=1\}} ~ \delta (M, e_{i,j}) ~|,\]
 \[ s.t. ~~ \sum_{i=1}^{m} \sum_{j=1}^{q_{i}}  a_{ij} x_{i,j} \leq b;
 ~~ \sum_{j=1}^{q_{i}} x_{ij} =  1;
%
%
 ~~ x_{ij} \in \{0, 1\}.\]
 Here
 ~ \(\sum_{i=1}^{m} \sum_{j=1}^{q_{i}} = \sum_{(i,j) \in S}\) ~.
 Evidently, this problem is similar to the above-mentioned combinatorial synthesis
 problem without compatibility of the selected items.
 Now let us consider a numerical example (Table 9).
 Some solutions for basic multiple choice problem
  (constraint: \(\sum_{j=1}^{q_{i}} x_{ij} =  1\))
 are:

 (a) \( S(b=7) = \{(1,2), (2,1), (3,2), (4,2) \}\),
 \(c(S(b=7)) = 16.1\);

 (b) \( S(b=8) = \{(1,2), (2,1), (3,3), (4,2) \}\),
  \(c(S(b=8)) = 17.1\).

 Here
 an estimate of computational complexity
 for a dynamic programming method
 is as follows:
 \[O(q_{1}~ \mu^{l,\eta}) +  O(q_{2}~ \mu^{l,2\eta}) + ... + O(q_{m}~ \mu^{l,(m-1)\eta})
 \leq O(m ~ \max_{i=\overline{1,m}} \{q_{i}\} ~  \mu^{l,(m-1)\eta}).\]

\begin{center}
\begin{picture}(61,80)
\put(04.5,75){\makebox(0,0)[bl] {Table 9.
 Multiple choice problem}}


\put(00,00){\line(1,0){61}} \put(00,62){\line(1,0){61}}
\put(10,68){\line(1,0){51}} \put(00,74){\line(1,0){61}}

\put(00,00){\line(0,1){74}} \put(10,00){\line(0,1){74}}

\put(22,00){\line(0,1){68}}


\put(34,00){\line(0,1){74}}


\put(61,00){\line(0,1){74}}


\put(01,69.5){\makebox(0,0)[bl]{\((i,j)\) }}


\put(11,69.4){\makebox(0,0)[bl]{Basic problem}}
\put(14,63.5){\makebox(0,0)[bl]{\(a_{ij}\) }}
\put(26,63.5){\makebox(0,0)[bl]{\(c_{ij}\)}}


\put(35,70){\makebox(0,0)[bl]{Assessment \(P^{3,3}\)}}
\put(45,62.7){\makebox(0,0)[bl]{\(e^{3,3}_{ij}\)}}



\put(01,56.3){\makebox(0,0)[bl]{\((1,1)\)}}

\put(14,57){\makebox(0,0)[bl]{\(1.3\)}}
\put(26,57){\makebox(0,0)[bl]{\(3.4\)}}



\put(42,56.3){\makebox(0,0)[bl]{\((1,2,0)\)}}


\put(01,51.3){\makebox(0,0)[bl]{\((1,2)\)}}

\put(14,52){\makebox(0,0)[bl]{\(3.1\)}}
\put(26,52){\makebox(0,0)[bl]{\(8.1\)}}



\put(42,51.3){\makebox(0,0)[bl]{\((3,0,0)\)}}


\put(01,46.3){\makebox(0,0)[bl]{\((1,3)\)}}

\put(14,47){\makebox(0,0)[bl]{\(0.7\)}}
\put(26,47){\makebox(0,0)[bl]{\(1.3\)}}



\put(42,46.3){\makebox(0,0)[bl]{\((0,1,2)\)}}


\put(01,41.3){\makebox(0,0)[bl]{\((2,1)\)}}

\put(14,42){\makebox(0,0)[bl]{\(2.0\)}}
\put(26,42){\makebox(0,0)[bl]{\(4.1\)}}



\put(42,41.3){\makebox(0,0)[bl]{\((2,1,0)\)}}


\put(01,36.3){\makebox(0,0)[bl]{\((2,2)\)}}

\put(14,37){\makebox(0,0)[bl]{\(1.3\)}}
\put(26,37){\makebox(0,0)[bl]{\(2.3\)}}



\put(42,36.3){\makebox(0,0)[bl]{\((0,2,1)\)}}


\put(01,31.3){\makebox(0,0)[bl]{\((3,1)\)}}

\put(14,32){\makebox(0,0)[bl]{\(3.0\)}}
\put(26,32){\makebox(0,0)[bl]{\(2.6\)}}



\put(42,31.3){\makebox(0,0)[bl]{\((0,3,0)\)}}


\put(01,26.3){\makebox(0,0)[bl]{\((3,2)\)}}

\put(14,27){\makebox(0,0)[bl]{\(0.6\)}}
\put(26,27){\makebox(0,0)[bl]{\(1.3\)}}



\put(42,26.3){\makebox(0,0)[bl]{\((0,1,2)\)}}


\put(01,21.3){\makebox(0,0)[bl]{\((3,3)\)}}

\put(14,22){\makebox(0,0)[bl]{\(1.6\)}}
\put(26,22){\makebox(0,0)[bl]{\(2.7\)}}



\put(42,21.3){\makebox(0,0)[bl]{\((0,3,0)\)}}


\put(01,16.3){\makebox(0,0)[bl]{\((4,1)\)}}

\put(14,17){\makebox(0,0)[bl]{\(2.0\)}}
\put(26,17){\makebox(0,0)[bl]{\(2.7\)}}



\put(42,16.3){\makebox(0,0)[bl]{\((0,3,0)\)}}


\put(01,11.3){\makebox(0,0)[bl]{\((4,2)\)}}

\put(14,12){\makebox(0,0)[bl]{\(1.3\)}}
\put(26,12){\makebox(0,0)[bl]{\(2.6\)}}



\put(42,11.3){\makebox(0,0)[bl]{\((0,3,0)\)}}


\put(01,06.3){\makebox(0,0)[bl]{\((4,3)\)}}

\put(14,07){\makebox(0,0)[bl]{\(3.3\)}}
\put(26,07){\makebox(0,0)[bl]{\(4.2\)}}



\put(42,06.3){\makebox(0,0)[bl]{\((2,1,0)\)}}


\put(01,01.3){\makebox(0,0)[bl]{\((4,4)\)}}

\put(14,02){\makebox(0,0)[bl]{\(2.3\)}}
\put(26,02){\makebox(0,0)[bl]{\(3.4\)}}



\put(42,01.3){\makebox(0,0)[bl]{\((1,2,0)\)}}

\end{picture}
\end{center}

  Some solutions for multiple choice problem with multiset estimates
 and
 integrated multiset estimate for the solution
  (constraint: \(\sum_{j=1}^{q_{i}} x_{ij} =  1\))  are:

 (a) \( S^{I}(b=7) = \{(1,2), (2,1), (3,2), (4,2) \}\),
  \(e^{3,12}(S^{I}(b=7)) = (5,5,2)\);

 (b) \( S^{I}(b=8) = \{(1,2), (2,1), (3,3), (4,2) \}\),
 \(e^{3,12}(S^{I}(b=8)) = (5,7,0)\); and

 (c) \( S^{I}(b=11.4) = \{(1,2), (2,1), (3,1), (4,3) \}\),
 \(e^{3,12}(S^{I}(b=11.4)) = (7,5,0)\).


  Some solutions for multiple choice problem with multiset estimates
 and ``generalized median'' estimate for the solution
  (constraint: \(\sum_{j=1}^{q_{i}} x_{ij} =  1\))  are:

 (a) \( S^{g}(b=7) = \{(1,2), (2,1), (3,2), (4,2) \}\),
  \(e^{3,3}(S^{g}(b=7)) = (1,2,0)\)
 (two ``generalized median'' estimates exist:
 \((1,2,0)\) and  \((0,3,0)\));

 (b) \( S^{g}(b=8) = \{(1,2), (2,1), (3,3), (4,2) \}\),
 \(e^{3,3}(S^{g}(b=8)) = (2,1,0)\)
 (two ``generalized median'' estimates exist:
 \((2,1,0)\) and  \((1,2,0)\));
 and

 (c) \( S^{g}(b=11.4) = \{(1,2), (2,1), (3,1), (4,3) \}\),
 \(e^{3,3}(S^{g}(b=11.4)) = (2,1,0)\).

 Some solutions for multiple choice problem with multiset estimates
 and ``set median'' estimate for the solution
  (constraint: \(\sum_{j=1}^{q_{i}} x_{ij} =  1\)) are:

 (a) \( S^{s}(b=7) = \{(1,2), (2,1), (3,2), (4,2) \}\),
 \(e^{3,3}(S^{s}(b=7)) = (2,1,0)\)

 (two ``median set'' estimates exist:
 \((0,3,0)\) and  \((2,1,0)\));

 (b) \( S^{s}(b=8) = \{(1,2), (2,1), (3,3), (4,2) \}\),
 \(e^{3,3}(S^{s}(b=8)) = (2,1,0)\)
 (two ``median set'' estimates exist:
 \((0,3,0)\) and  \((2,1,0)\));
 and

  (c) \( S^{s}(b=11.4) = \{(1,2), (2,1), (3,1), (4,3) \}\),
 \(e^{3,3}(S^{s}(b=11.4)) = (2,1,0)\).

 In other cases
 when the condition
 in knapsack-like problem
  is an inequality
 (e.g., \(  ~~\sum_{j=1}^{q_{i}} x_{ij} \leq  1 \)),
 it is necessary to consider
 an integrated multiset estimate
 for the solution.
%
%
 Here it is reasonable to describe a special approach to integration of
 different assessment multiset problems:
 \[\{ P^{l,\eta_{1}},...,P^{l,\eta_{i}},...,P^{l,\eta_{m}} \} \Longrightarrow P^{l,\eta} \]
 where \(\eta = \max_{i=\overline{1,n}} \{\eta_{i}\}\).
 First, poset for assessment problem  \(P^{l,\eta} \)
 is extended by addition of posets for problems
 ~\(P^{l,\eta_{i}} ~~ \forall \eta_{i} < \eta \).
 Secondly, the  preference rules of the following type are added
 (for comparison of multiset estimates from different posets above):
%
%
%
 \[ (\overbrace{0,...,0}^{p_{1}},1,\overbrace{0,...0}^{p_{2}})
 \succ
 (\overbrace{0,...,0}^{p_{1}+1},\beta,\overbrace{\zeta_{1},...,\zeta_{p_{3}}}^{p_{3}}),\]
 where \(p_{1} + 1 + p_{2} \leq \l\),
 \(p_{1} + 2 + p_{3} \leq \l\), \(\beta \geq 1 \),
 \(\zeta_{\kappa} \geq  0  ~~\forall \kappa = \overline{1,p_{3}}\).
 The described approach is targeted to comparison of multiset
 estimates with different numbers of elements: \(\eta_{1} \neq
 \eta_{2}\). This is significant for basic knapsack problem and
 multiple choice problem because the number of elements in
 different solutions may be different.

~~

 {\bf Example 20.}

 (a) \(e^{3,2}_{1} = (1,1,0)\),
  \(e^{3,3}_{2} = (0,0,3)\), ~~ \(e^{3,2}_{1}  \succ e^{3,3}_{2}\).

 (b) \(e^{3,1}_{3} = (1,0,0)\),
  \(e^{3,3}_{4} = (0,3,0)\), ~~ \(e^{3,1}_{3}  \succ e^{3,3}_{4}\).

 (c) \(e^{3,1}_{5} = (0,1,0)\),
  \(e^{3,5}_{6} = (0,0,5)\), ~~ \(e^{3,1}_{5}  \succ e^{3,5}_{6}\).

 (d) \(e^{3,1}_{7} = (1,0,0)\),
  \(e^{3,5}_{8} = (0,3,2)\), ~~ \(e^{3,1}_{7}  \succ e^{3,5}_{8}\).

 (e) \(e^{3,2}_{9} = (0,2,0)\),
  \(e^{3,5}_{10} = (0,1,5)\), ~~ \(e^{3,1}_{9}  \succ e^{3,3}_{10}\).

~~

  Further, basic multiple choice problem with multiset estimates is:
%
%
 \[ \max~ e(S) = \biguplus_{(i,j) \in S=\{(i,j)| x_{i,j}=1\}} ~  e_{i,j} ,\]
 \[ s.t. ~~ \sum_{i=1}^{m} \sum_{j=1}^{q_{i}}   a_{ij} x_{i,j} \leq b;
 ~~ \sum_{j=1}^{q_{i}} x_{ij} \leq  1;
%
%
 ~~ x_{ij} \in \{0, 1\}.\]
 Here solutions for multiple choice problem with multiset estimates
 and the integrated estimate for the solution
  (constraint: \(\sum_{j=1}^{q_{i}} x_{ij} \leq  1\))  are
  (Table 9):

 (a) \( S(b=5.1) = \{(1,2), (2,1) \}\),
 \(e^{I}(S(b=5.1)) = (5,1,0)\);

 (b) \( S(b=8.4) = \{(1,2), (2,1), (4,3) \}\),
  \(e^{I}(S(b=8.4)) = (7,2,0)\).

  Note,
   \(e^{I}(S(b=8.4)) \succ   e^{I}(S(b=5.1))  \).
 Here
 an estimate of computational complexity
 for a dynamic programming method
 is as follows:
 \[O(q_{1}~ \mu^{l,\eta}) +  O(q_{2}~ (\mu^{l,2\eta} + \mu^{l,\eta} )) + ...
 + O(q_{m}~ (\mu^{l,(m-1)\eta} + \mu^{l,(m-2)\eta} + ... + \mu^{l,\eta} )
 \leq \]
 \[O(m^{2} ~ \max_{i=\overline{1,m}} \{q_{i}\} ~  \mu^{l,(m-1)\eta}).\]
%
%
%
 Knapsack problem with multiset estimates
 and the integrated estimate for the solution is
 (solution \(S = \{ i | x_{i} = 1\}\)):
%
 \[ \max~ e(S) = \biguplus_{ i \in S=\{ i | x_{i}=1\}} ~  e_{i},\]
 \[  s.t. ~~ \sum_{i=1}^{m}    a_{i} x_{i} \leq b;
%
%
 ~~ x_{i} \in \{0, 1\}.\]
 A numerical example for knapsack problem is presented
 in Table 10.

\begin{center}
\begin{picture}(61,58)
\put(07.5,53){\makebox(0,0)[bl] {Table 10.
 Knapsack problem}}


\put(00,00){\line(1,0){61}} \put(00,37){\line(1,0){61}}
\put(10,44){\line(1,0){51}} \put(00,51){\line(1,0){61}}

\put(00,00){\line(0,1){51}} \put(10,00){\line(0,1){51}}

\put(22,00){\line(0,1){44}}


\put(34,00){\line(0,1){51}}


\put(61,00){\line(0,1){51}}


\put(04.6,45.5){\makebox(0,0)[bl]{\(i\) }}


\put(11,46){\makebox(0,0)[bl]{Basic problem}}
\put(15,38.5){\makebox(0,0)[bl]{\(a_{i}\) }}
\put(27,38.5){\makebox(0,0)[bl]{\(c_{i}\)}}


\put(35,46.4){\makebox(0,0)[bl]{Assessment \(P^{3,3}\)}}
\put(45,38.2){\makebox(0,0)[bl]{\(e^{3,3}_{i}\)}}



\put(04.2,32){\makebox(0,0)[bl]{\(1\)}}

\put(14,32){\makebox(0,0)[bl]{\(1.3\)}}
\put(26,32){\makebox(0,0)[bl]{\(3.4\)}}



\put(42,31.3){\makebox(0,0)[bl]{\((1,1,1)\)}}


\put(04.2,27){\makebox(0,0)[bl]{\(2\)}}

\put(14,27){\makebox(0,0)[bl]{\(3.1\)}}
\put(26,27){\makebox(0,0)[bl]{\(7.9\)}}



\put(42,26.3){\makebox(0,0)[bl]{\((3,0,0)\)}}


\put(04.2,22){\makebox(0,0)[bl]{\(3\)}}

\put(14,22){\makebox(0,0)[bl]{\(0.7\)}}
\put(26,22){\makebox(0,0)[bl]{\(1.3\)}}



\put(42,21.3){\makebox(0,0)[bl]{\((0,1,2)\)}}


\put(04.2,17){\makebox(0,0)[bl]{\(4\)}}

\put(14,17){\makebox(0,0)[bl]{\(2.0\)}}
\put(26,17){\makebox(0,0)[bl]{\(4.1\)}}



\put(42,16.3){\makebox(0,0)[bl]{\((1,2,0)\)}}


\put(04.2,12){\makebox(0,0)[bl]{\(5\)}}

\put(14,12){\makebox(0,0)[bl]{\(1.3\)}}
\put(26,12){\makebox(0,0)[bl]{\(2.3\)}}



\put(42,11.3){\makebox(0,0)[bl]{\((0,2,1)\)}}


\put(04.2,7){\makebox(0,0)[bl]{\(6\)}}

\put(14,07){\makebox(0,0)[bl]{\(3.0\)}}
\put(26,07){\makebox(0,0)[bl]{\(5.6\)}}



\put(42,6.3){\makebox(0,0)[bl]{\((2,1,0)\)}}


\put(04.2,2){\makebox(0,0)[bl]{\(7\)}}

\put(14,2){\makebox(0,0)[bl]{\(1.3\)}}
\put(26,2){\makebox(0,0)[bl]{\(2.8\)}}



\put(42,1.3){\makebox(0,0)[bl]{\((0,3,0)\)}}

\end{picture}
\end{center}

 Some solutions for basic knapsack problem
  (constraint: \(\sum_{j=1}^{q_{i}} x_{ij} \leq  1\)) are:

 (a) \( S(b=5.1) = \{1,2,5 \}\),
 \(c(S(b=5.1)) = 13.2\);

 (b) \( S(b=5.7) = \{1,2,7 \}\),
 \(c(S(b=5.7)) = 14.1\);

 (c) \( S(b=7.7) = \{1,2,4,7 \}\),
  \(c(S(b=7.7)) = 18.2\);
 and

 (d) \( S(b=8.4) = \{1,2,3,4,7 \}\),
  \(c(S(b=8.4)) = 19.5\).

    Here
   \(e^{I}(S(b=8.4)) >   e^{I}(S(b=7.7)) >
     e^{I}(S(b=5.7)) >   e^{I}(S(b=5.1)) \).

 Some solutions for knapsack problem with multiset estimates
 and the integrated estimate for the solution
  (constraint: \(\sum_{j=1}^{q_{i}} x_{ij} \leq  1\)) are:

 (a) \( S(b=5.1) = \{1,2,5 \}\),
 \(e^{I}(S(b=5.1)) = (4,3,2)\);

 (b) \( S(b=5.7) = \{1,2,7 \}\),
  \(e^{I}(S(b=5.7)) = (4,4,1)\);

 (c) \( S(b=7.7) = \{1,2,4,7 \}\),
   \(e^{I}(S(b=7.7)) = (5,6,1)\);
   and

 (c) \( S(b=8.4) = \{1,2,3,4,7 \}\),
  \(e^{I}(S(b=8.4)) = (5,7,3)\).

  Note,
   \(e^{I}(S(b=8.4)) \succ   e^{I}(S(b=7.7)) \succ
     e^{I}(S(b=5.7)) \succ   e^{I}(S(b=5.1)) \).
 Here
 an estimate of computational complexity
 for a dynamic programming method
 is as follows:
 \[O(\mu^{l,\eta}) +  O((\mu^{l,2\eta} + \mu^{l,\eta} )) + ...
 + O((\mu^{l,(m-1)\eta} + \mu^{l,(m-2)\eta} + ... + \mu^{l,\eta} )
 \leq
   O(m^{2}   \mu^{l,(m-1)\eta}).\]
 Clearly, multiset estimates can be used for
 resource constraint(s) as well.



\section{Multiset Estimates and Multi-attribute Alternatives}

 Let \(\A = \{A_{1},...,A_{i},...,A_{n}\}\)
 be the set of alternatives,
  \(C = \{C_{1},...,C_{j},...,C_{m}\}\)
 be the set of attributes (criteria),
 \(e^{l,\eta}_{i,j}\)
 be the multiset estimate of alternative \(A_{i}\)
 upon criterion \(C_{j}\)
 (i.e., assessment problem
 \(P^{1,\eta}\)
  is used at all processing stages).
 Thus, the estimate vector for alternative \(A_{i}\) is:
 \[ ( e^{l,\eta}_{i,1} , ... ,   e^{l,\eta}_{i,j} , ...,e^{l,\eta}_{i,m}   ).\]

 {\bf Definition 1.}
 The aggregated multiset estimate for alternative \(A_{i}\) is
 (``median'' \(M\) is considered as ``generalized median'' or ``set median''):
%
%
%
 \[e^{l,\eta}_{M} (A_{i}) = M =   \underset{M \in D}{\operatorname{argmin}}~
 | \biguplus_{j=1}^{m} ~ \delta (M, e^{l,\eta}_{i,j} ) ~|.\]
%
%
%
%


 {\bf Definition 2.}
 The vector estimate for median alternative  \(A_{M}\) is defined as follows
 (``median'' \(M\) is considered as ``generalized median'' or ``set median''):
%
%
%
 \[ ( e^{l,\eta}_{M,1} , ... ,   e^{l,\eta}_{M,j} , ...,e^{l,\eta}_{M,m}   ),\]
%
 where
 \[e^{l,\eta}_{M,j} =   \underset{M \in D}{\operatorname{argmin}}~
 | \biguplus_{i=1}^{m} ~ \delta (M, e^{l,\eta}_{i,j}) ~|.\]
%
%
%
%

  The aggregated multiset estimate for alternative \(A_{i}\)
 (\(e^{l,\eta}_{M} (A_{i}) \))
 and
 the median alternative  \(A_{M}\)
 can be used at
 preliminary stages in combinatorial synthesis.


 Application of multiset estimates in real-world
 problems requires taking into account the following basic requirements:

 {\it 1.} correspondence to applied problem (and evaluation processes),

 {\it 2.} easy to use for domain experts,

 {\it 3.} limited computing complexity of the used computer
 procedures.

 In many applications, the suggested multiset estimates can be
 used as a simple approximation of traditional fuzzy-set based
 estimates (or histogram-like estimates).
 Let us consider assessment problem \(P^{6,4}\) as an example.
 Fig. 24 illustrates
 several estimates for problem
 \(P^{6,4}\) (assessment over scale \([1,2,3,4,5,6]\) with four
 elements.

 Here
``multiset coefficient'' or ``multiset number'' is
 (i.e., the number of estimates):
 \[ \mu^{4,6} =  \left( \left(  \begin{matrix}  6 \\
   4 \end{matrix} \right) \right) =
   \left(  \begin{matrix}  6+4-1 \\  4 \end{matrix} \right) = 126.  \]
 Evidently, the number (\(126\)) is the parameter for computational complexity
 and corresponding processing procedures will be sufficiently
 simple (``effective'').

\begin{center}
\begin{picture}(78,96)

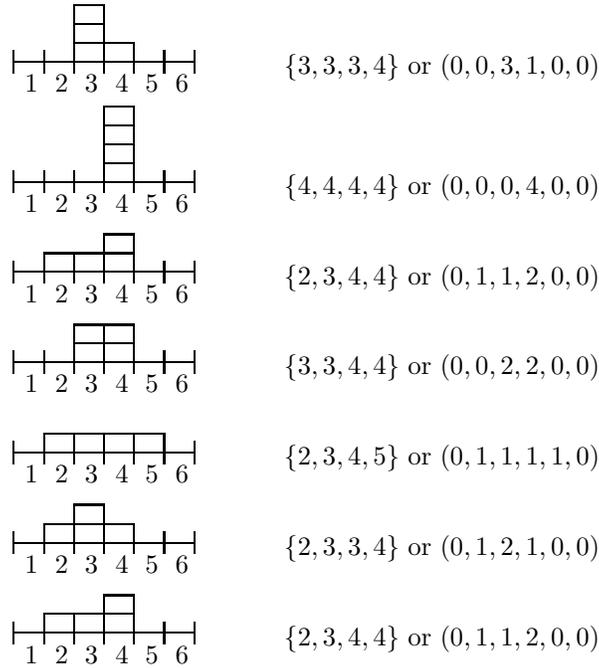
\put(00.5,00){\makebox(0,0)[bl] {Fig. 24. Examples of estimates
  (assessment \(P^{6,4}\))}}




\put(36,83){\makebox(0,0)[bl]{\(\{3,3,3,4\}\) or \((0,0,3,1,0,0)\)
}}

\put(08,85.5){\line(0,1){7.5}} \put(12,85.5){\line(0,1){7.5}}
\put(16,85.5){\line(0,1){2.5}}

\put(08,88){\line(1,0){08}} \put(08,90.5){\line(1,0){4}}
\put(08,93){\line(1,0){04}}

\put(00,85.5){\line(1,0){24}}

\put(00,84){\line(0,1){3}} \put(04,84){\line(0,1){3}}
\put(08,84){\line(0,1){3}} \put(12,84){\line(0,1){3}}
\put(16,84){\line(0,1){3}} \put(20,84){\line(0,1){3}}
\put(24,84){\line(0,1){3}}

\put(01.5,81.5){\makebox(0,0)[bl]{\(1\)}}
\put(05.5,81.5){\makebox(0,0)[bl]{\(2\)}}
\put(09.5,81.5){\makebox(0,0)[bl]{\(3\)}}
\put(13.5,81.5){\makebox(0,0)[bl]{\(4\)}}
\put(17.5,81.5){\makebox(0,0)[bl]{\(5\)}}
\put(21.5,81.5){\makebox(0,0)[bl]{\(6\)}}




\put(36,67){\makebox(0,0)[bl]{\(\{4,4,4,4\}\) or \((0,0,0,4,0,0)\)
}}

\put(12,69.5){\line(0,1){10}} \put(16,69.5){\line(0,1){10}}

\put(12,72){\line(1,0){04}} \put(12,74.5){\line(1,0){4}}
\put(12,77){\line(1,0){04}} \put(12,79.5){\line(1,0){4}}

\put(00,69.5){\line(1,0){24}}

\put(00,68){\line(0,1){3}} \put(04,68){\line(0,1){3}}
\put(08,68){\line(0,1){3}} \put(12,68){\line(0,1){3}}
\put(16,68){\line(0,1){3}} \put(20,68){\line(0,1){3}}
\put(24,68){\line(0,1){3}}

\put(01.5,65.5){\makebox(0,0)[bl]{\(1\)}}
\put(05.5,65.5){\makebox(0,0)[bl]{\(2\)}}
\put(09.5,65.5){\makebox(0,0)[bl]{\(3\)}}
\put(13.5,65.5){\makebox(0,0)[bl]{\(4\)}}
\put(17.5,65.5){\makebox(0,0)[bl]{\(5\)}}
\put(21.5,65.5){\makebox(0,0)[bl]{\(6\)}}




\put(36,55){\makebox(0,0)[bl]{\(\{2,3,4,4\}\) or \((0,1,1,2,0,0)\)
}}

\put(04,57.5){\line(0,1){2.5}} \put(08,57.5){\line(0,1){2.5}}
\put(04,60){\line(1,0){08}}

\put(12,57.5){\line(0,1){05}} \put(16,57.5){\line(0,1){05}}
\put(12,60){\line(1,0){04}} \put(12,62.5){\line(1,0){4}}

\put(00,57.5){\line(1,0){24}}

\put(00,56){\line(0,1){3}} \put(04,56){\line(0,1){3}}
\put(08,56){\line(0,1){3}} \put(12,56){\line(0,1){3}}
\put(16,56){\line(0,1){3}} \put(20,56){\line(0,1){3}}
\put(24,56){\line(0,1){3}}

\put(01.5,53.5){\makebox(0,0)[bl]{\(1\)}}
\put(05.5,53.5){\makebox(0,0)[bl]{\(2\)}}
\put(09.5,53.5){\makebox(0,0)[bl]{\(3\)}}
\put(13.5,53.5){\makebox(0,0)[bl]{\(4\)}}
\put(17.5,53.5){\makebox(0,0)[bl]{\(5\)}}
\put(21.5,53.5){\makebox(0,0)[bl]{\(6\)}}





\put(36,43){\makebox(0,0)[bl]{\(\{3,3,4,4\}\) or \((0,0,2,2,0,0)\)
}}

\put(08,45.5){\line(0,1){05}} \put(12,45.5){\line(0,1){05}}
\put(16,45.5){\line(0,1){05}}

\put(08,48){\line(1,0){8}} \put(08,50.5){\line(1,0){8}}

\put(00,45.5){\line(1,0){24}}

\put(00,44){\line(0,1){3}} \put(04,44){\line(0,1){3}}
\put(08,44){\line(0,1){3}} \put(12,44){\line(0,1){3}}
\put(16,44){\line(0,1){3}} \put(20,44){\line(0,1){3}}
\put(24,44){\line(0,1){3}}

\put(01.5,41.5){\makebox(0,0)[bl]{\(1\)}}
\put(05.5,41.5){\makebox(0,0)[bl]{\(2\)}}
\put(09.5,41.5){\makebox(0,0)[bl]{\(3\)}}
\put(13.5,41.5){\makebox(0,0)[bl]{\(4\)}}
\put(17.5,41.5){\makebox(0,0)[bl]{\(5\)}}
\put(21.5,41.5){\makebox(0,0)[bl]{\(6\)}}





\put(36,31){\makebox(0,0)[bl]{\(\{2,3,4,5\}\) or \((0,1,1,1,1,0)\)
}}

\put(04,33.5){\line(0,1){2.5}} \put(08,33.5){\line(0,1){02.5}}
\put(12,33.5){\line(0,1){02.5}} \put(16,33.5){\line(0,1){02.5}}
\put(20,33.5){\line(0,1){02.5}}

\put(04,36){\line(1,0){16}}

\put(00,33.5){\line(1,0){24}}

\put(00,32){\line(0,1){3}} \put(04,32){\line(0,1){3}}
\put(08,32){\line(0,1){3}} \put(12,32){\line(0,1){3}}
\put(16,32){\line(0,1){3}} \put(20,32){\line(0,1){3}}
\put(24,32){\line(0,1){3}}

\put(01.5,29.5){\makebox(0,0)[bl]{\(1\)}}
\put(05.5,29.5){\makebox(0,0)[bl]{\(2\)}}
\put(09.5,29.5){\makebox(0,0)[bl]{\(3\)}}
\put(13.5,29.5){\makebox(0,0)[bl]{\(4\)}}
\put(17.5,29.5){\makebox(0,0)[bl]{\(5\)}}
\put(21.5,29.5){\makebox(0,0)[bl]{\(6\)}}





\put(36,19){\makebox(0,0)[bl]{\(\{2,3,3,4\}\) or \((0,1,2,1,0,0)\)
}}

\put(04,21.5){\line(0,1){02.5}} \put(08,21.5){\line(0,1){05}}
\put(12,21.5){\line(0,1){05}} \put(16,21.5){\line(0,1){02.5}}

\put(04,24){\line(1,0){12}} \put(08,26.5){\line(1,0){4}}

\put(00,21.5){\line(1,0){24}}

\put(00,20){\line(0,1){3}} \put(04,20){\line(0,1){3}}
\put(08,20){\line(0,1){3}} \put(12,20){\line(0,1){3}}
\put(16,20){\line(0,1){3}} \put(20,20){\line(0,1){3}}
\put(24,20){\line(0,1){3}}

\put(01.5,17.5){\makebox(0,0)[bl]{\(1\)}}
\put(05.5,17.5){\makebox(0,0)[bl]{\(2\)}}
\put(09.5,17.5){\makebox(0,0)[bl]{\(3\)}}
\put(13.5,17.5){\makebox(0,0)[bl]{\(4\)}}
\put(17.5,17.5){\makebox(0,0)[bl]{\(5\)}}
\put(21.5,17.5){\makebox(0,0)[bl]{\(6\)}}






\put(36,7){\makebox(0,0)[bl]{\(\{2,3,4,4\}\) or \((0,1,1,2,0,0)\)
}}

\put(04,9.5){\line(0,1){02.5}} \put(08,9.5){\line(0,1){02.5}}
\put(12,9.5){\line(0,1){05}} \put(16,9.5){\line(0,1){05}}

\put(04,12){\line(1,0){12}} \put(12,14.5){\line(1,0){4}}

\put(00,9.5){\line(1,0){24}}

\put(00,08){\line(0,1){3}} \put(04,08){\line(0,1){3}}
\put(08,08){\line(0,1){3}} \put(12,08){\line(0,1){3}}
\put(16,08){\line(0,1){3}} \put(20,08){\line(0,1){3}}
\put(24,08){\line(0,1){3}}

\put(01.5,5.5){\makebox(0,0)[bl]{\(1\)}}
\put(05.5,5.5){\makebox(0,0)[bl]{\(2\)}}
\put(09.5,5.5){\makebox(0,0)[bl]{\(3\)}}
\put(13.5,5.5){\makebox(0,0)[bl]{\(4\)}}
\put(17.5,5.5){\makebox(0,0)[bl]{\(5\)}}
\put(21.5,5.5){\makebox(0,0)[bl]{\(6\)}}


\end{picture}
\end{center}

  \section{Conclusion}

 In the article, an approach to assessment of alternatives based on
 assignment of elements into an ordinal scale
 has been suggested.
 This kind of assessment corresponds to usage of
 special multiset based estimates (cardinality of
 multiset-estimate is a constant).
 Concurrently, the assessment approach provides
 evaluation from the viewpoint of uncertainty.
%
 Our evaluation process is based on an assignment procedure:
 for each alternative several elements (e.g., 1,2,3)
 are assigned
 to levels of an ordinal scale [1,2,...,l].
%
 Here, the case of one element corresponds
 to well-known ordinal assessment.
 Several basic assessment problems are described:
 (i) traditional assessment on ordinal scale \([1,2,3]\)
 ~ (\(P^{3,1}\));
 (ii) traditional assessment on ordinal scale \([1,2,3,4]\)
 ~ (\(Pi^{4,1}\));
 (iii)  assessment on ordinal scale \([1,2,3]\)
 based on assignment of two elements
 ~ (\(P^{3,2}\));
 (iv)  assessment on ordinal scale \([1,2,3]\)
 based on assignment of three elements
 ~ (\(P^{3,3}\));
 and
 (v)  assessment on ordinal scale \([1,2,3]\)
 based on assignment of four elements ~ (\(P^{3,4}\)).
 For each assessment problem above,
 the following issues are examined:
 assessment scale,
 order over the scale components (poset).
%
%
 In addition, ``interval multiset estimates''
 are suggested.

 For the suggested  multiset estimates,
 operations over the estimates are examined:
 (a) integration,
 (b) proximity,
 (c) comparison,
 (d) aggregation (e.g., by computing a median estimate),
 and
 (e) alignment.
%
%
%
  The suggested assessment approach is used for
  combinatorial synthesis
  (morphological approach, knapsack-like problems).
 The assessment approach,
 multiset-estimates and corresponding problems
 are illustrated by numerical examples.

%
%

 In general, it is necessary to point out the following:

  {\it Note 1.}
  The suggested assessment approach is based on
  expert judgment and/or computation procedures
  (including interactive mode and information visualization).

 {\it Note 2.}
 The considered simplified versions
 of the assessment problem based on small numbers of
 elements and levels of used ordinal scale
 are very useful for
 data presentation/visualization
 and
 are often sufficient for many applications.
%

 {\it Note 3.}
 The described assessment methods are very understandable and useful
 for domain experts.
 A similar assessment scale
 is widely used in financial engineering for
 evaluation of financial institutions
 (a version of the scale):~
 \('AAA'\), \('AAB'\), \('ABB'\), \('BBB'\), \('BBC'\), \('BCD'\), etc.
 This corresponds to assessment problem \(P^{3,4}\)
  with scale \([A,B,C,D]\)
 (\( A \succ B \succ  C \succ  D \)).

  {\it Note 4.}
  The usage of the suggested assessment approach in composition
  problem
  leads to the extended version of HMMD method
  \cite{lev98}.

 {\it Note 5.}
 The suggested approach can be considered as an
 approximation of traditional fuzzy set based assessment procedures.
 As a result, combinatorial estimates processing procedures
 based on posets (or lattices) can be used (presentation, computation).


%
 In the future, the following research directions can be considered:

 {\it 1.}
 Usage of the suggested multiset assessment approach in decision support systems
 (including special visualization support subsystem and
 special human-computer interface).

  {\it 2.}
 Comparison of the suggested approach and approaches,
 based on fuzzy sets
 (including aspects of computation and human-computer
 interaction).

 {\it 3.} Usage of the suggested multiset approach
  for assessment of elements/alternatives
 in decision making problems:
 (i) sorting problems
 (e.g., \cite{levmih88},\cite{pet08},\cite{roy96},\cite{zap02});
%
 (ii) clustering, classification
 (e.g.,
 \cite{jain99},\cite{lev07clust},\cite{ped05},\cite{pet08},\cite{van77},\cite{mirkin05},\cite{mur93}).

 {\it 4.}
 Analyzing the usage of the suggested multiset assessment
 in design and decision making procedures:
 (a) quality of the obtained solutions,
 (b) complexity of the solving schemes.

 {\it 5.}
 Special studies of algorithms and their computational complexity
 for considered combinatorial synthesis
 (e.g., combinatorial synthesis based on morphological approach
 and combinatorial synthesis based on knapsack-like problems)
 (e.g., \cite{lev98},\cite{lev06},\cite{lev09},\cite{lev12morph})
 based on the suggested multiset assessment.
 Here
 the attention can be targeted
 to dynamic programming procedures.
 Note,
 analogues of FPTAS for knapsack-like problems
 (including intervals-based methods)
 (e.g., \cite{gens78},\cite{kellerer03},\cite{lev81},\cite{pruhs07},\cite{sahni75},\cite{sahni76},\cite{sahni78},\cite{war87},\cite{woe00})
 can be successfully
 used for combinatorial synthesis (knapsack-like problems)
 in case of integrated multiset estimates for solutions.
 On the other hand,
 various hybrid heuristics and man-machine procedures
 have to be widely used
 for the suggested versions
 of combinatorial synthesis problems.

  {\it 6.}
  Usage of the suggested approach
  for assessment of solution components
 in combinatorial optimization
 (e.g., scheduling problems, travelling salesman problem,
 routing problems,
 tree spanning problems,
 assignment problems,
 graph coloring problems)
 (e.g., \cite{gar79},\cite{lev98},\cite{papa06}).

 {\it 7.}
 Conducting some special psychological studies
 of the suggested multiset assessment approach.



\thebibliography{300}

 \bibitem {alk11} S. Alkhazaleh, A.R. Salleh, N. Hassan,
 Soft multisets theory.
 {\it Applied Mathematical Sciences},
 5(72), 3561-3573, 2011.

  \bibitem {gar79} M.R. Garey, D.S. Johnson,
  {\it Computers and Intractability. The
  Guide to the Theory of NP-Completeness}, San Francisco,
  W.H.Freeman and Company, 1979.

 \bibitem {gens78} G.V. Gens, E.V. Levner,
 Approximation algorithms for certain universal problems in scheduling theory.
 {\it Engineering Cybernetics}, 16(6), 31-36, 1978.


  \bibitem {jain99} A.K. Jain, M.N. Murty, P.J. Flynn,
 Data clustering: a review.
 {\it ACM Computing Surveys}, 31(3), 264–323, 1999.

 \bibitem {kellerer03} H. Kellerer, R. Mansini, U. Pferschy,
 M.G. Speranza,
 An efficient fully polynomial approximation scheme for the
 Sub-Sum Problem.
 {\it J. of Computer and System Sciences},
 66(2), 349-370, 2003.

  \bibitem {kellerer04} H. Kellerer, U. Pferschy, D. Pisinger,
 {\it Knapsack Problems}, Berlin, Springer, 2004.

 \bibitem {knuth98} D.E. Knuth,
 {\it The Art of Computer Programming.
 Vol. 2, Seminumerical Algorithms.}
 Addison Wesley, Reading, 1998.

 \bibitem {lev81} M.Sh. Levin,
 An extremal problem of organization of data.
 {\it Engineering Cybernetics}, 19(5), 87-95, 1981.

 \bibitem {lev96} M.Sh. Levin,
  Hierarchical morphological multicriteria design of decomposable systems.
  {\it Concurrent Engineering: Research and Applications}, 4(2),
  111-117, 1996.

 \bibitem {lev98a} M.Sh. Levin,
 Towards comparison of decomposable systems.
 In:
  {\it Data Science, Classification, and Related Methods},
  Springer, Tokyo, 154-161, 1998.

  \bibitem{lev98} M.Sh. Levin,
 {\it Combinatorial Engineering of Decomposable Systems}.
 Dordrecht, Kluwer Academic Publishers, 1998.

  \bibitem {lev01} M.Sh. Levin,
  System synthesis with morphological clique problem:
  fusion of subsystem evaluation decisions,
  {\it Inform. Fusion}, 2(3), 225-237, 2001.


 \bibitem {lev02} M.Sh. Levin,
  Towards combinatorial planning of human-computer systems,
  {\it Applied Intelligence}, 16(3), 235-247, 2002.


 \bibitem {lev05} M.Sh. Levin,
 Modular system synthesis: example for composite packaged software,
 {\it IEEE Trans. on SMC - Part C}, 35(4), 544-553, 2005.

 \bibitem {lev06} M.Sh. Levin,
 {\em Composite Systems Decisions}, New York, Springer, 2006.



 \bibitem {lev07clust} M.Sh. Levin,
 Towards hierarchical clustering,
 In: V. Diekert, M. Volkov, A. Voronkov, (Eds.),
 {\it CSR 2007}, LNCS 4649, Springer, 205-215, 2007.


 \bibitem {lev09} M.Sh. Levin,
 Combinatorial optimization in system configuration design,
 {\it Automation and Remote Control}, 70(3), 519-561, 2009.





   \bibitem{lev11agg} M.Sh. Levin,
  Aggregation of composite solutions: strategies, models, examples.
   Electronic
  preprint. 72 pp., Nov. 29, 2011.
  http://arxiv.org/abs/1111.6983 [cs.SE]

   \bibitem{lev12morph} M.Sh. Levin,
 Morphological methods for design of
 modular systems (a survey)
  Electronic  preprint. 20 pp., Jan. 9, 2012.
  http://arxiv.org/abs/1201.1712 [cs.SE]

%

 \bibitem {levmih88}  M.Sh. Levin, A.A. Mikhailov,
 {\it Fragments of Objects Set Stratification Technology},
  Preprint, Moscow, Inst. for Systems
 Studies (RAS), 60 p., 1988 (in Russian)

 \bibitem {mirkin05} B.G. Mirkin,
 {\it Clustering for Data Mining: A Data Recovery Approach}.
 Chapman \& Hall / CRC, 2005.

 \bibitem {mur93} F. Murtagh,
  A survey of recent advances in hierarchical clustering algorithms.
  {\it The Computer Journal}, 26(), 354–359, 1993.

 \bibitem {papa06} C.H. Papadimitriou, K. Steiglitz,
 {\it Combinatorial Optimization: algorithms and complexity}.
 Dover Pubns, 1998.

 \bibitem {ped05} W. Pedrycz,
 {\it Knowledge-Based Clustering}. Wiley, 2005.

 \bibitem {pet08} A.B. Petrovsky,
  Clustering and sorting multi-attribute objects in multiset
  metric space.
  {\it 4th Int. Conf. Intelligent Systems IS 2008},
   1144-1148, 2008.

 \bibitem {pruhs07} K. Pruhs, G.J. Woeginger,
 Approximation schemes for a class of subset selection problems.
 {\it Theoretical Computer Science},
 382(2), 151-156, 2007.


  \bibitem {roy96} B. Roy,
  {\it Multicriteria Methodology for Decision Aiding},
  Dordrecht, Kluwer Academic Publishers, 1996.

 \bibitem {sahni75} S. Sahni,
 Approximate algorithms for 0/1 knapsack problem.
 {\it J. of the ACM},
 22(1), 115-124, 1975.

  \bibitem {sahni76} S. Sahni,
 Algorithms for scheduling independent tasks.
 {\it J. of the ACM}, 23(1), 116-127, 1976.

  \bibitem {sahni78} S. Sahni, E. Horowitz,
  Combinatorial problems: reducibility and approximation.
 {\it Oper. Res.},
 26(5), 718-759, 1978.

  \bibitem {solnon07}
  C. Solnon, J.-M. Jolion,
 Generalized vs set median string for histogram based distances:
 Algorithms and classification results in image domain.
 In: F. Escolano, M. Vento (Eds.),
 {\it Proc. of the 6th Workshop on Graph Based Representation
 in Pattern Recognition GbPrP 2007}, LNCS 4538, Springer,
 404-414, 2007.

  \bibitem {syr01} A. Syropoulos,
 Mathematics of multisets.
 In:
 C.S. Calude et. al. (Eds.),
 {\it Multiset Processing: Mathematical Computer Science,
 and Molecular Computing Points of View.}
 LNCS 2235, Springer, 347-358, 2001.

  \bibitem {van77} J. Van Ryzin, (ed.),
 {\it Classification and Clustering}. New York, Academic Press,
 1977.

 \bibitem {war87} A. Warburton,
 Approximation of pareto optima in multiobjective shortest path
 problem.
 {\it Oper. Res.}, 35, 70-79, 1987.

 \bibitem {woe00} G.J. Woeginger,
 When does a dynamic programming formulation guarantee
 the existence of a fully polynomial approximation scheme (FPTAS)?
 {\it INFORMS J. on Computing},
 12(1), 57-75, 2000.

 \bibitem {yager86} R.R. Yager,
 On the theory of bags.
 {\it Int. J. of General Systems}, 13(1), 23-37, 1986.

 \bibitem {zadeh65} L.A. Zadeh,
 Fuzzy sets.
 {\it Information and Control}, 8(3), 338-353, 1965.

%
%

 \bibitem {zap02} C. Zopounidis, M. Doumpos,
 Multicriteria classification and sorting methods:
  A literature review.
 {\it Eur. J. of Oper. Res.}, 138(2), 229-246, 2002.

 \bibitem {zim87} H.J. Zimmerman,
 {\it Fuzzy Sets, Decision Making, and Expert Systems.}
 Berlin, Springer, 1987.
\end{document}